\documentclass[prd,aps,10pt,nofootinbib,twocolumn,superscriptaddress,preprintnumbers,balancelastpage,longbibliography]{revtex4-1}

\usepackage{amsmath,amssymb}	
\usepackage{mathtools}
\usepackage{fontawesome}
\usepackage[dvipsnames]{xcolor}
\usepackage{hyperref}
\usepackage{xspace}
\usepackage{fancyhdr}
\usepackage{braket}
\usepackage{graphicx}
\usepackage{siunitx}

\definecolor{linkcolor}{rgb}{0.7752941176470588, 0.22078431372549023, 0.2262745098039215}

\newcommand{\nbicon}{{\color{linkcolor}\faFileCodeO}\xspace}
\newcommand{\nblink}[1]{\href{https://github.com/smsharma/dark-photons-perturbations/blob/apr-2020/notebooks/#1.ipynb}{\nbicon}}
\newcommand{\githubmaster}{\href{https://github.com/smsharma/dark-photons-perturbations/}{\faGithub}\xspace}

\newcommand{\dd}{\mathrm{d}}
\newcommand{\mAp}{m_{A^\prime}}

\definecolor{deepgreen}{rgb}{0.2,0.8,0.2}

\hypersetup{colorlinks=true,
linkcolor=linkcolor,
citecolor=linkcolor,
urlcolor=linkcolor,
linktocpage=true,
pdfproducer=medialab,
}

\defcitealias{OurLongPaper}{Paper~II}

\DeclareSIUnit \h {\ensuremath{\mathit{h}}}
\DeclareSIUnit\electronvolt{e\kern-.05em V}
\DeclareSIUnit\parsec{pc}

\begin{document}

\title{Dark Photon Oscillations in Our Inhomogeneous Universe}

\author{Andrea Caputo}
\email{andrea.caputo@uv.es}
\thanks{ORCID: \href{https://orcid.org/0000-0003-1122-6606}{0000-0003-1122-6606}}
\affiliation{Instituto de F\'{i}sica Corpuscular, CSIC-Universitat de Valencia, Apartado de Correos 22085, E-46071, Spain}

\author{Hongwan Liu}
\email{hongwanl@princeton.edu}
\thanks{ORCID: \href{https://orcid.org/0000-0003-2486-0681}{0000-0003-2486-0681}}
\affiliation{Center for Cosmology and Particle Physics, Department of Physics, New York University, New York, NY 10003, USA}
\affiliation{Department of Physics, Princeton University, Princeton, NJ 08544, USA}

\author{Siddharth Mishra-Sharma}
\email{sm8383@nyu.edu}
\thanks{ORCID: \href{https://orcid.org/0000-0001-9088-7845}{0000-0001-9088-7845}}
\affiliation{Center for Cosmology and Particle Physics, Department of Physics, New York University, New York, NY 10003, USA}

\author{Joshua T. Ruderman}
\email{ruderman@nyu.edu}
\thanks{ORCID: \href{https://orcid.org/0000-0001-6051-9216}{0000-0001-6051-9216}}
\affiliation{Center for Cosmology and Particle Physics, Department of Physics, New York University, New York, NY 10003, USA}

\date{\protect\today}

\begin{abstract}
A dark photon kinetically mixing with the ordinary photon represents one of the simplest viable extensions to the Standard Model, and would induce oscillations with observable imprints on cosmology. Oscillations are resonantly enhanced if the dark photon mass equals the ordinary photon plasma mass, which tracks the free electron number density.  Previous studies have assumed a homogeneous Universe; in this Letter, we introduce for the first time an analytic formalism for treating resonant oscillations in the presence of inhomogeneities of the photon plasma mass.  We apply our formalism to determine constraints from Cosmic Microwave Background photons oscillating into dark photons, and from heating of the primordial plasma due to dark photon dark matter converting into low-energy photons. Including the effect of inhomogeneities demonstrates that prior homogeneous constraints are not conservative, and simultaneously extends current experimental limits into a vast new parameter space. \githubmaster
\end{abstract}

\maketitle

\noindent
{\bf Introduction.---}A minimal extension of the Standard Model (SM) is a dark photon, $A'$, kinetically mixing~\cite{Holdom:1985ag} with the ordinary photon, $\gamma$. 
Kinetic mixing is one of a few portals allowing new physics to couple to the Standard Model through a dimensionless interaction that can be manifest at low energies.  Further motivations for dark photons is that they may constitute dark matter~\cite{Redondo:2008ec,Nelson:2011sf,Arias:2012az,Fradette:2014sza,An:2014twa,Graham:2015rva,Agrawal:2018vin,Dror:2018pdh,Co:2018lka,Bastero-Gil:2018uel,Long:2019lwl} and are ubiquitous in theories beyond the SM~\cite{Dienes:1996zr,Goodsell:2009xc,Goodsell:2010ie,Goodsell:2009pi,Abel:2003ue,Abel:2006qt,Abel:2008ai}. 
Very light dark photons decouple from experiments as a positive power of $\mAp / E_{\rm exp}$, where $\mAp$ is the mass of the dark photon and $E_{\rm exp}$ is the experimental energy scale.  Because of this decoupling behavior, light dark photons with sizable interactions are consistent with current experimental constraints.

Kinetic mixing induces oscillations of photons into dark photons, $\gamma \rightarrow A'$, as well as the reverse process, $A' \rightarrow \gamma$.  In the early Universe, given a redshift $z$ and position $\vec x$, the photon has a plasma mass, $m_\gamma(z, \vec x)$, that tracks the free electron number density,  $n_\mathrm{e}(z, \vec x)$.  
The oscillation probability is resonantly enhanced if the plasma mass equals the mass of the dark photon, $m_\gamma(z, \vec x) \approx \mAp$. We consider massive dark photons, with a mass in the interval $10^{-14} \lesssim \mAp \lesssim \SI{e-9}{\eV}$, the homogeneous (spatially averaged) value of the plasma mass, $\overline {m}_\gamma(z)$, crosses the dark photon mass after recombination.  In this regime there are powerful constraints~\cite{Mirizzi:2009iz,Kunze:2015noa} from $\gamma \rightarrow A'$ oscillations distorting the shape of the Cosmic Microwave Background (CMB) spectrum measured by FIRAS~\cite{Fixsen:1996nj}.  Dark matter composed of dark photons is also constrained by heating of the primordial plasma from resonant  $A' \rightarrow \gamma$ oscillations~\cite{McDermott:2019lch}, producing low-energy photons that are efficiently absorbed by baryons.  Additional constraints on dark photon dark matter are considered by Refs.~\cite{Arias:2012az,Dubovsky:2015cca,Kovetz:2018zes,Wadekar:2019xnf}.

\begin{figure*}[htbp]
    \centering
    \includegraphics[width=0.99\textwidth]{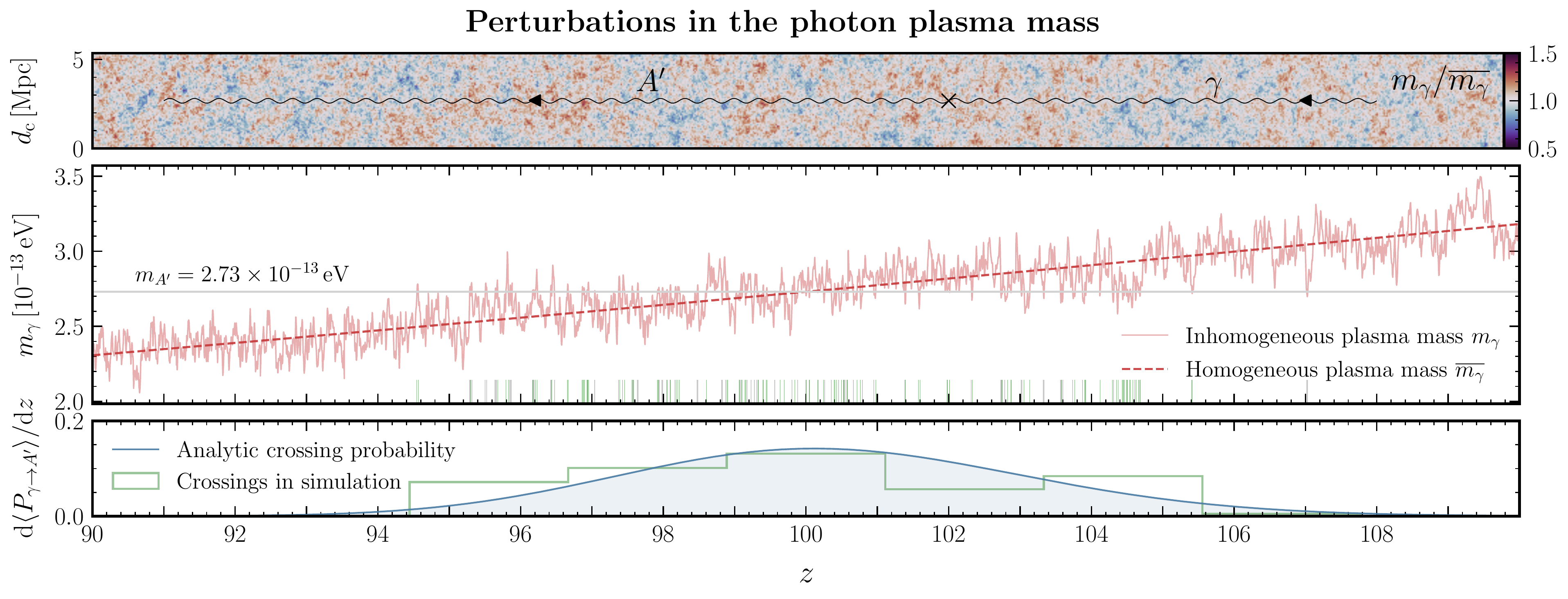}
    \caption{
    ({\it Top}) A simulated realization of the plasma mass in a box of comoving thickness 5\,Mpc, centered around $z=100$. An example photon path with conversion to a dark photon at the redshift marked ``$\times$'' is shown.  ({\it Middle}) A line-of-sight section through the perturbed plasma mass (solid red), as might be encountered by a traversing CMB photon, compared to the homogeneous plasma mass (dashed red). For a dark photon mass of $m_{A'} = \SI{2.73e-13}{\eV}$, the corresponding homogeneous transition occurs at $z\simeq 100$ where the plasma mass reaches $m_{A'}$ (gray). Multiple level crossings 
    are possible after accounting for perturbations---individual crossings in this realization are shown as the vertical green lines. ({\it Bottom}) The corresponding analytical differential conversion probability (blue) and a histogram of the crossing density corresponding to the specific realization shown (green).~\nblink{03_simulations}
    }
    \label{fig:simulations}
\end{figure*}

Previous studies of cosmological dark photon oscillations have assumed a homogeneous plasma mass, \emph{i.e.},\ $m_\gamma(z, \vec x) \approx \overline {m}_\gamma(z)$.  However, the plasma mass has perturbations that track inhomogeneities in the electron number density, which are a predicted consequence of the growth of structure in the early Universe.  Consider a photon that propagates along a worldline through the primordial plasma.  In the homogeneous limit, this photon may experience a level crossing at a specific redshift, $z_{\rm res}$, when $\overline {m}_\gamma(z_{\rm res}) \approx \mAp$.   In reality, a photon's path traverses regions with overdensities and underdensities, and may pass through many different level crossings at redshifts that differ from $z_{\rm res}$. Fig.~\ref{fig:simulations} shows a simulation of this process for a dark photon mass with $z_{\rm res} \approx 100$; perturbations in the plasma mass induce resonant conversions over a wide range of redshifts, $90 \lesssim z \lesssim 110$.  The effect of inhomogeneities is especially dramatic for dark photons with masses $\mAp \lesssim \SI{e-14}{\eV}$, which experience no level crossings in the homogeneous limit, but in reality can experience level crossings in regions with lower-than-average electron number density.

This Letter is part of a pair of companion papers in which we initiate the study of resonant oscillations between photons and dark photons in the presence of inhomogeneities in the photon plasma mass.  We introduce an analytic formalism for calculating the probability that photons or dark photons oscillate as they travel through the inhomogeneous plasma.  As applications of our formalism, we revisit bounds from the CMB spectrum on photons oscillating to dark photons, $\gamma \rightarrow A'$, and bounds from energy injection due to dark photon dark matter oscillating into ordinary photons, $A' \rightarrow \gamma $.  We find that these bounds require significant revision: compared to the homogeneous limit, perturbations both induce new resonances in underdensities and overdensities, extending these bounds into a vast new parameter space, and can also wash out resonances, making the sensitivity derived in the homogeneous approximation an overestimate for certain dark photon masses. The homogeneous limit is therefore not a conservative approximation of our inhomogeneous Universe.

Dark photons with masses $10^{-15} \lesssim \mAp \lesssim \SI{e-9}{\eV}$ are the target of several planned experiments using resonant detectors. DM Radio targets dark photon dark matter~\cite{Chaudhuri:2014dla,Silva-Feaver:2016qhh}, while Dark SRF~\cite{HarnikSRF,GrassellinoSRF} aims to produce and detect dark photons without assuming a cosmic abundance~\cite{Graham:2014sha}. 

In our companion paper,~\citetalias{OurLongPaper}, we examine the physics of oscillations in detail, giving a derivation of our formalism together with a complete description of the cosmological inputs that are required to derive the limits shown here. We also validate our analytical results with simulations of $\gamma \to A'$ oscillations.

The remainder of this Letter is organized as follows.  We begin by reviewing $\gamma \leftrightarrow A'$ oscillations.  We then introduce our analytic formalism for treating these oscillations in the presence of perturbations of the photon plasma mass.  Next, we apply our formalism to determine the constraints on $\gamma \rightarrow A'$ oscillations from FIRAS data.  We then show how inhomogeneities extend constraints on energy injection from dark photon dark matter to new dark photon masses.  Our conclusions highlight additional possible applications and extensions of our formalism. 
Throughout this work, we use units with $\hbar = c = k_\mathrm{B} = 1$, and the \textit{Planck} 2018 cosmology~\cite{Aghanim:2019ame}. For reproducibility, we provide links in the figure captions (\nbicon) pointing to the code used to generate them.

\noindent
{\bf Resonant photon-dark photon oscillations.---}We consider the following photon-dark photon Lagrangian, 
\begin{equation}
\mathcal{L}_{\gamma A'}=-\frac{1}{4}F_{\mu \nu}^{2}-\frac{1}{4}(F_{\mu \nu}^{\prime})^{2}-\frac{\epsilon}{2} F^{\mu \nu} F_{\mu \nu}^{\prime}+\frac{1}{2} \mAp^{2}(A_{\mu}^{\prime})^{2} \, ,
\end{equation}
where  $\epsilon$ is a dimensionless measure of kinetic mixing with typical ``natural'' values in the range $10^{-13}$--$10^{-2}$\,eV~\cite{Dienes:1996zr,Abel:2003ue,Abel:2006qt,Abel:2008ai,Gherghetta:2019coi}. $A'$ is the dark photon field, with $F$ and $F'$ representing the field strength tensor for the photon and dark photon respectively.

The propagation of CMB photons in the primordial plasma leads to in-medium effects that are described by a mass term, $m_\gamma$, in the photon dispersion relation.  There are positive and negative contributions to $m_{\gamma}^{2}$ from scattering off free electrons and neutral atoms, respectively~\cite{Kunze:2015noa,Mirizzi:2009iz}:
\begin{multline}
    {m_{\gamma}^2(z, \vec{x})} \simeq \, \SI{1.4e-21}{\eV\squared} \left(\frac{n_\mathrm{e}(z, \vec{x})}{\SI{}{\per\centi\meter\cubed}}\right)  \label{eq:m_gamma_sq}  \\
      - \SI{8.4e-24}{\eV\squared} \left( \frac{\omega(z)}{\SI{}{\eV}} \right)^2 \left(\frac{n_{\mathrm{HI}}(z, \vec{x})}{\SI{}{\per\centi\meter\cubed}}\right)  \,, 
\end{multline}
where $\omega(z)$ is the photon energy, and  $n_\mathrm{e}(z, \vec{x})$ and $n_\mathrm{HI}(z, \vec{x})$ represent the local free electron and neutral hydrogen densities. We model the evolution of cosmological quantities using CLASS~\cite{Blas:2011rf} interfaced with \texttt{HyRec}~\cite{AliHaimoud:2010dx}. For $\epsilon \ll 1$, $\gamma \to A'$ conversion is a resonant process that is efficient only when the dark photon mass is equal to the plasma mass; in this limit, we can apply the Landau-Zener approximation for non-adiabatic transitions~\cite{Mirizzi:2009iz, Kuo:1989qe, Parke:1986jy,OurLongPaper},
\begin{equation}
    P_{\gamma \rightarrow A'} \simeq \sum_i \frac{\pi \mAp^{2} \epsilon^{2}}{\omega(t_i)}\left|\frac{\dd\ln m_{\gamma}^{2}(t)}{\dd t}\right|_{t=t_i}^{-1} \!\!\!,
    \label{eq:Prob_homo}
\end{equation}
where $i$ indexes times $t_i$ when $m_\gamma^2(t_i) = \mAp^2$ and therefore the resonance condition is met. Eq.~\eqref{eq:Prob_homo} describes the probability that a photon will convert along its path, which depends on $m_\gamma(t)$ along this path. Eq.~\eqref{eq:Prob_homo} assumes $P_{\gamma \rightarrow A'} \ll 1$, which applies throughout this work. Similar results have also been derived in the context of neutrino oscillations in supernovae~\cite{Dasgupta:2005wn,Friedland:2006ta,Fogli:2006xy}.

\noindent
{\bf The effect of inhomogeneities.---}Inhomogeneities in the photon plasma mass substantially affect the conversion probability of photons into dark photons and vice versa, allowing for efficient oscillations over a range of cosmic times rather than at a single epoch.

In the presence of plasma mass inhomogeneities, we need to take the average of Eq.~\eqref{eq:Prob_homo} over different photon paths to account for transitions in locally overdense and underdense regions. This problem reduces to integrating over $m_\gamma^2$ at each point in time, weighted by the probability density function of finding a region with plasma mass $m_\gamma^2$. Our formalism draws from Rice's formula for the average number of level crossings of a random field~\cite{1944BSTJ...23..282R,lindgren2012stationary}. In~\citetalias{OurLongPaper}, we derive the following differential conversion probability
\begin{multline} 
    \label{eq:prob}
    \frac{\dd\langle P_{\gamma \to A'} \rangle}{\dd z} =  \frac{\pi \mAp^2 \epsilon^2}{\omega(t)} \left|\frac{\dd t}{\dd z}\right| \\ \times  \int \dd m_\gamma^2  \, f(m_\gamma^2;t) \, \delta_\mathrm{D}(m_\gamma^2 - \mAp^2) \, m_\gamma^2  \,,
\end{multline}
where $f(m_\gamma^2;t)$ is the probability density function (PDF) of $m_\gamma^2$ at time $t$, and $\delta_\mathrm{D}$ is the Dirac delta distribution. Neglecting perturbations in the free electron fraction $x_\mathrm{e}$ (see~\citetalias{OurLongPaper} for a discussion on why this assumption is valid here), Eq.~\eqref{eq:m_gamma_sq} shows that $m_\gamma^2(z,\vec{x}) \propto n_\mathrm{b}(z,\vec{x})$, where $n_\mathrm{b}$ is the baryon number density; this implies that
\begin{alignat}{1}
    f(m_\gamma^2; t) = \mathcal{P}(\delta_\mathrm{b};t)/\overline{m_\gamma^2} \,,
    \label{eq:one_point_pdf}
\end{alignat}
where $\mathcal{P}(\delta_\mathrm{b};t)$ is the one-point PDF of baryon density fluctuations $\delta_\mathrm{b} \equiv (n_\mathrm{b} - \overline{n}_\mathrm{b}) / \overline{n}_\mathrm{b}$ and $\overline{m_\gamma^2}$ the average squared plasma mass. Eq.~\eqref{eq:one_point_pdf} therefore ties the physics of $\gamma \leftrightarrow A'$ directly to a cosmological observable. The proportionality $m_\gamma^2 \propto n_\mathrm{b}$ together with the definition of $\delta_\mathrm{b}$ implies that that $1 + \delta_\mathrm{b} = m_\gamma^2 / \overline{m_\gamma^2}$.

\begin{figure}[htbp]
    \centering
    \includegraphics[width=0.47\textwidth]{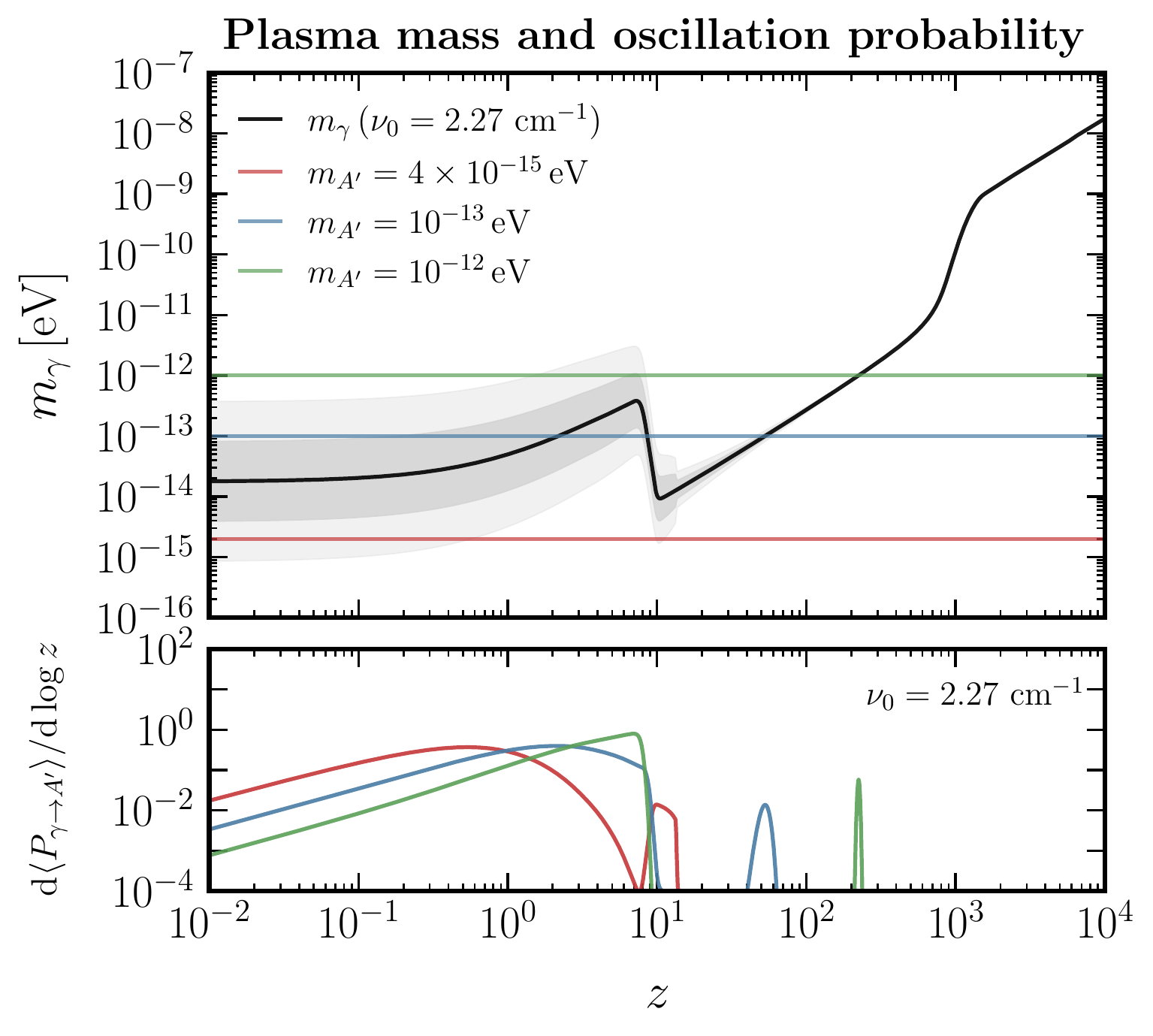}
    \caption{({\it Top}) The photon plasma mass as a function of redshift corresponding to the lowest-frequency FIRAS band ($\nu_0=\,$\SI{2.27}{\per\centi\meter}), which dominates the total conversion probability. The middle 68 and 95\% containment of plasma mass fluctuations in the log-normal prescription is shown in dark and light gray, respectively. Horizontal lines correspond to the fiducial mass points $\mAp=4\times10^{-15}$ (red), $10^{-13}$ (blue), and \SI{e-12}{\eV} (green), respectively. ({\it Bottom}) The differential resonant transition probability for this frequency as a function of redshift for the fiducial masses, normalized to unity total probability, showing efficient conversion probability over a wide range of redshifts.~\nblink{09_plasma_mass_plot}} 
    \label{fig:plasma_mass}
\end{figure}
\begin{figure*}[!htbp]
    \centering
    \includegraphics[width=0.49\textwidth]{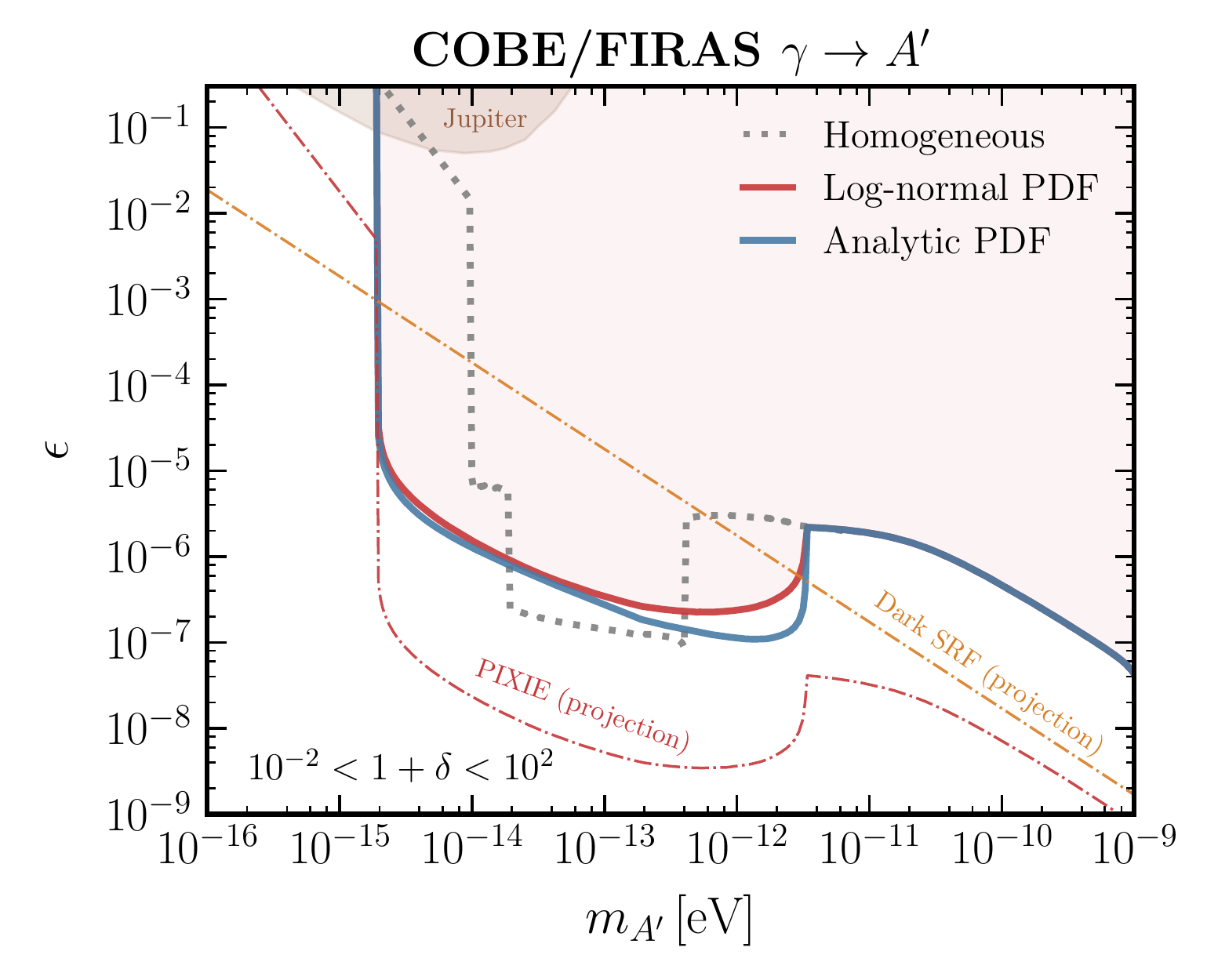}
    \includegraphics[width=0.49\textwidth]{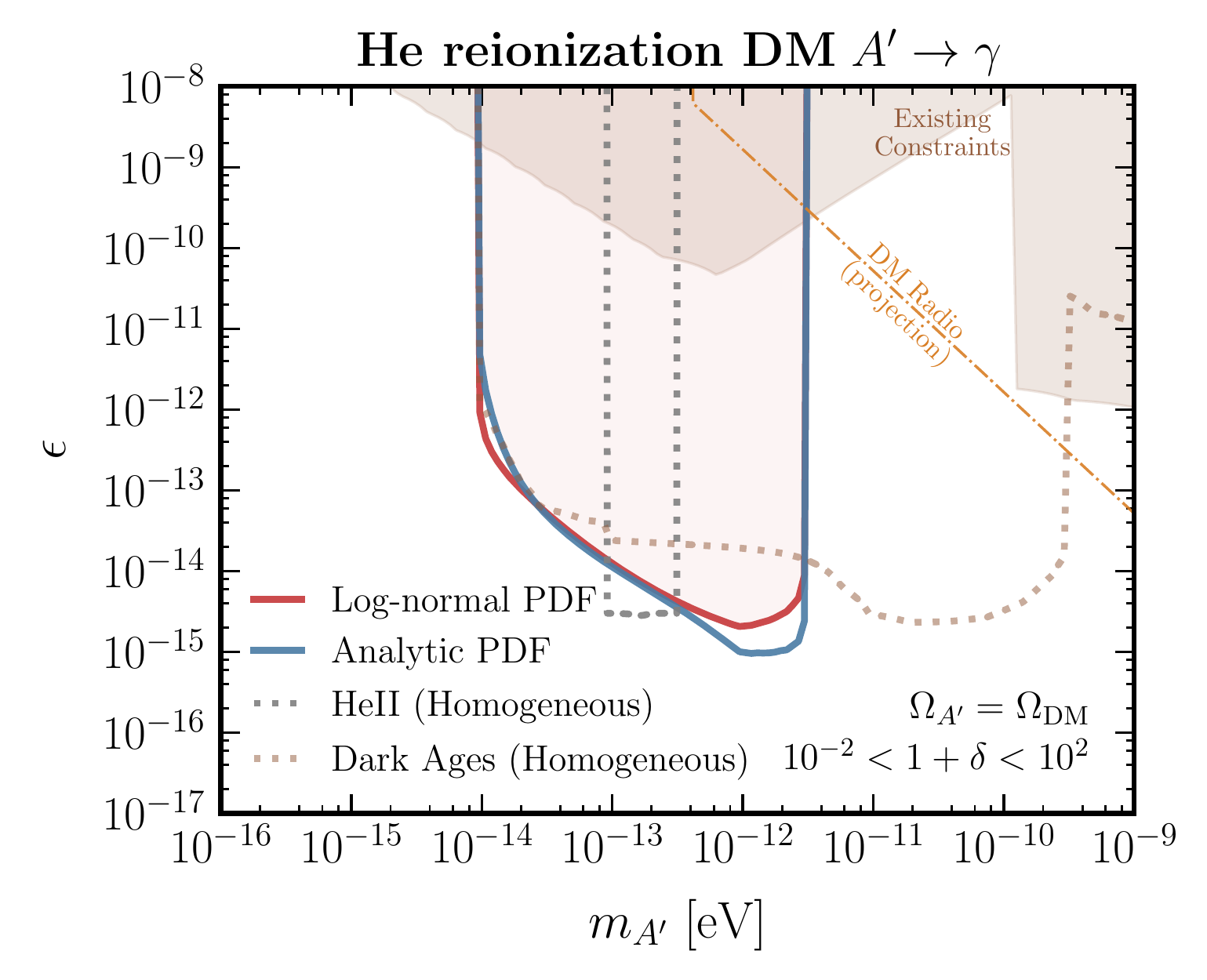}
    \caption{\emph{(Left)} The 95\% confidence level constraints on the kinetic mixing parameter $\epsilon$ as a function of dark photon mass $\mAp$, assuming log-normal (red) or analytic (blue) PDFs; the shaded region is ruled out by the more conservative of the two PDF choices.  We also show the reach of the proposed PIXIE satellite~\cite{Kogut:2011xw} (dot-dashed red) assuming a log-normal PDF\@.  For comparison we show the previous limit assuming a homogeneous plasma (dotted gray), a constraint from the magnetic field of Jupiter~\cite{Davis:1975mn,Ahlers:2008qc} (shaded brown), and the projected reach of the Dark SRF experiment~\cite{HarnikSRF,GrassellinoSRF} (dot-dashed orange), {which would be complementary to our cosmological constraints}.~\nblink{02_firas_dp_limits}
\emph{(Right)} Constraints on dark photon dark matter from anomalous heating of the IGM during the epoch of HeII reionization, for the same PDFs. Prior constraints (shaded brown) come from non-resonant heating of the IGM~\cite{McDermott:2019lch} and heating of the gas in the dwarf galaxy Leo T~\cite{Wadekar:2019xnf}.  We also show the projected reach of DM Radio Stage 3~\cite{Chaudhuri:2014dla,Silva-Feaver:2016qhh,Battaglieri:2017aum} (dot-dashed orange). Limits from changes to the dark matter density and from IGM heating during the dark ages assuming a homogeneous plasma mass have been derived in Ref.~\cite{McDermott:2019lch} (dotted orange).~\nblink{05_dp_dm_He_limits}} 
    \label{fig:limit}
\end{figure*}

Eq.~\eqref{eq:prob} is one of our main results, and we consider a few different possibilities for the one-point PDF $\mathcal{P}(\delta_\mathrm{b}; t)$ in order to estimate the theoretical uncertainty associated with the nonlinear distribution of matter at low redshifts $z \lesssim 6$. First, we consider a log-normal distribution, which has long been used as a simple model for the distribution of the low-redshift matter density~\cite{1934ApJ....79....8H, Coles:1991if, Kayo:2001gu, Wild:2004me}. To inform the spectrum of fluctuations for this distribution, we use the baryonic power spectra $P_\mathrm{bb}(k)$ derived from hydrodynamic simulations~\cite{Nelson:2018uso,McAlpine:2015tma,McCarthy:2016mry,Genel:2014lma} and extracted in Refs.~\cite{Foreman:2019ahr,vanDaalen:2019pst}.
Second, we adopt an analytical prescription~\cite{Ivanov:2018lcg}, which extends the spherical collapse model~\cite{Valageas:2001zr, Valageas:2001td} to 
perform a first-principles computation of the nonlinear matter PDF\@.

In the literature, $\mathcal{P}(\delta_\mathrm{b};t)$ is typically defined as a function of a smoothing scale $R$ over which densities are averaged in order to match observations and simulation results; furthermore, the width of the distribution can exhibit a log-divergence in $R$ if $P_\mathrm{bb}(k) \propto k^{-3}$ at large $k$. In our work, we assume that baryonic structures are suppressed on scales smaller than the baryonic Jeans scale $R_\mathrm{J} \sim \SI{10}{\kilo\parsec}$. In practice, the log-normal PDF is computed with a $P_\mathrm{bb}(k)$ which has a cutoff at $k_\mathrm{J} \sim 1/R_\mathrm{J}$ derived from CLASS, while our analytic PDF is obtained with a smoothing scale $R = R_\mathrm{J}$. A complete description of our PDFs and the Jeans scale is given in~\citetalias{OurLongPaper}.

The differential transition probability (normalized to unity) for a few benchmark dark photon mass points $\mAp=\SI{2e-15}{}, \SI{e-13}{}$, and \SI{e-12}{\eV} is shown in the bottom panel of Fig.~\ref{fig:plasma_mass}. For $\mAp = \SI{e-12}{\eV}$ there is a narrow resonance corresponding to a transition in the limit of a homogeneous plasma at $z\sim 200$.  An additional broad resonance at $z\sim 6$ is also present, corresponding to conversions in overdensities in the plasma mass post-reionization. Note that Eq.~\eqref{eq:Prob_homo} implies that later resonances typically contribute more to the total conversion probability. For $\mAp = \SI{2e-15}{\eV}$, no resonance exists in the homogeneous limit; remarkably, however, fluctuations in the plasma mass result in resonant transitions over a broad range of redshifts at $z\lesssim20$ due to underdensities in the plasma mass.
This opens up the possibility of probing dark photon masses $\mAp \lesssim \SI{e-14}{\eV}$ through previously-neglected cosmological conversions. 

\noindent
{\bf Dark photon oscillations in the CMB spectrum.---}We first apply our formalism to analyze the intensity of the CMB as measured by the FIRAS instrument aboard COBE~\cite{Fixsen:1996nj} for evidence of deviations from a blackbody spectrum due to $\gamma \to A'$ oscillations. Notably in this case the dark photon does not need to be the dark matter. The spectrum of the FIRAS data is fit by the nearly perfectly Planckian spectrum $B_{\omega}$ with temperature $T_\mathrm{CMB} = \SI{2.725}{\kelvin}$~\cite{Mather:1998gm}. For a given dark photon model specified by its mass $\mAp$ and mixing parameter $\epsilon$, the spectral distortion to the CMB spectrum will be given by $I_{\omega_0}\left(m_{A'}, \epsilon\right)=B_{\omega_0}\left(1- \left< P_{\gamma \rightarrow A'} \right>\right)$, where $\left\langle P_{\gamma \rightarrow A'} \right\rangle$ is the conversion probability for the given model corresponding to the present-day frequency $\omega_0$, obtained by integrating Eq.~\eqref{eq:prob}. Details of the data analysis are presented in the \hyperref[supp]{Supplemental Material}.

Erring on the conservative side, we do not consider fluctuations outside of the range $10^{-2} < 1 + \delta_\mathrm{b} < 10^2$, and as such our results do not rely on conversions in the tails of the PDF where uncertainties are large. Additionally, for all cases considered here conversions in the redshift range $6 < z < 20$ have been excised, providing a conservative result while being agnostic to the uncertainties arising from the complex physics of reionization in this epoch. We explore the effects of these choices in the \hyperref[supp]{Supplemental Material}.

We observe no significant evidence for a signal. In the left panel of Fig.~\ref{fig:limit} we show our fiducial constraints at the 95\% confidence level on the dark photon mixing parameter $\epsilon$ for a range of dark photon masses $\mAp$. We show constraints using both the log-normal and analytic description of the PDF\@. We also show the projected limits for a future measurement of the CMB spectrum such as the proposed PIXIE satellite~\cite{Kogut:2011xw} using the putative specifications from Ref.~\cite{Kunze:2015noa}. The traditional constraint assuming a homogeneous plasma mass as a function of redshift is also shown for comparison, together with constraints projected by the resonant cavity-based Dark SRF experiment~\cite{HarnikSRF,GrassellinoSRF} and existing constraints from an analysis of the magnetic field of Jupiter~\cite{Davis:1975mn}.  There are bounds from black hole superradiance for values of $\mAp$ that overlap our bounds assuming $\epsilon = 0$~\cite{Pani:2012vp,Baryakhtar:2017ngi,Cardoso:2018tly}, but it is unknown if these apply when $\epsilon > 0$ implying interactions of  $A'$ with plasma around the black hole.

\noindent
{\bf Dark photon dark matter.---}So far, we have studied the resonant conversion of CMB photons into relativistic ($v \simeq c$) dark photons. A cold, nonrelativistic ($v \ll c$) population of dark photons can also be produced nonthermally in the early Universe, and is a good candidate for dark matter~\cite{Redondo:2008ec,Arias:2012az, Agrawal:2018vin,Long:2019lwl,Co:2018lka}. Additional constraints apply in this case; in particular, Ref.~\cite{McDermott:2019lch} proposed using measurements of the temperature of the intergalactic medium (IGM) around the epoch of HeII reionization ($2 \lesssim z \lesssim 6$)~\cite{Becker:2010cu,Bolton:2013cba,Boera:2014sia,Rorai:2017qft,Hiss:2017qyw,Walther:2018pnn} to constrain the dark photon dark matter scenario. These measurements show that during HeII reionization, the total heat input per baryon is on the order of \SI{1}{\eV}. $A'\to\gamma$ conversion for light $A'$s produce soft photons that are absorbed efficiently through free-free absorption, leading to an anomalous heating of the IGM\@. The derived bound in the homogeneous limit extends over a limited mass range, precisely where the dark photon mass matches the homogeneous plasma mass in the narrow redshift range $2 \lesssim z \lesssim 6$. Our formalism accounting for inhomogeneities extends this treatment to a wider range of dark photon masses. 

The total energy injected per unit baryon $\langle E_{A' \to \gamma} \rangle$ from the dark matter can be computed as
\begin{multline}  \label{eq:prob_dm}
     \frac{\dd \langle E_{A' \to \gamma} \rangle}{\dd z} = \pi \mAp \epsilon^2 \frac{\overline{\rho}_{A'}}{\overline{n}_{\mathrm{b}}} \left|\frac{\dd t}{\dd z}\right| \\ 
     \times \int \dd m_\gamma^2 \, \frac{m_\gamma^2}{\overline{m_\gamma^2}(t)} f(m_\gamma^2;t) \, \delta_\mathrm{D}(m_\gamma^2 - \mAp^2) \, m_\gamma^2   \,,
\end{multline}
where $\bar{\rho}_{A'} / \bar{n}_\mathrm{b}$ is the ratio of the homogeneous dark matter energy density to baryon number density, which is a time-independent quantity.\footnote{We assume that the deposited heat from $A'$ conversions is shared equally among all baryons: for further discussion of this assumption, we refer the reader to the \hyperref[supp]{Supplemental Material}.}  
The total energy injected is then obtained by performing an integral over $2 < z < 6$. Considering the same PDFs discussed in the previous section and imposing $\langle E_{A' \to \gamma} \rangle < \SI{1}{\eV}$, we derive the constraints shown in the right panel of Fig.~\ref{fig:limit}, with the homogeneous limit shown for comparison. Also shown is the parameter space covered by existing constraints~\cite{McDermott:2019lch,Wadekar:2019xnf}, as well as the projected constraints from DM Radio Stage~3~\cite{Chaudhuri:2014dla,Silva-Feaver:2016qhh,Battaglieri:2017aum}.  Finally, our limits can be rescaled as a function of the maximum $\langle E_{A' \to \gamma} \rangle$ allowed and $\rho_{A'}$ by noting that $\langle E_{A' \to \gamma} \rangle \propto \epsilon^2 \rho_{A'}$. 

\noindent
{\bf Conclusions.---}We have introduced a framework for treating oscillations between dark photons and ordinary photons as they traverse the inhomogeneous plasma of our Universe.  Our main results are Eqs.~\eqref{eq:prob} and \eqref{eq:prob_dm}.  A complete discussion and derivation of these results appear in~\citetalias{OurLongPaper}.  We have applied this framework to determine constraints from CMB photons oscillating into dark photons (Fig.~\ref{fig:limit}, left panel) and from energy injection from dark photon dark matter (Fig.~\ref{fig:limit}, right panel).  Prior studies have assumed the homogeneous limit and require significant revision because inhomogeneities both extend the mass reach, and either strengthen or weaken the sensitivity for masses constrained in the homogeneous limit.

We anticipate broader applications of our framework.  Perturbations in the photon plasma mass will modify resonant oscillations of photons into axion-like-particles, which can occur in the presence of primordial magnetic fields~\cite{Mirizzi:2009nq} or dark magnetic fields~\cite{Choi:2019jwx}.  Here we have considered oscillations of dark photon dark matter, but dark photons (or axion-like-particles) can also resonantly inject photons that impact 21\,cm observations~\cite{Pospelov:2018kdh,Moroi:2018vci,Choi:2019jwx}.  We have here considered global (sky-averaged) effects, but photon-to-dark photon oscillations in an inhomogeneous background will imprint anisotropies in the CMB that may be testable by \emph{Planck}~\cite{Aghanim:2019ame} and/or next-generation probes of CMB anisotropies~\cite{Abazajian:2016yjj,Ade:2018sbj}.

Additional details of the data analysis performed and a discussion of systematic effects is presented in the \hyperref[supp]{Supplemental Material}. 
A much more in-depth discussion of our formalism, the choice of one-point PDFs, the construction of the baryon power spectrum, and a verification of our formalism with simulations are all discussed in~\citetalias{OurLongPaper}. 
The code used to obtain the results in this paper,~\citetalias{OurLongPaper}, as well as digitized constraints are available at \url{https://github.com/smsharma/dark-photons-perturbations}~\cite{andrea_caputo_2020_4081407}.

\vspace{.3cm}

\noindent
{\bf Acknowledgements.---}We thank Yacine Ali-Ha\"{i}moud, Masha Baryakhtar, Asher Berlin, Julien Lesgourgues, Sam McDermott, Alessandro Mirizzi, Julian Mu\~{n}oz, Stephen Parke, Maxim Pospelov, Josef Pradler, Javier Redondo, Roman Scoccimarro, Alfredo Urbano, Edoardo Vitagliano, Sam Witte, and Chih-Liang Wu for helpful conversations. We thank Marcel van Daalen for providing baryonic power spectra from high-resolution BAHAMAS simulations. We are especially grateful to Misha Ivanov for many enlightening discussions regarding the analytic PDF of density fluctuations utilized in this work. AC acknowledges support from the ``Generalitat Valenciana'' (Spain) through the ``plan GenT'' program (CIDEGENT/2018/019), as well as national grants FPA2014-57816-P, FPA2017-85985-P, and the European projects H2020-MSCA-ITN-2015//674896-ELUSIVES. HL is supported by the DOE under contract DESC0007968. SM and JTR are supported by the NSF CAREER grant PHY-1554858 and NSF grant PHY-1915409. SM is additionally supported by NSF grant PHY-1620727 and the Simons Foundation. 
JTR acknowledges hospitality from  the Aspen Center for Physics, which is supported by the NSF grant PHY-1607611.
This work made use of the NYU IT High Performance Computing resources, services, and staff expertise. The authors are pleased to acknowledge that the work reported on in this paper was substantially performed using the Princeton Research Computing resources at Princeton University which is a consortium of groups including the Princeton Institute for Computational Science and Engineering and the Princeton University Office of Information Technology's Research Computing department. This research has made use of NASA's Astrophysics Data System. We acknowledge the use of the Legacy Archive for Microwave Background Data Analysis (LAMBDA), part of the High Energy Astrophysics Science Archive Center (HEASARC). HEASARC/LAMBDA is a service of the Astrophysics Science Division at the NASA Goddard Space Flight Center. This research made use of the \texttt{astropy}~\cite{Price-Whelan:2018hus,Robitaille:2013mpa}, CAMB~\cite{Lewis:1999bs,Lewis:2002ah}, CLASS~\cite{Blas:2011rf}, \texttt{HyRec}~\cite{AliHaimoud:2010dx}, \texttt{IPython}~\cite{PER-GRA:2007}, Jupyter~\cite{Kluyver2016JupyterN}, \texttt{matplotlib}~\cite{Hunter:2007}, \texttt{nbodykit}~\cite{Hand:2017pqn}, \texttt{NumPy}~\cite{numpy:2011}, \texttt{seaborn}~\cite{seaborn}, \texttt{pandas}~\cite{pandas:2010}, \texttt{SciPy}~\cite{2020SciPy-NMeth}, and \texttt{tqdm}~\cite{da2019tqdm}  software packages.  

\appendix

\bigskip
\bigskip
\begin{center}{\textbf{SUPPLEMENTAL MATERIAL}} \end{center}
\label{supp}
\smallskip

This Supplemental Material is organized as follows. First, we provide details of the COBE/FIRAS data analysis performed. Second, we provide a discussion of the systematic effects associated with the results presented in the main text. Finally, we explore alternative assumptions about the energy deposition mechanism responsible for the dark photon dark matter bounds presented in the main text.

\section{COBE/FIRAS data analysis}
\label{app:firas_data}

\begin{figure}[tbp]
    \centering
    \includegraphics[width=0.47\textwidth]{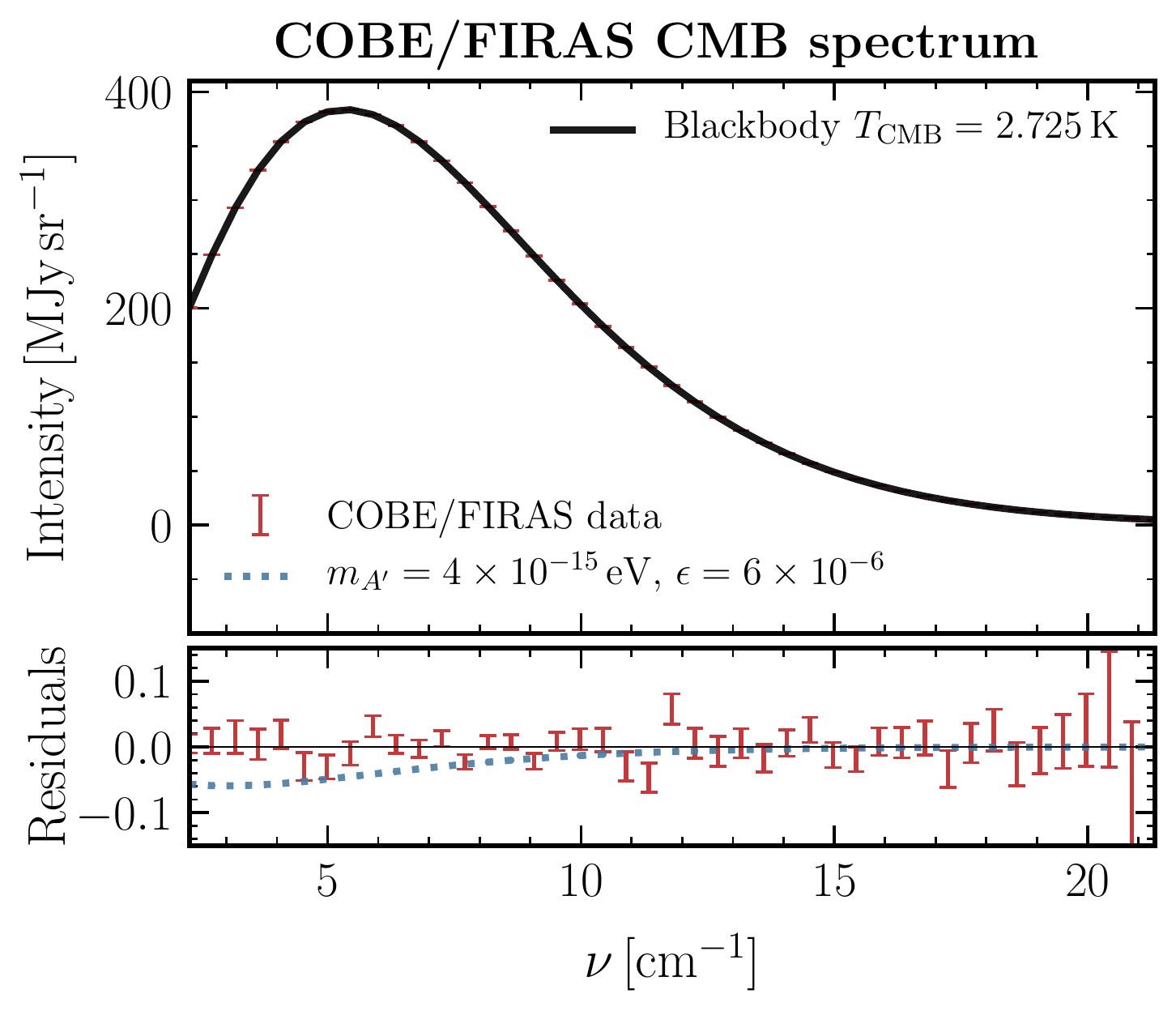}
    \caption{The CMB blackbody spectrum (solid black) best-fit to the FIRAS data (red data points) with $T_\mathrm{CMB}=2.725$\,K\@. Residuals from the blackbody spectrum are shown in the bottom panel, additionally illustrating the spectral distortions induced by photon-dark photon mixing with $\epsilon=6\times 10^{-6}$ and $\mAp = \SI{4e-15}{\eV}$, close to the detectability threshold of the present analysis (dotted blue).~\nblink{08_firas_plot}} 
    \label{fig:firas_blackbody}
\end{figure}

We utilize the low frequency FIRAS monopole data,\footnote{Available at \url{https://lambda.gsfc.nasa.gov/product/cobe/firas_monopole_get.cfm}.} consisting of 43 linearly spaced data points spanning the frequency range $\nu_0 = 2.27$--\SI{21.33}{\per\cm} and construct data covariance matrices following Ref.~\cite{Fixsen:1996nj}. The spectrum of the FIRAS data is illustrated in Fig.~\ref{fig:firas_blackbody}, fit by the nearly perfectly Planckian spectrum
\begin{equation}
B_{\omega_0}=\frac{\omega_0^{3}}{2 \pi^{2}}\left[\exp \left(\frac{\omega_0}{T_\mathrm{CMB}}\right)-1\right]^{-1}
\end{equation}
with temperature $T_\mathrm{CMB} \simeq \SI{2.725}{\kelvin}$, and $\omega_0 = 2\pi\nu_0$ the angular frequency. Residuals between the FIRAS data and this spectrum are shown in the bottom panel. For a given dark photon model specified by its mass $\mAp$ and mixing parameter $\epsilon$, the spectral distortion to the CMB spectrum will be given by $I_{\omega_0}\left(\mAp, \epsilon;T_\mathrm{CMB}\right)=B_{\omega_0}\left(1- \left< P_{\gamma \rightarrow A'} \right>\right)$, where $\left\langle P_{\gamma \rightarrow A'} \right\rangle$ is the conversion probability for the given model corresponding to the present-day frequency $\omega_0$, obtained by integrating the differential probability given in the Letter, 
\begin{multline} 
    \label{eq:prob}
    \frac{\dd\langle P_{\gamma \to A'} \rangle}{\dd z} =  \frac{\pi \mAp^2 \epsilon^2}{\omega(t)} \left|\frac{\dd t}{\dd z}\right| \\
    \times \int \dd m_\gamma^2  \, f(m_\gamma^2;t) \, \delta_\mathrm{D}(m_\gamma^2 - \mAp^2) \, m_\gamma^2  \,.
\end{multline}
An illustration of the distortion to the blackbody spectrum induced by dark photons of mass $\mAp = \SI{6e-15}{\eV}$ with mixing $\epsilon=4\times10^{-6}$ is shown in dashed blue in the bottom panel of Fig.~\ref{fig:firas_blackbody}.

We construct a Gaussian log-likelihood as
\begin{equation}
    \ln\mathcal L(d|\mAp, \epsilon)  = \max_{T_\mathrm{CMB}}\left[-\frac{1}{2}\Delta \vec I^T\,\mathsf C_{I_d}^{-1}\,\Delta \vec I\right],
\end{equation}
where $\Delta \vec I = \left(\vec I \left(\mAp, \epsilon;T_\mathrm{CMB}\right) - \vec I_d\right)$ is the residual between the distorted CMB spectrum $\vec I \left(\mAp, \epsilon;T_\mathrm{CMB}\right) = \left\{I_{\omega_1}, I_{\omega_2},\ldots\right\}$ and the FIRAS data vector $\vec I_d$, and $\mathsf C_{I_d}$ is the data covariance matrix. We treat the CMB temperature as a nuisance parameter and profile over it by maximizing the log-likelihood for $T_\mathrm{CMB}$ at each $\{\mAp, \epsilon\}$ point. We define our test-statistic as 
\begin{equation}
\mathrm{TS}(\mAp, \epsilon) = 2\left[\ln\mathcal L(d|\mAp, \epsilon) - \ln\mathcal L(d|\mAp, \hat\epsilon)\right], 
\end{equation}
where $\hat\epsilon$ is the value of $\epsilon$ that maximizes the log-likelihood for a given $\mAp$, and obtain our limit by finding the value of $\epsilon$ at which $\mathrm{TS} = -2.71$ corresponding to 95\% containment for the one-sided $\chi^2$ distribution.

Although we consider the incoming photons to have energies corresponding to the specified FIRAS frequency bands, in reality FIRAS has a finite spectral resolution resulting in a spread in energies over a finite range of the order of the bin size. We check the impact of this binning on the constraints presented by modeling the response of each frequency bin as a Gaussian centered on the central frequency with standard deviation corresponding to the bin size~\cite{Fixsen:1998js}. We find that this choice has a percent-level impact on the computed inhomogeneous oscillation probabilities in the lowest frequency bin, with smaller errors for larger frequencies. Since the mixing angle constraint scales as the square root of the oscillation probability, our constraints are not qualitatively impacted by finite binning effects.

Nevertheless, resonances relying on particular values of $\omega_0$ can cause local enhancements in the \emph{homogeneous} constraints at masses $\mAp \gtrsim \SI{4e-13}{\eV}$ due to individual crossings with small characteristic derivatives $\dd\ln m_{\gamma}^{2}(t)/\dd t$. These sharp enhancements are likely artifacts of finite spectral binning, and we thus smooth the homogeneous constraints with a Savitzky-Golay filter above this mass.

\section{Systematics and additional results}
\label{app:systematics}

The behavior of $\gamma \leftrightarrow A'$ conversions in the inhomogeneous Universe depends critically on the distribution of density perturbations as a function of redshift. While significant uncertainty exists for this distribution, we have already shown in the Letter that using two radically different approaches to computing the probability density function (PDF) of the photon plasma mass $f(m_\gamma^2;t)$ does not lead to qualitative differences in our results. In this section, we discuss several other possible sources of uncertainty, more consistency checks of our fiducial limits, and additional results that are more optimistic or are of pedagogical interest. 

\subsection{Probability density functions}

\begin{figure}[tbp]
    \centering
    \includegraphics[width=0.47\textwidth]{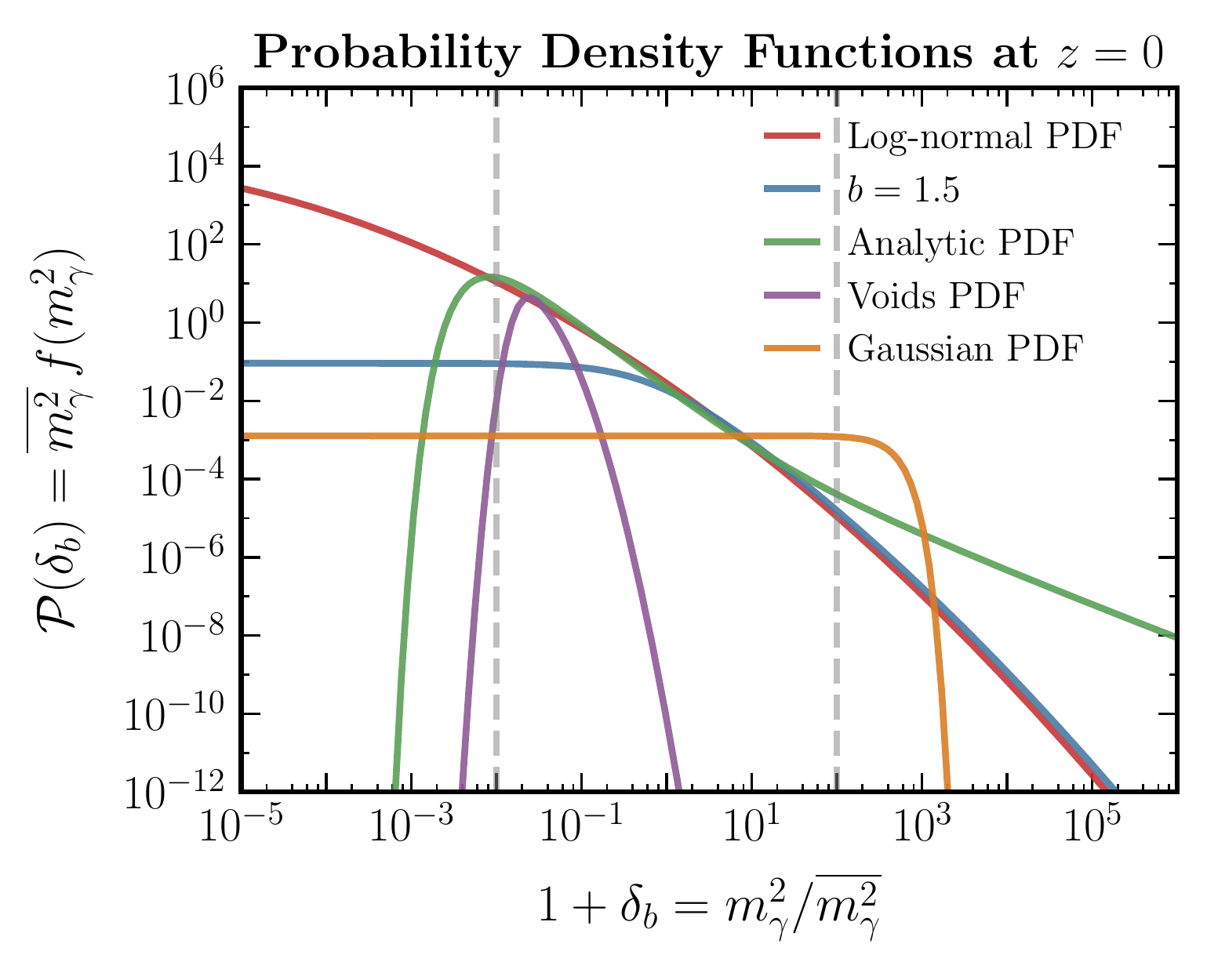}
    \caption{One-point PDFs $\mathcal{P}(\delta_\mathrm{b})$ at $z = 0$. We show the fiducial log-normal (red) and analytic (green) PDFs, together with several other PDFs considered in the Supplemental Material, including a log-normal PDF with bias $b = 1.5$ (blue), a PDF constructed from a model of voids (purple)~\cite{Adermann:2018jba} and a Gaussian PDF (orange). Also shown are the fiducial $10^{-2} < 1+\delta < 10^2$ boundaries (dashed gray). At $z = 0$, the homogeneous plasma mass is $\overline{m_\gamma^2} = \SI{1.9e-14}{\eV}$.~\nblink{07_pdf_plot}} 
    \label{fig:PDFs}
\end{figure}

The two different prescriptions for the one-point PDF used to construct $f(m_\gamma^2;t)$ are the log-normal distribution and an analytic distribution based on ideas presented in Refs.~\cite{Valageas:2001zr,Bernardeau:2015khs,Uhlemann:2015npz,BetancortRijo:2001ge,Lam:2007qw,Ivanov:2018lcg}. The log-normal distribution is a phenomenological PDF that can take the nonlinear baryon power spectra from simulations into account, while the analytic distribution has a theoretical basis in structure formation theory, but only models the matter distribution without baryonic effects. Fig.~\ref{fig:PDFs} demonstrates that these two PDFs have a similar behavior in the range $10^{-2} < 1 + \delta < 10^2$ despite being very different approaches, giving us confidence that considering fluctuations only in this range is a reasonable choice, and explaining why the limits derived from our two fiducial prescriptions are similar.

We have also considered three further models for the PDFs, which we believe are useful checks for our results, but are unlikely to be more accurate than our two fiducial approaches. In this section, we provide a brief description of these PDFs; for more details, we refer the reader to~\citetalias{OurLongPaper}.

\begin{figure*}[tbp]
    \centering
    \includegraphics[width=0.47\textwidth]{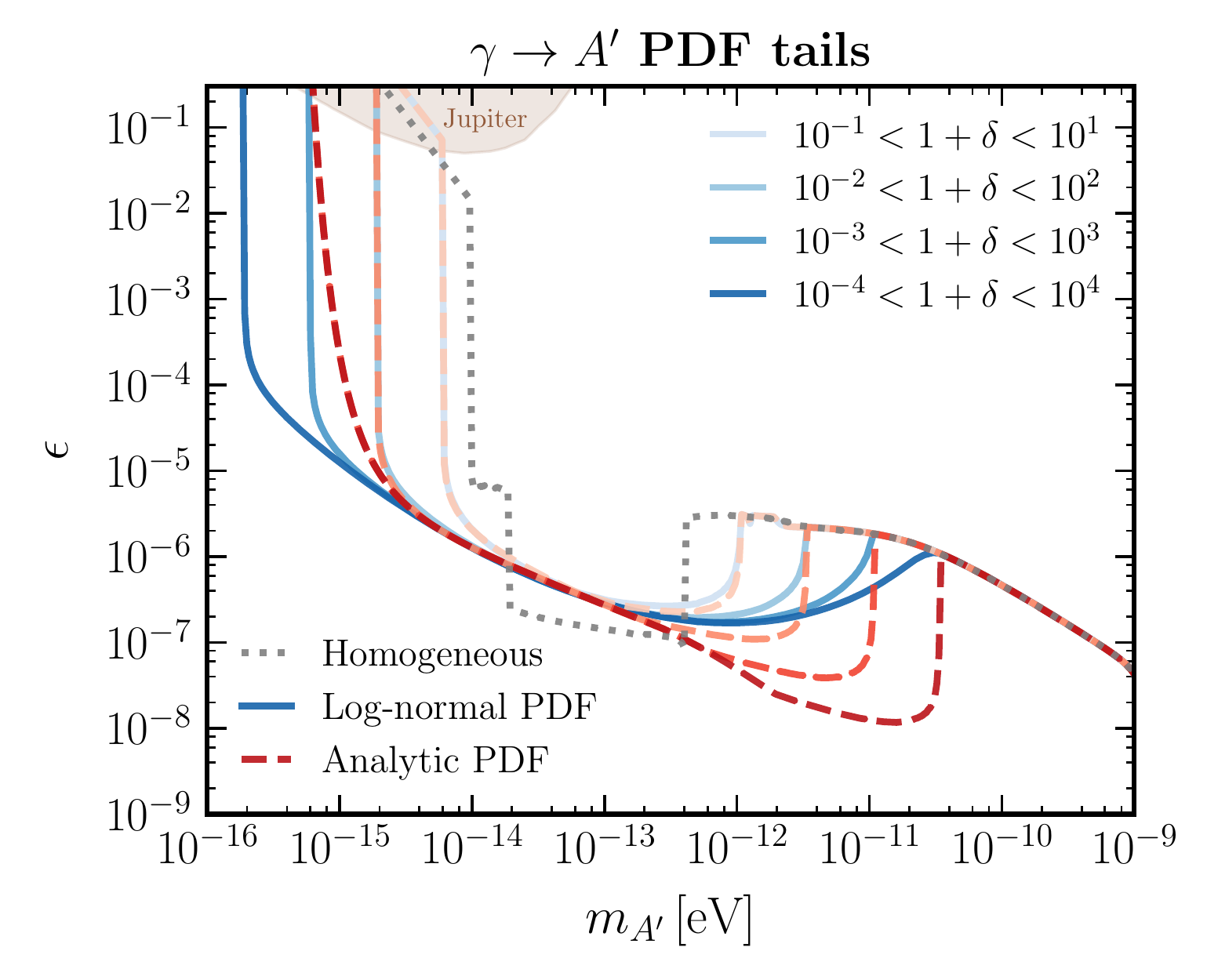}
    \includegraphics[width=0.47\textwidth]{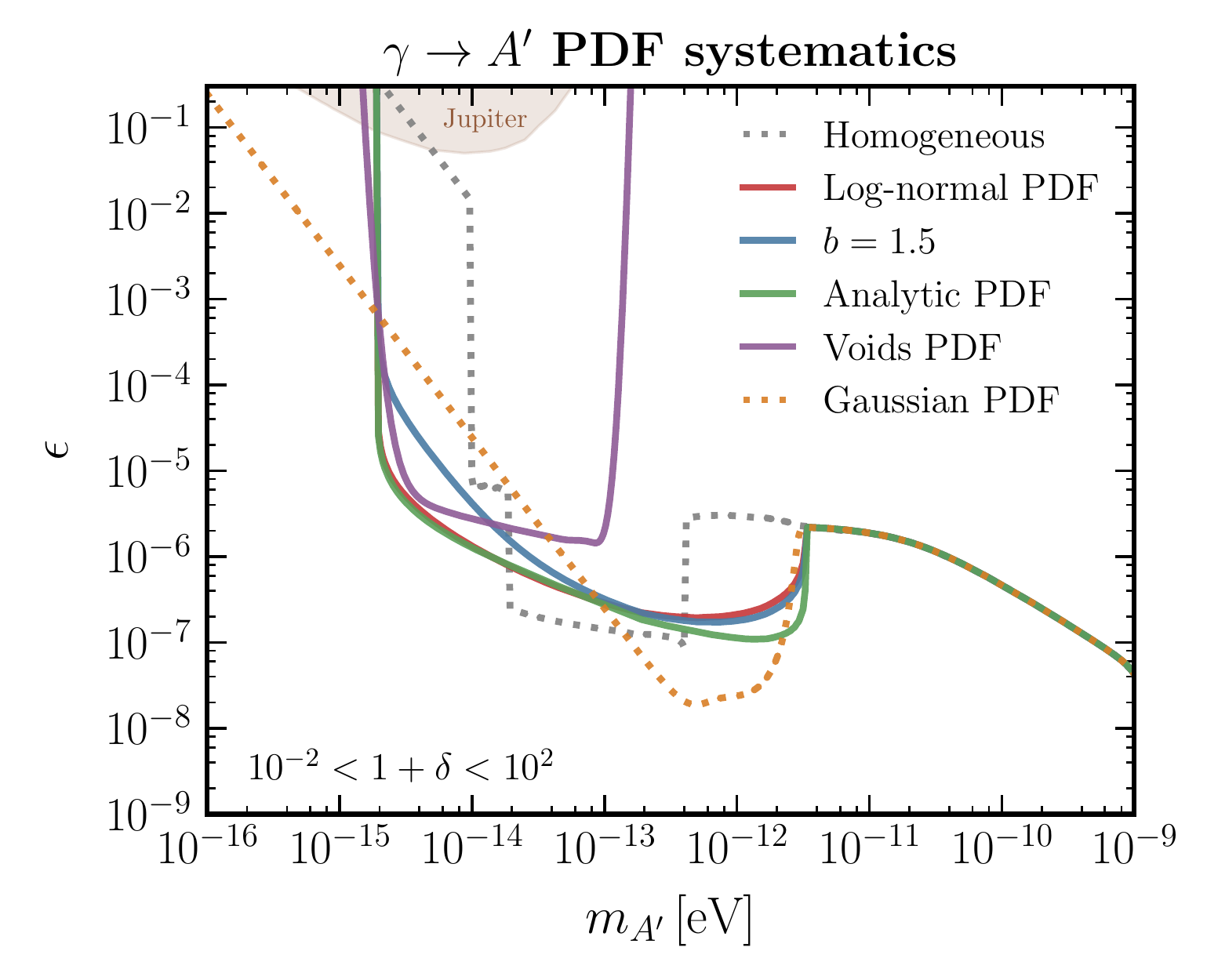}
    \includegraphics[width=0.47\textwidth]{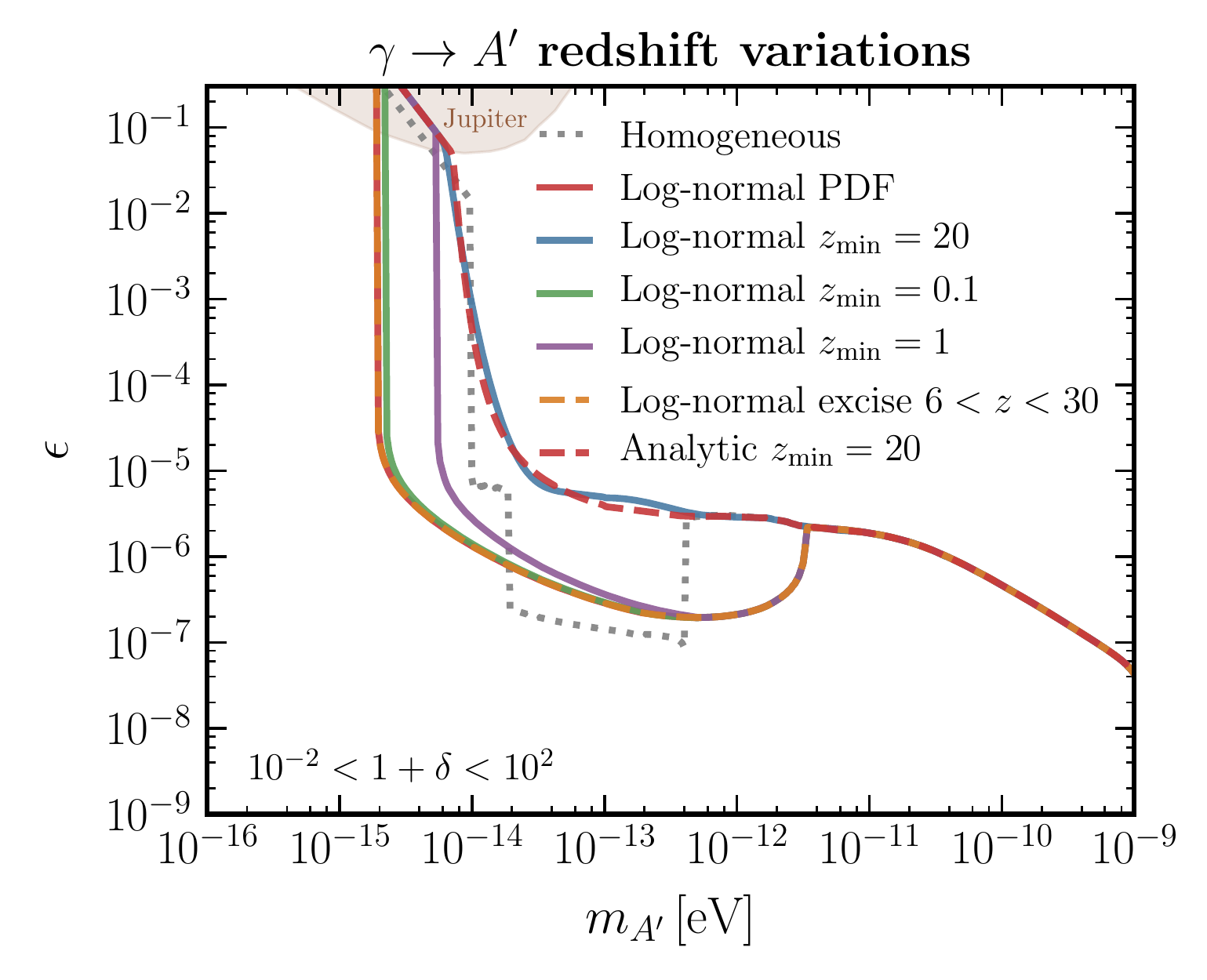}
    \caption{Systematic variations on dark photon constraints from $\gamma\to A'$ oscillations. \emph{(Top left)} The effect of truncating the 1-point PDF at different cutoffs in $1 + \delta$, shown for the log-normal(analytic) PDF in solid blue(dashed red). Progressively darker lines corresponding to the inclusion of larger underdensities and overdensities, from 10 times larger and smaller than the mean plasma mass to $10^4$ times larger and smaller than the plasma mass, respectively. \emph{(Top right)} Sensitivity of the constraints to the choice of the PDF parameterization. Shown is our fiducial constraint with the log-normal PDF (solid red) and the analytic PDF (solid green). Constraints with the inclusion of a linear bias $b=1.5$ between the dark matter and baryons in the log-normal prescription (solid blue), the inferred PDF of underdensities in voids from Ref.~\cite{Adermann:2018jba} (solid purple), and with a Gaussian PDF (dotted orange) are also displayed. \emph{(Bottom)} Effect of excluding conversions over different redshift ranges on the constraints are presented. Excising a larger redshift range $6 < z < 30$ (dashed orange) has no effect on the fiducial limits, obtained by excising $6 < z < 20$ (solid red). Constraints relying solely on conversions at redshifts above $z > 0.1(1)$ are shown in solid green(purple), and those relying on conversions at linear cosmological epochs ($z \gtrsim 20$) are shown for the log-normal and analytic PDFs in solid blue and dashed red, respectively.~\nblink{02_firas_dp_limits}} 
    \label{fig:limit_dp_systematics}
\end{figure*}
\begin{figure*}[tbp]
    \centering
    \includegraphics[width=0.47\textwidth]{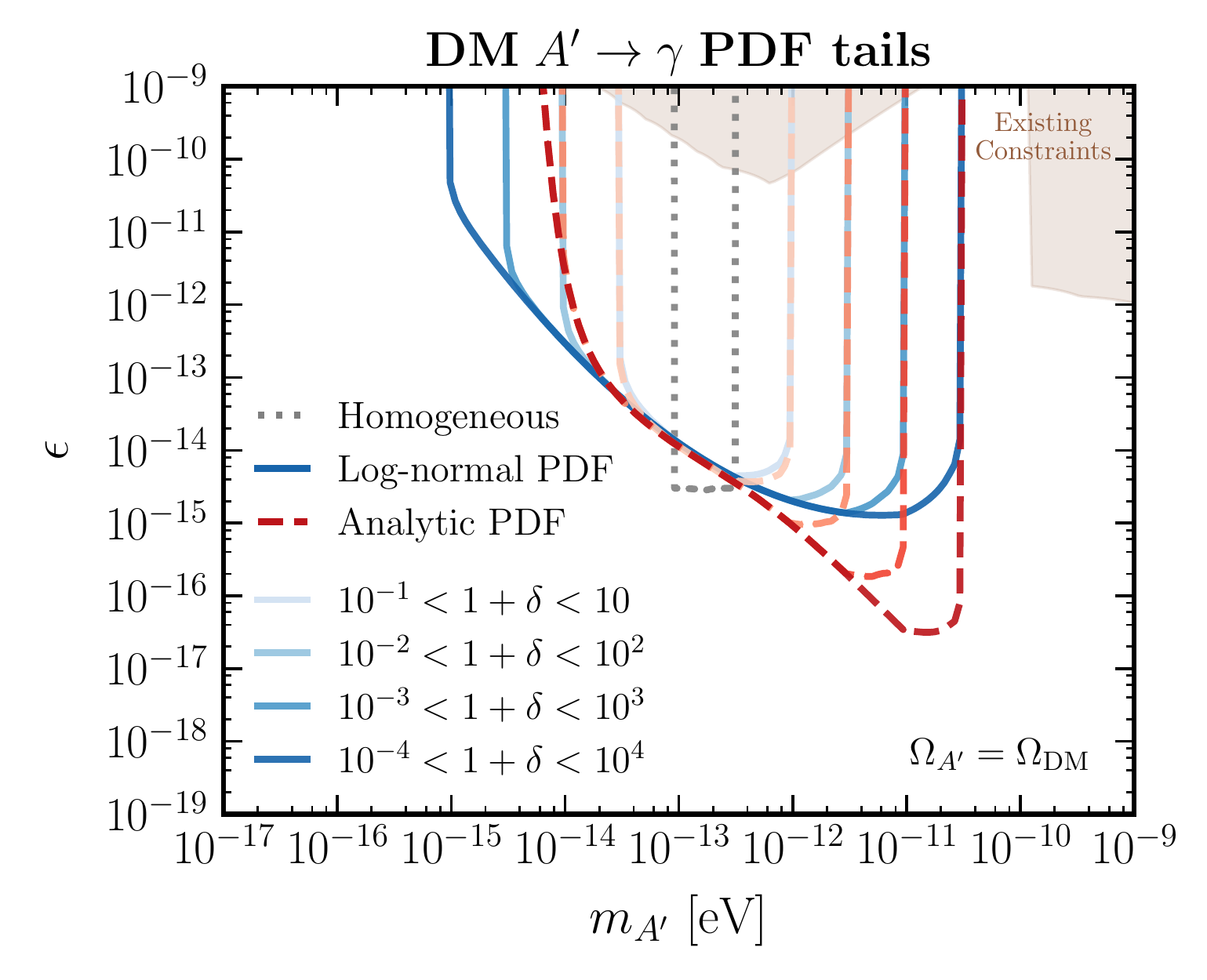}
    \includegraphics[width=0.47\textwidth]{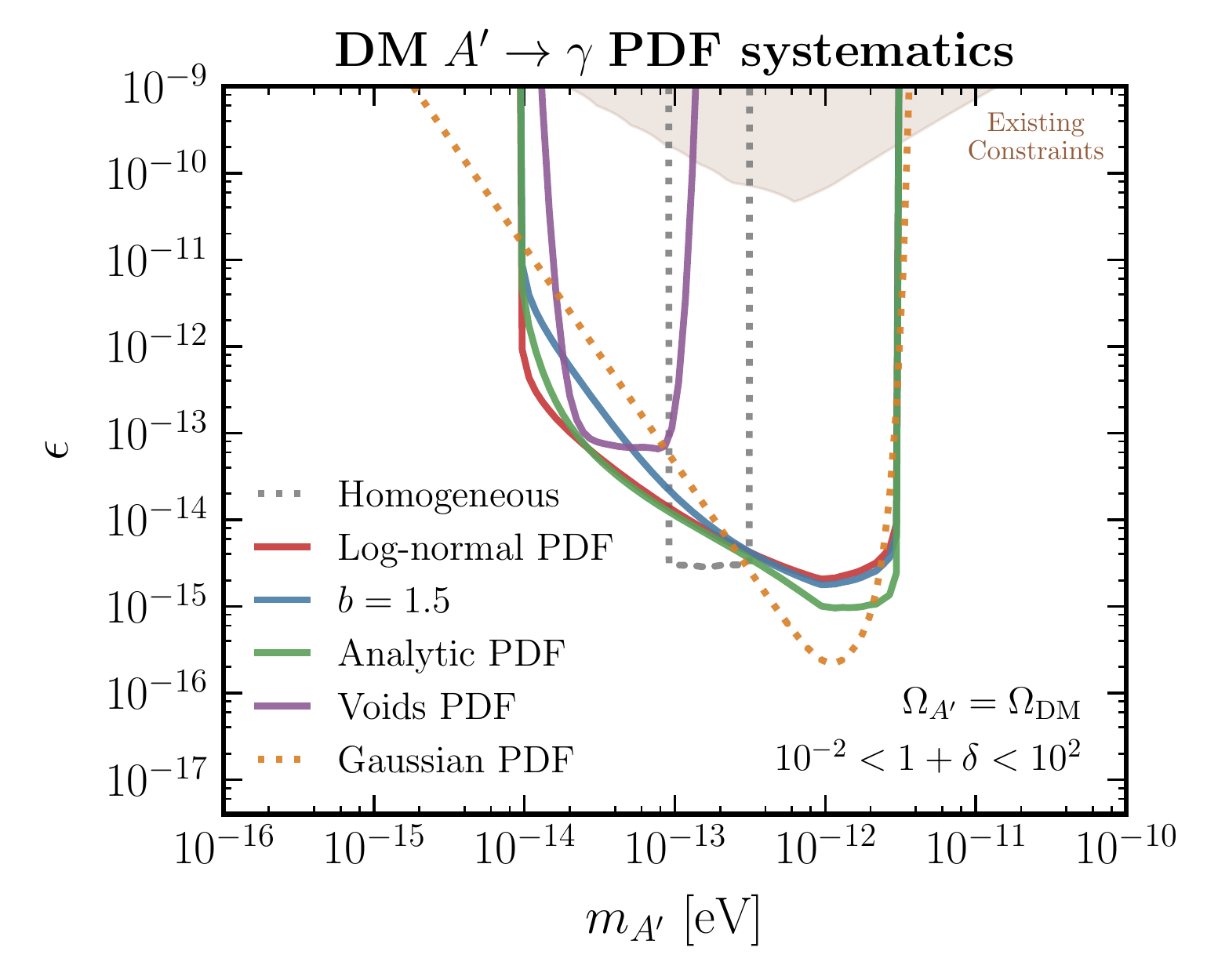}
    \caption{Systematic variations on dark photon dark matter constraints from $A'\to \gamma$ effects on IGM heating during HeII reionization. \emph{(Left)} The effect of truncating the 1-point PDF at different cutoffs in $1 + \delta$, shown for the log-normal (solid blue) and analytic (dashed red) PDFs. Progressively darker lines corresponding to the inclusion of larger underdensities and overdensities, from 10 times larger and smaller than the mean plasma mass to $10^4$ times larger and smaller than the plasma mass, respectively. \emph{(Right)} Sensitivity of the constraints to the choice of the PDF parameterization. Shown is our fiducial constraint with the log-normal PDF (solid red), as well as constraints with the inclusion of a linear bias $b=1.5$ between the dark matter and baryons in the log-normal prescription (solid blue), the analytic PDF (solid green), the inferred PDF of underdensities in voids from Ref.~\cite{Adermann:2018jba} (solid purple), and a Gaussian PDF (dotted orange).~\nblink{05_dp_dm_He_limits}} 
    \label{fig:limit_dp_dm_systematics}
\end{figure*}

\noindent
{\bf Log-normal with bias.---}For our fiducial log-normal PDF, we use the nonlinear baryonic power spectrum as an input to determine $f(m_\gamma^2;t)$. Another approach in the literature is to add a bias parameter $b$, a constant ratio between baryonic fluctuations and matter fluctuations~\cite{Dekel:1998eq},  to the log-normal distribution. This has been shown to be effective in modeling the one-point PDF extracted from galaxy count surveys~\cite{Wild:2004me,Hurtado-Gil:2017dbm}. 

As an independent check of the nonlinear baryonic power spectrum that we obtained from simulations, we use the log-normal with bias PDF together with the Halofit nonlinear \textit{matter} power spectrum provided by CLASS~\cite{Mead:2016zqy} with a bias parameter $b$ within the range of fit values obtained in Ref.~\cite{Hurtado-Gil:2017dbm}. This serves as an independent way of modeling the PDF without relying on the simulation results that we use for the fiducial log-normal distribution. We stress however that this model is unphysical, since negative fluctuations in the baryon density are mathematically allowed, calling into question the accuracy of the PDF for underdensities. 

\noindent
{\bf Voids.---}The simulation and characterization of voids in our Universe has been the subject of ongoing interest ~\cite{Zeldovich:1982zz,Pisani:2019cvo,Einasto:2011eu,Jennings:2013nsa,Chan:2014qka,Plionis:2001gr,Schuster:2019hyl}, and can inform the PDF $f(m_\gamma^2;t)$ for values of $m_\gamma^2$ significantly below $\overline{m_\gamma^2}$. To construct $f(m_\gamma^2;t)$ from these studies, we rely on the simulation results in Ref.~\cite{Adermann:2018jba}, which provides a PDF for the mean density of voids. Together with the PDF for the volume of voids and the number of voids in their simulation, we find that voids typically occupy $f_\text{void} \sim 10\%$ of their simulation volume, and we rescale the void density PDF by this number to obtain a model of the PDF of finding a void of a certain mean density in our Universe and hence $f(m_\gamma^2;t)$. 

This approach gives an estimate of $f(m_\gamma^2;t)$ only for values of $m_\gamma^2$ below the homogeneous value and should not be used for overdensities. Even so, it is highly conservative for two reasons: first, not all underdensities are local minima in position space, which is the working definition of a void, and second, we do not account for the density profile of the void, which neglects the fact that the centers of voids are likely to be significantly less dense than the mean density. Nevertheless, comparing this PDF with our fiducial choices can give us confidence in our modeling of underdensities.

\noindent
{\bf Gaussian.---}Fluctuations in densities deep in the linear regime ($z \gg 200$) are well-described by a Gaussian random field. In this limit, $f(m_\gamma^2;t)$ takes on a particularly simple form, making it useful for an intuitive understanding of our results. With a Gaussian PDF, the differential conversion probability in Eq.~\eqref{eq:prob} is given by (see~\citetalias{OurLongPaper})
\begin{multline}
    \frac{\dd \langle P_{\gamma \to A'} \rangle_\text{G}}{\dd z} = \frac{\pi \mAp^4 \epsilon^2}{\overline{m_\gamma^2}(z) \omega(z)} \left|\frac{\dd t}{\dd z}\right| \\
    \times \frac{1}{\sqrt{2\pi \sigma_\text{b}^2}} \exp \left[- \frac{( \mAp^2 / \overline{m_\gamma^2} - 1 )^2}{2 \sigma_\text{b}^2}\right],
    \label{eq:dP_dz}
\end{multline}    
where $\sigma_\text{b}$ is the variance of baryon fluctuations. At late times, $\sigma_\text{b}$ attains values of one or larger, defining the nonlinear regime. Fluctuations with $\delta < -1$ can occur with significant probability. Hence, the Gaussian distribution is not suitable for describing fluctuations at low redshifts on which our results critically depend; results using the Gaussian PDF should be taken as pedagogically interesting, but incorrect.

Fig.~\ref{fig:PDFs} shows a plot of the different PDFs discussed here. Within the range of $10^{-2} < 1 + \delta < 10^2$, we can see that the two fiducial PDFs agree very well; outside this range, however, significant differences develop across all PDFs. Despite being highly conservative, the PDF constructed from voids generally agrees well with both the analytic and log-normal PDFs in the range $10^{-2} < 1 + \delta < 10^{-1}$, while the log-normal with bias PDF shows good agreement with the fiducial log-normal PDF for $1 + \delta \gtrsim 1$, even though they use different power spectra as inputs. We stress that we expect neither the void PDF nor the log-normal PDF with bias to agree with our fiducial results outside of the respective ranges specified here. 

The top-right panel of Fig.~\ref{fig:limit_dp_systematics} and the right panel of Fig.~\ref{fig:limit_dp_dm_systematics} show the limits on $\epsilon$ obtained for $\gamma \to A'$ oscillations and dark matter $A' \to \gamma$ oscillations, respectively. The limits are qualitatively similar between the different PDFs in the relevant ranges of $1+\delta$, providing some reassurance that our fiducial choice is reasonably conservative. 

\begin{figure*}[tbp]
    \centering
    \includegraphics[width=0.47\textwidth]{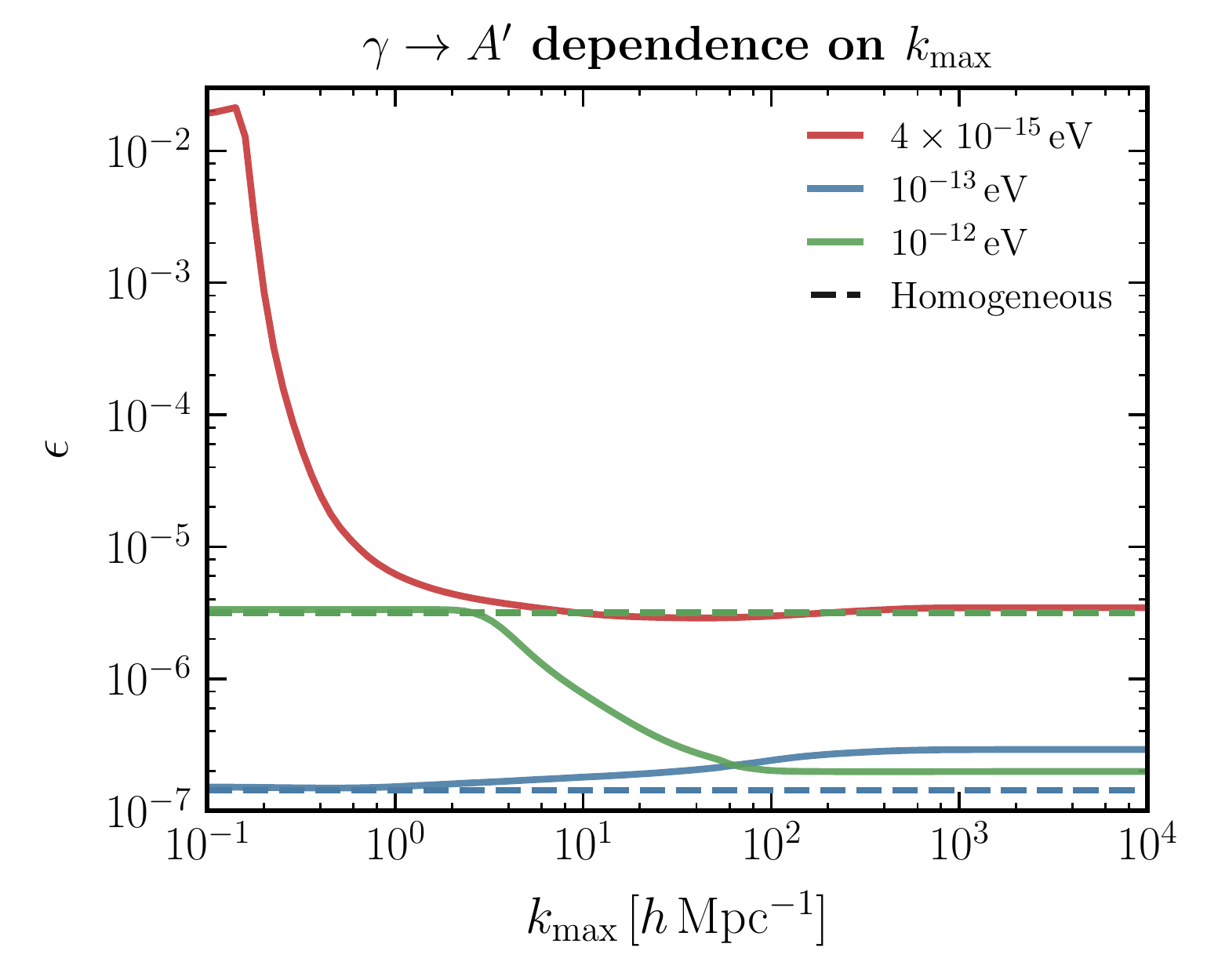}
    \includegraphics[width=0.47\textwidth]{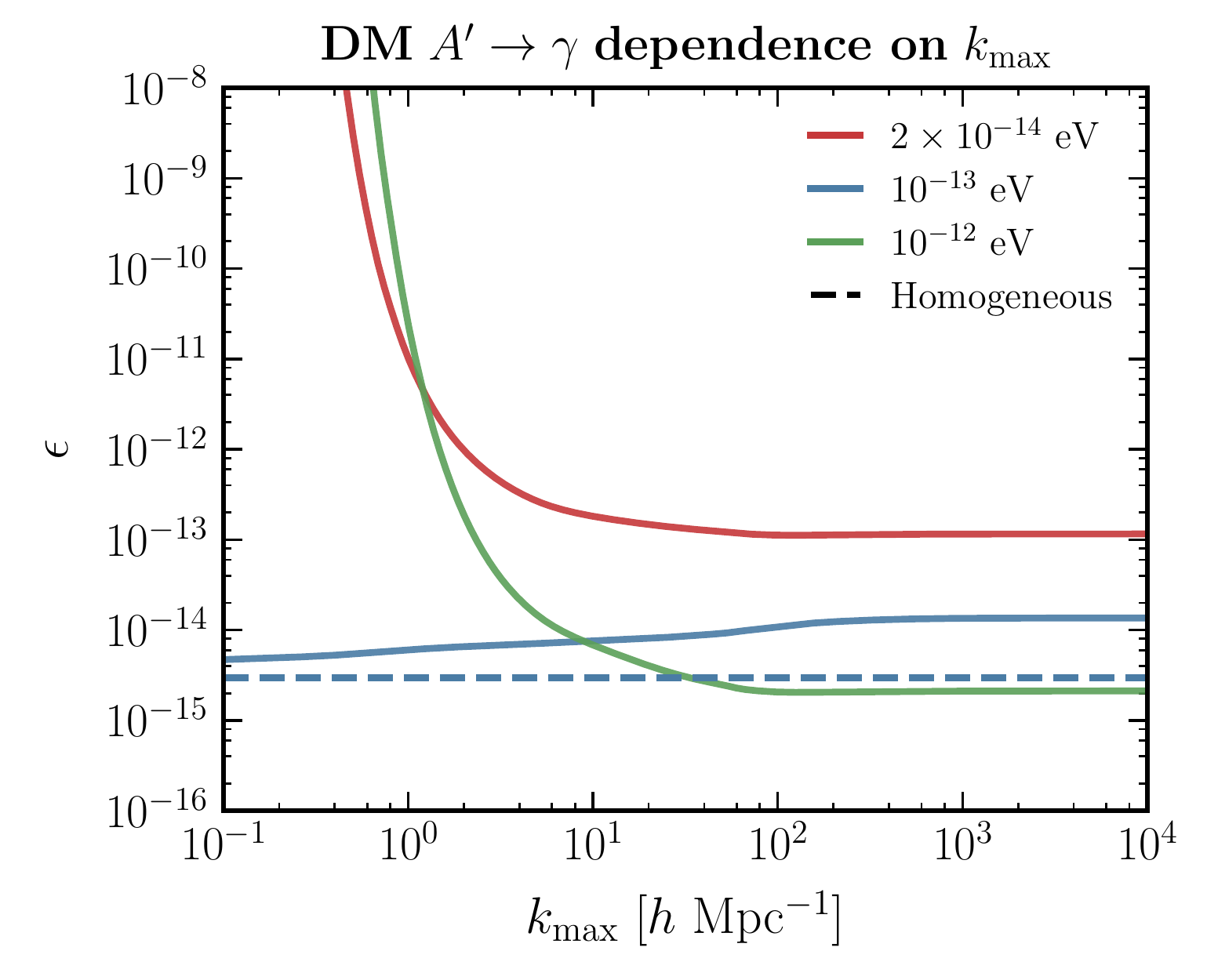}
    \caption{Effect of imposing a maximum cutoff on the scale of perturbations, $k_\mathrm{max}$, on the dark photon constraints for the log-normal PDF. In all cases, constraints stabilize around the baryon Jeans scale, $k_\mathrm{J}\sim \SI{300}{\h\per\mega\parsec}$. \emph{(Left)} Constraints on dark photons from $\gamma\to A'$, shown for benchmark masses $\mAp = \SI{2e-15}{}$ (solid red), $\SI{e-13}{}$ (solid red) and $\SI{e-12}{\eV}$ (solid green), respectively. The constraint in the homogeneous limit is shown for the latter two benchmark mass points (dashed), while for $\mAp = \SI{2e-15}{\eV}$ no conversions are accessible in the homogeneous limit.~\nblink{02_firas_dp_limits} \emph{(Right)} Constraints on dark photon dark matter from $A'\to \gamma$, shown for benchmark masses $\mAp = \SI{2e-14}{\eV}$ (solid red), $\SI{e-13}{\eV}$ (solid blue), and $\SI{e-12}{\eV}$ (solid green), respectively. The constraint in the homogeneous limit is shown for $\mAp = \SI{e-13}{\eV}$ (dashed blue), while no conversions are accessible in the homogeneous limit for the other two benchmark mass points shown.~\nblink{05_dp_dm_He_limits} } 
    \label{fig:limit_k_max_dep}
\end{figure*}

\subsection{Baryonic power spectrum}

The log-normal distribution for $f(m_\gamma^2;t)$ is fully characterized by two statistics: the mean, given by the homogeneous value $\overline{m_\gamma^2}$, and a variance in log-space $\Sigma^2$; the latter is defined through the usual variance of the baryon density fluctuations $\sigma_\text{b}^2$ as $\Sigma^2 \equiv \ln(1 + \sigma_\text{b}^2)$. Since $\sigma_\text{b}^2$ is directly proportional to the integral over the baryonic power spectrum (which can usually be taken to describe fluctuation in the free electron number density, as discussed in~\citetalias{OurLongPaper}), uncertainties in the baryonic power spectrum translate  into uncertainties in $f(m_\gamma^2;t)$.

To assess how significant these uncertainties are, we adopt two extremal prescriptions for the baryonic power spectrum, PS$_{\min}$ and PS$_{\max}$, corresponding to reasonable lower and upper bounds; PS$_{\min}$ leads to a narrower distribution $f(m_\gamma^2;t)$, while PS$_{\max}$ leads to a broader one. These are described in detail in~\citetalias{OurLongPaper}, and take into account typical uncertainties between different hydrodynamic simulations~\cite{Nelson:2018uso,McAlpine:2015tma,McCarthy:2016mry,Genel:2014lma}. In the main Letter, we always show the more conservative bound of the ones obtained with the two prescriptions. A maximum difference of $\mathcal O (15\%)$ in the mixing parameter constraint is obtained between the two different power spectrum prescriptions, which is expected since $\Sigma^2$ only has a log-dependence on the integral of the power spectrum.

\subsection{PDF tails}

If the log-normal or the analytic PDF continues to be a good description out to larger upward or downward fluctuations in $m_\gamma^2$, we can extend the acceptable range of $1+\delta$ for these PDFs. In the top left panel of Fig.~\ref{fig:limit_dp_systematics} and the left panel of Fig.~\ref{fig:limit_dp_dm_systematics} we show the effect of truncating the one-point PDF at different cutoffs in $1 + \delta$ for the $\gamma\to A'$ dark photon and $A' \to \gamma$ dark photon dark matter cases, respectively. These are shown for the log-normal PDF (solid blue) and analytic PDF (dashed red). Progressively darker lines corresponding to the inclusion of larger underdensities and overdensities, from 10 times larger and smaller than the mean plasma mass to $10^4$ times larger and smaller than the plasma mass, respectively.

Since the analytic PDF shows a strong cut-off in the probability of fluctuations below $10^{-2}$, extending the range of $1+\delta$ does not significantly improve our lower mass reach. For the log-normal distribution, however, an order of magnitude improvement in mass reach can be obtained with $10^{-4} < 1+\delta < 10^4$ as compared to our fiducial results. Both PDFs promise significant improvements at $m_{A'} > \SI{e-13}{\eV}$. This underscores the fact that pinning down the PDF of plasma fluctuations at the tails can significantly improve on the fiducial constraints presented in this work.

\subsection{Additional redshift variations}

We show in the bottom panel of Fig.~\ref{fig:limit_dp_systematics} the effect of excluding conversions over redshift ranges different from the ones considered in the main Letter. Excising a larger redshift range of $6 < z < 30$ (dashed orange) has no effect on the fiducial limits, obtained after excising $6 < z < 20$. This shows that our results are robust to the details of reionization; in particular, an earlier onset to reionization as allowed by \emph{Planck} 2018 data with a \texttt{FlexKnot} reionization parameterization~\cite{Aghanim:2018eyx} does not lead to any change in our limits.

Constraints relying solely on conversions before the deeply nonlinear cosmological epoch ($z \gtrsim 20$)  are also shown in Fig.~\ref{fig:limit_dp_systematics} for both the log-normal (solid blue) and analytic (dashed red) PDFs. Similar results are obtained with either prescription, as expected---the spectrum of fluctuations in the linear regime become increasingly Gaussian and are not subject to large systematics.

We further show constraints relying on conversions above $z > 0.1$ (solid green) and $z > 1$ (solid purple). The former has minimal impact on our fiducial constraints while the latter, in co\"ordination with our cut on the allowed fluctuation size $10^{-2} < 1 + \delta < 10^2$, restricts conversions for the lowest masses accessible to our fiducial analysis. \\

\subsection{Dependence of limits on smallest scale}

An understanding of the various scales at which fluctuations influence constraints from conversions in inhomogeneities is crucial. We show in Fig.~\ref{fig:limit_k_max_dep} the constraints in the fiducial log-normal prescription as a function of maximum cutoff on the scale of perturbations, $k_\mathrm{max}$, for a few benchmark masses. In all cases, constraints stabilize around the baryon Jeans scale, $k_\mathrm{J}\sim \SI{300}{\h\per\mega\parsec}$. In the left panel, we show results for dark photon constraints from $\gamma\to A'$, shown for benchmark masses $\mAp = 2\times10^{-15}, 10^{-13}$, and \SI{e-12}{\eV}, respectively. The constraint in the homogeneous limit is shown for the latter two benchmark mass points, while for $\mAp = \SI{2e-15}{\eV}$ no conversions are accessible in the homogeneous limit. In the right panel we show dark photon dark matter constraints from $A' \to \gamma$, shown for benchmark masses $\mAp = 2\times10^{-14}, 10^{-13}$, and \SI{e-12}{\eV} in solid red, blue, and green lines, respectively. The constraint in the homogeneous limit is shown for $\mAp = \SI{e-13}{\eV}$, while no conversions are accessible in the homogeneous limit for the other two benchmark mass points shown.

\section{Note added---energy deposition assumptions}
\label{app:energy_dep_assumptions}

\begin{figure}[htbp]
    \centering
    \includegraphics[width=0.47\textwidth]{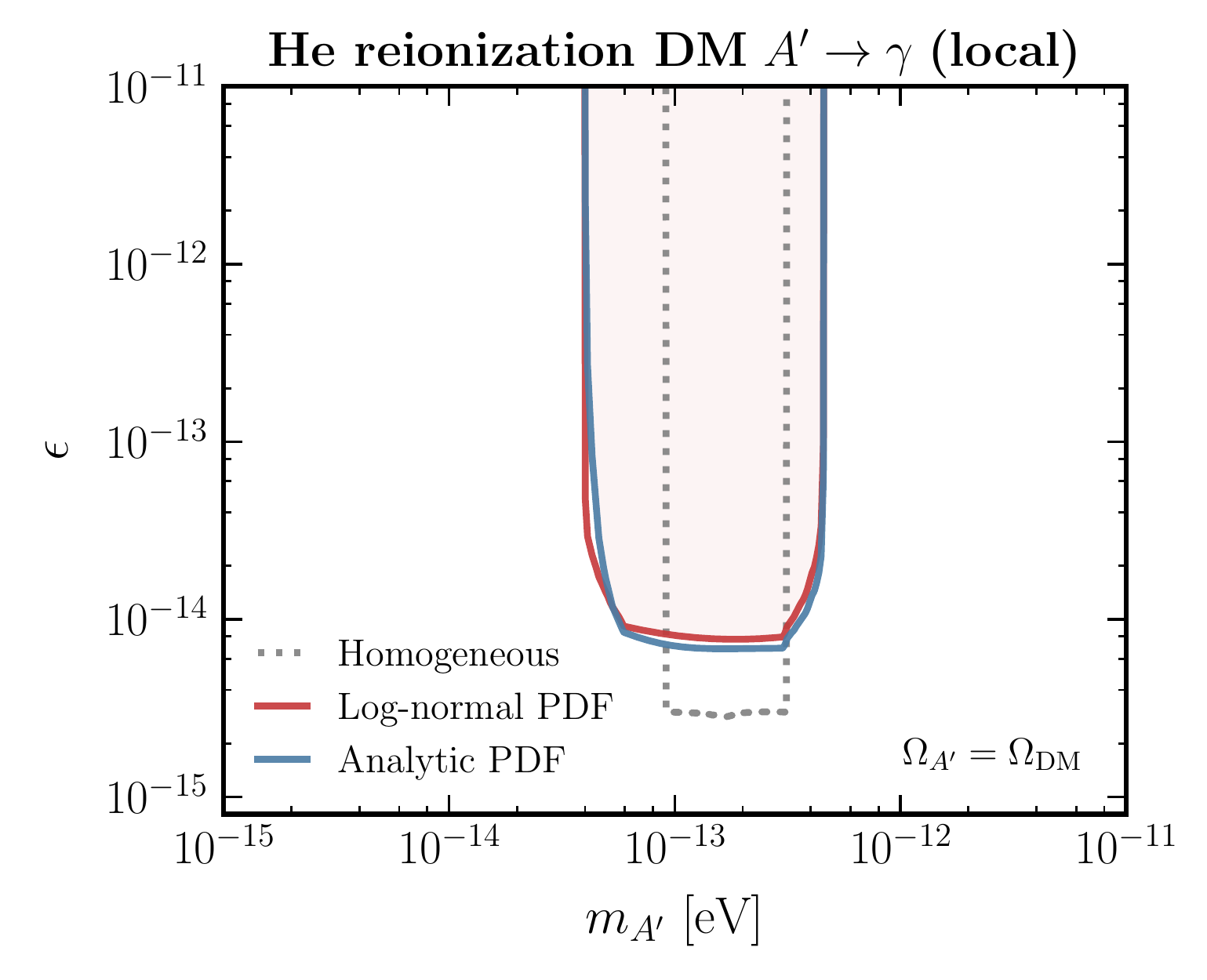}
    \caption{Constraints on dark photon dark matter from anomalous heating of the IGM during the epoch of HeII reionization with the local heating prescription, assuming log-normal (red) or analytic (blue) PDFs. Homogeneous constraints derived in Ref.~\cite{McDermott:2019lch} are also shown (dotted gray).~\nblink{11_dp_dm_He_local_limits}} 
    \label{fig:limit_DP_DM_local}
\end{figure}

Recently, Ref.~\cite{Witte:2020rvb} also presented constraints on dark photon dark matter from Ly$\alpha$ measurements of the IGM temperature during HeII reionization, taking into account inhomogeneities using a similar formalism. 
However, the authors claim that $A'$ conversions cause only local heating of the IGM, in contrast to the implicit assumption made in the main Letter that the energy deposited from $A' \to \gamma$ conversion is distributed evenly across the Universe. 
Understanding how the energy transport actually proceeds is complicated and beyond the scope of our work; for now, we simply present our constraints assuming that heating is local under some heuristic assumptions made by Ref.~\cite{Witte:2020rvb}, leaving a detailed comparison to~\citetalias{OurLongPaper}. 
Assuming that heating is only local, the total energy injected per unit baryon in the local region of conversion $\langle E_{A'\to \gamma} \rangle_\text{local}$ is
\begin{multline}
    \frac{\dd \langle E_{A' \to \gamma} \rangle_\text{local}}{\dd z} = \pi m_{A'} \epsilon^2 \frac{\overline{\rho}_{A'}}{\overline{n}_\text{b}} \left| \frac{\dd t}{\dd z} \right| \\
    \times \int \dd m_\gamma^2 \, f(m_\gamma^2;t) \delta_\text{D}(m_\gamma^2 - m_{A'}^2) m_\gamma^2 \,,
\end{multline}
where we have divided the integrand by a factor of $1 + \delta_\text{b}$, since the local baryon density is $(1 + \delta_\text{b}) \overline{n}_\text{b}$. 
Following Ref.~\cite{Witte:2020rvb}, we only consider regions of $\delta_\text{b}$ where the optical depth $\tau(z, \delta_\text{b})$ at redshift $z$ for Ly$\alpha$ radiation is given by $0.05 \leq \exp\left[-\tau(z, \delta_\text{b}) \right] \leq 0.95$, since Ly$\alpha$ flux power spectrum measurements contain no temperature information when Ly$\alpha$ photons are hardly absorbed or almost completely absorbed~\cite{Becker:2010cu}. 
Our constraints based on these assumptions are shown in Fig.~\ref{fig:limit_DP_DM_local}, and agree well with the equivalent results in Ref.~\cite{Witte:2020rvb}. 
We stress however that these results rely on a number of assumptions that are ultimately heuristic in nature, a point we discuss further in~\citetalias{OurLongPaper}.

\bibliographystyle{apsrev4-1}
\bibliography{firas-perturbations}

\begin{thebibliography}{108}%
\makeatletter
\providecommand \@ifxundefined [1]{%
 \@ifx{#1\undefined}
}%
\providecommand \@ifnum [1]{%
 \ifnum #1\expandafter \@firstoftwo
 \else \expandafter \@secondoftwo
 \fi
}%
\providecommand \@ifx [1]{%
 \ifx #1\expandafter \@firstoftwo
 \else \expandafter \@secondoftwo
 \fi
}%
\providecommand \natexlab [1]{#1}%
\providecommand \enquote  [1]{``#1''}%
\providecommand \bibnamefont  [1]{#1}%
\providecommand \bibfnamefont [1]{#1}%
\providecommand \citenamefont [1]{#1}%
\providecommand \href@noop [0]{\@secondoftwo}%
\providecommand \href [0]{\begingroup \@sanitize@url \@href}%
\providecommand \@href[1]{\@@startlink{#1}\@@href}%
\providecommand \@@href[1]{\endgroup#1\@@endlink}%
\providecommand \@sanitize@url [0]{\catcode `\\12\catcode `\$12\catcode
  `\&12\catcode `\#12\catcode `\^12\catcode `\_12\catcode `\%12\relax}%
\providecommand \@@startlink[1]{}%
\providecommand \@@endlink[0]{}%
\providecommand \url  [0]{\begingroup\@sanitize@url \@url }%
\providecommand \@url [1]{\endgroup\@href {#1}{\urlprefix }}%
\providecommand \urlprefix  [0]{URL }%
\providecommand \Eprint [0]{\href }%
\providecommand \doibase [0]{http://dx.doi.org/}%
\providecommand \selectlanguage [0]{\@gobble}%
\providecommand \bibinfo  [0]{\@secondoftwo}%
\providecommand \bibfield  [0]{\@secondoftwo}%
\providecommand \translation [1]{[#1]}%
\providecommand \BibitemOpen [0]{}%
\providecommand \bibitemStop [0]{}%
\providecommand \bibitemNoStop [0]{.\EOS\space}%
\providecommand \EOS [0]{\spacefactor3000\relax}%
\providecommand \BibitemShut  [1]{\csname bibitem#1\endcsname}%
\let\auto@bib@innerbib\@empty
\bibitem [{\citenamefont {Holdom}(1986)}]{Holdom:1985ag}%
  \BibitemOpen
  \bibfield  {author} {\bibinfo {author} {\bibfnamefont {B.}~\bibnamefont
  {Holdom}},\ }\href {\doibase 10.1016/0370-2693(86)91377-8} {\bibfield
  {journal} {\bibinfo  {journal} {Phys. Lett.}\ }\textbf {\bibinfo {volume}
  {166B}},\ \bibinfo {pages} {196} (\bibinfo {year} {1986})}\BibitemShut
  {NoStop}%
\bibitem [{\citenamefont {Redondo}\ and\ \citenamefont
  {Postma}(2009)}]{Redondo:2008ec}%
  \BibitemOpen
  \bibfield  {author} {\bibinfo {author} {\bibfnamefont {J.}~\bibnamefont
  {Redondo}}\ and\ \bibinfo {author} {\bibfnamefont {M.}~\bibnamefont
  {Postma}},\ }\href {\doibase 10.1088/1475-7516/2009/02/005} {\bibfield
  {journal} {\bibinfo  {journal} {JCAP}\ }\textbf {\bibinfo {volume} {0902}},\
  \bibinfo {pages} {005} (\bibinfo {year} {2009})},\ \Eprint
  {http://arxiv.org/abs/0811.0326} {arXiv:0811.0326 [hep-ph]} \BibitemShut
  {NoStop}%
\bibitem [{\citenamefont {Nelson}\ and\ \citenamefont
  {Scholtz}(2011)}]{Nelson:2011sf}%
  \BibitemOpen
  \bibfield  {author} {\bibinfo {author} {\bibfnamefont {A.~E.}\ \bibnamefont
  {Nelson}}\ and\ \bibinfo {author} {\bibfnamefont {J.}~\bibnamefont
  {Scholtz}},\ }\href {\doibase 10.1103/PhysRevD.84.103501} {\bibfield
  {journal} {\bibinfo  {journal} {Phys. Rev.}\ }\textbf {\bibinfo {volume}
  {D84}},\ \bibinfo {pages} {103501} (\bibinfo {year} {2011})},\ \Eprint
  {http://arxiv.org/abs/1105.2812} {arXiv:1105.2812 [hep-ph]} \BibitemShut
  {NoStop}%
\bibitem [{\citenamefont {Arias}\ \emph {et~al.}(2012)\citenamefont {Arias},
  \citenamefont {Cadamuro}, \citenamefont {Goodsell}, \citenamefont {Jaeckel},
  \citenamefont {Redondo},\ and\ \citenamefont {Ringwald}}]{Arias:2012az}%
  \BibitemOpen
  \bibfield  {author} {\bibinfo {author} {\bibfnamefont {P.}~\bibnamefont
  {Arias}}, \bibinfo {author} {\bibfnamefont {D.}~\bibnamefont {Cadamuro}},
  \bibinfo {author} {\bibfnamefont {M.}~\bibnamefont {Goodsell}}, \bibinfo
  {author} {\bibfnamefont {J.}~\bibnamefont {Jaeckel}}, \bibinfo {author}
  {\bibfnamefont {J.}~\bibnamefont {Redondo}}, \ and\ \bibinfo {author}
  {\bibfnamefont {A.}~\bibnamefont {Ringwald}},\ }\href {\doibase
  10.1088/1475-7516/2012/06/013} {\bibfield  {journal} {\bibinfo  {journal}
  {JCAP}\ }\textbf {\bibinfo {volume} {1206}},\ \bibinfo {pages} {013}
  (\bibinfo {year} {2012})},\ \Eprint {http://arxiv.org/abs/1201.5902}
  {arXiv:1201.5902 [hep-ph]} \BibitemShut {NoStop}%
\bibitem [{\citenamefont {Fradette}\ \emph {et~al.}(2014)\citenamefont
  {Fradette}, \citenamefont {Pospelov}, \citenamefont {Pradler},\ and\
  \citenamefont {Ritz}}]{Fradette:2014sza}%
  \BibitemOpen
  \bibfield  {author} {\bibinfo {author} {\bibfnamefont {A.}~\bibnamefont
  {Fradette}}, \bibinfo {author} {\bibfnamefont {M.}~\bibnamefont {Pospelov}},
  \bibinfo {author} {\bibfnamefont {J.}~\bibnamefont {Pradler}}, \ and\
  \bibinfo {author} {\bibfnamefont {A.}~\bibnamefont {Ritz}},\ }\href {\doibase
  10.1103/PhysRevD.90.035022} {\bibfield  {journal} {\bibinfo  {journal} {Phys.
  Rev.}\ }\textbf {\bibinfo {volume} {D90}},\ \bibinfo {pages} {035022}
  (\bibinfo {year} {2014})},\ \Eprint {http://arxiv.org/abs/1407.0993}
  {arXiv:1407.0993 [hep-ph]} \BibitemShut {NoStop}%
\bibitem [{\citenamefont {An}\ \emph {et~al.}(2015)\citenamefont {An},
  \citenamefont {Pospelov}, \citenamefont {Pradler},\ and\ \citenamefont
  {Ritz}}]{An:2014twa}%
  \BibitemOpen
  \bibfield  {author} {\bibinfo {author} {\bibfnamefont {H.}~\bibnamefont
  {An}}, \bibinfo {author} {\bibfnamefont {M.}~\bibnamefont {Pospelov}},
  \bibinfo {author} {\bibfnamefont {J.}~\bibnamefont {Pradler}}, \ and\
  \bibinfo {author} {\bibfnamefont {A.}~\bibnamefont {Ritz}},\ }\href {\doibase
  10.1016/j.physletb.2015.06.018} {\bibfield  {journal} {\bibinfo  {journal}
  {Phys. Lett.}\ }\textbf {\bibinfo {volume} {B747}},\ \bibinfo {pages} {331}
  (\bibinfo {year} {2015})},\ \Eprint {http://arxiv.org/abs/1412.8378}
  {arXiv:1412.8378 [hep-ph]} \BibitemShut {NoStop}%
\bibitem [{\citenamefont {Graham}\ \emph {et~al.}(2016)\citenamefont {Graham},
  \citenamefont {Mardon},\ and\ \citenamefont {Rajendran}}]{Graham:2015rva}%
  \BibitemOpen
  \bibfield  {author} {\bibinfo {author} {\bibfnamefont {P.~W.}\ \bibnamefont
  {Graham}}, \bibinfo {author} {\bibfnamefont {J.}~\bibnamefont {Mardon}}, \
  and\ \bibinfo {author} {\bibfnamefont {S.}~\bibnamefont {Rajendran}},\ }\href
  {\doibase 10.1103/PhysRevD.93.103520} {\bibfield  {journal} {\bibinfo
  {journal} {Phys. Rev.}\ }\textbf {\bibinfo {volume} {D93}},\ \bibinfo {pages}
  {103520} (\bibinfo {year} {2016})},\ \Eprint
  {http://arxiv.org/abs/1504.02102} {arXiv:1504.02102 [hep-ph]} \BibitemShut
  {NoStop}%
\bibitem [{\citenamefont {Agrawal}\ \emph {et~al.}(2020)\citenamefont
  {Agrawal}, \citenamefont {Kitajima}, \citenamefont {Reece}, \citenamefont
  {Sekiguchi},\ and\ \citenamefont {Takahashi}}]{Agrawal:2018vin}%
  \BibitemOpen
  \bibfield  {author} {\bibinfo {author} {\bibfnamefont {P.}~\bibnamefont
  {Agrawal}}, \bibinfo {author} {\bibfnamefont {N.}~\bibnamefont {Kitajima}},
  \bibinfo {author} {\bibfnamefont {M.}~\bibnamefont {Reece}}, \bibinfo
  {author} {\bibfnamefont {T.}~\bibnamefont {Sekiguchi}}, \ and\ \bibinfo
  {author} {\bibfnamefont {F.}~\bibnamefont {Takahashi}},\ }\href {\doibase
  10.1016/j.physletb.2019.135136} {\bibfield  {journal} {\bibinfo  {journal}
  {Phys. Lett.}\ }\textbf {\bibinfo {volume} {B801}},\ \bibinfo {pages}
  {135136} (\bibinfo {year} {2020})},\ \Eprint
  {http://arxiv.org/abs/1810.07188} {arXiv:1810.07188 [hep-ph]} \BibitemShut
  {NoStop}%
\bibitem [{\citenamefont {Dror}\ \emph {et~al.}(2019)\citenamefont {Dror},
  \citenamefont {Harigaya},\ and\ \citenamefont {Narayan}}]{Dror:2018pdh}%
  \BibitemOpen
  \bibfield  {author} {\bibinfo {author} {\bibfnamefont {J.~A.}\ \bibnamefont
  {Dror}}, \bibinfo {author} {\bibfnamefont {K.}~\bibnamefont {Harigaya}}, \
  and\ \bibinfo {author} {\bibfnamefont {V.}~\bibnamefont {Narayan}},\ }\href
  {\doibase 10.1103/PhysRevD.99.035036} {\bibfield  {journal} {\bibinfo
  {journal} {Phys. Rev.}\ }\textbf {\bibinfo {volume} {D99}},\ \bibinfo {pages}
  {035036} (\bibinfo {year} {2019})},\ \Eprint
  {http://arxiv.org/abs/1810.07195} {arXiv:1810.07195 [hep-ph]} \BibitemShut
  {NoStop}%
\bibitem [{\citenamefont {Co}\ \emph {et~al.}(2019)\citenamefont {Co},
  \citenamefont {Pierce}, \citenamefont {Zhang},\ and\ \citenamefont
  {Zhao}}]{Co:2018lka}%
  \BibitemOpen
  \bibfield  {author} {\bibinfo {author} {\bibfnamefont {R.~T.}\ \bibnamefont
  {Co}}, \bibinfo {author} {\bibfnamefont {A.}~\bibnamefont {Pierce}}, \bibinfo
  {author} {\bibfnamefont {Z.}~\bibnamefont {Zhang}}, \ and\ \bibinfo {author}
  {\bibfnamefont {Y.}~\bibnamefont {Zhao}},\ }\href {\doibase
  10.1103/PhysRevD.99.075002} {\bibfield  {journal} {\bibinfo  {journal} {Phys.
  Rev.}\ }\textbf {\bibinfo {volume} {D99}},\ \bibinfo {pages} {075002}
  (\bibinfo {year} {2019})},\ \Eprint {http://arxiv.org/abs/1810.07196}
  {arXiv:1810.07196 [hep-ph]} \BibitemShut {NoStop}%
\bibitem [{\citenamefont {Bastero-Gil}\ \emph {et~al.}(2019)\citenamefont
  {Bastero-Gil}, \citenamefont {Santiago}, \citenamefont {Ubaldi},\ and\
  \citenamefont {Vega-Morales}}]{Bastero-Gil:2018uel}%
  \BibitemOpen
  \bibfield  {author} {\bibinfo {author} {\bibfnamefont {M.}~\bibnamefont
  {Bastero-Gil}}, \bibinfo {author} {\bibfnamefont {J.}~\bibnamefont
  {Santiago}}, \bibinfo {author} {\bibfnamefont {L.}~\bibnamefont {Ubaldi}}, \
  and\ \bibinfo {author} {\bibfnamefont {R.}~\bibnamefont {Vega-Morales}},\
  }\href {\doibase 10.1088/1475-7516/2019/04/015} {\bibfield  {journal}
  {\bibinfo  {journal} {JCAP}\ }\textbf {\bibinfo {volume} {1904}},\ \bibinfo
  {pages} {015} (\bibinfo {year} {2019})},\ \Eprint
  {http://arxiv.org/abs/1810.07208} {arXiv:1810.07208 [hep-ph]} \BibitemShut
  {NoStop}%
\bibitem [{\citenamefont {Long}\ and\ \citenamefont
  {Wang}(2019)}]{Long:2019lwl}%
  \BibitemOpen
  \bibfield  {author} {\bibinfo {author} {\bibfnamefont {A.~J.}\ \bibnamefont
  {Long}}\ and\ \bibinfo {author} {\bibfnamefont {L.-T.}\ \bibnamefont
  {Wang}},\ }\href {\doibase 10.1103/PhysRevD.99.063529} {\bibfield  {journal}
  {\bibinfo  {journal} {Phys. Rev.}\ }\textbf {\bibinfo {volume} {D99}},\
  \bibinfo {pages} {063529} (\bibinfo {year} {2019})},\ \Eprint
  {http://arxiv.org/abs/1901.03312} {arXiv:1901.03312 [hep-ph]} \BibitemShut
  {NoStop}%
\bibitem [{\citenamefont {Dienes}\ \emph {et~al.}(1997)\citenamefont {Dienes},
  \citenamefont {Kolda},\ and\ \citenamefont {March-Russell}}]{Dienes:1996zr}%
  \BibitemOpen
  \bibfield  {author} {\bibinfo {author} {\bibfnamefont {K.~R.}\ \bibnamefont
  {Dienes}}, \bibinfo {author} {\bibfnamefont {C.~F.}\ \bibnamefont {Kolda}}, \
  and\ \bibinfo {author} {\bibfnamefont {J.}~\bibnamefont {March-Russell}},\
  }\href {\doibase 10.1016/S0550-3213(97)80028-4,
  10.1016/S0550-3213(97)00173-9} {\bibfield  {journal} {\bibinfo  {journal}
  {Nucl. Phys.}\ }\textbf {\bibinfo {volume} {B492}},\ \bibinfo {pages} {104}
  (\bibinfo {year} {1997})},\ \Eprint {http://arxiv.org/abs/hep-ph/9610479}
  {arXiv:hep-ph/9610479 [hep-ph]} \BibitemShut {NoStop}%
\bibitem [{\citenamefont {Goodsell}\ \emph {et~al.}(2009)\citenamefont
  {Goodsell}, \citenamefont {Jaeckel}, \citenamefont {Redondo},\ and\
  \citenamefont {Ringwald}}]{Goodsell:2009xc}%
  \BibitemOpen
  \bibfield  {author} {\bibinfo {author} {\bibfnamefont {M.}~\bibnamefont
  {Goodsell}}, \bibinfo {author} {\bibfnamefont {J.}~\bibnamefont {Jaeckel}},
  \bibinfo {author} {\bibfnamefont {J.}~\bibnamefont {Redondo}}, \ and\
  \bibinfo {author} {\bibfnamefont {A.}~\bibnamefont {Ringwald}},\ }\href
  {\doibase 10.1088/1126-6708/2009/11/027} {\bibfield  {journal} {\bibinfo
  {journal} {JHEP}\ }\textbf {\bibinfo {volume} {11}},\ \bibinfo {pages} {027}
  (\bibinfo {year} {2009})},\ \Eprint {http://arxiv.org/abs/0909.0515}
  {arXiv:0909.0515 [hep-ph]} \BibitemShut {NoStop}%
\bibitem [{\citenamefont {Goodsell}\ and\ \citenamefont
  {Ringwald}(2010)}]{Goodsell:2010ie}%
  \BibitemOpen
  \bibfield  {author} {\bibinfo {author} {\bibfnamefont {M.}~\bibnamefont
  {Goodsell}}\ and\ \bibinfo {author} {\bibfnamefont {A.}~\bibnamefont
  {Ringwald}},\ }\href {\doibase 10.1002/prop.201000026} {\bibfield  {journal}
  {\bibinfo  {journal} {Fortsch. Phys.}\ }\textbf {\bibinfo {volume} {58}},\
  \bibinfo {pages} {716} (\bibinfo {year} {2010})},\ \Eprint
  {http://arxiv.org/abs/1002.1840} {arXiv:1002.1840 [hep-th]} \BibitemShut
  {NoStop}%
\bibitem [{\citenamefont {Goodsell}(2009)}]{Goodsell:2009pi}%
  \BibitemOpen
  \bibfield  {author} {\bibinfo {author} {\bibfnamefont {M.}~\bibnamefont
  {Goodsell}},\ }in\ \href {\doibase 10.3204/DESY-PROC-2009-05/goodsell\_mark}
  {\emph {\bibinfo {booktitle} {{5th Patras Workshop on Axions, WIMPs and
  WISPs}}}}\ (\bibinfo {year} {2009})\ pp.\ \bibinfo {pages} {165--168},\
  \Eprint {http://arxiv.org/abs/0912.4206} {arXiv:0912.4206 [hep-th]}
  \BibitemShut {NoStop}%
\bibitem [{\citenamefont {Abel}\ and\ \citenamefont
  {Schofield}(2004)}]{Abel:2003ue}%
  \BibitemOpen
  \bibfield  {author} {\bibinfo {author} {\bibfnamefont {S.~A.}\ \bibnamefont
  {Abel}}\ and\ \bibinfo {author} {\bibfnamefont {B.~W.}\ \bibnamefont
  {Schofield}},\ }\href {\doibase 10.1016/j.nuclphysb.2004.02.037} {\bibfield
  {journal} {\bibinfo  {journal} {Nucl. Phys.}\ }\textbf {\bibinfo {volume}
  {B685}},\ \bibinfo {pages} {150} (\bibinfo {year} {2004})},\ \Eprint
  {http://arxiv.org/abs/hep-th/0311051} {arXiv:hep-th/0311051 [hep-th]}
  \BibitemShut {NoStop}%
\bibitem [{\citenamefont {Abel}\ \emph
  {et~al.}(2008{\natexlab{a}})\citenamefont {Abel}, \citenamefont {Jaeckel},
  \citenamefont {Khoze},\ and\ \citenamefont {Ringwald}}]{Abel:2006qt}%
  \BibitemOpen
  \bibfield  {author} {\bibinfo {author} {\bibfnamefont {S.~A.}\ \bibnamefont
  {Abel}}, \bibinfo {author} {\bibfnamefont {J.}~\bibnamefont {Jaeckel}},
  \bibinfo {author} {\bibfnamefont {V.~V.}\ \bibnamefont {Khoze}}, \ and\
  \bibinfo {author} {\bibfnamefont {A.}~\bibnamefont {Ringwald}},\ }\href
  {\doibase 10.1016/j.physletb.2008.03.076} {\bibfield  {journal} {\bibinfo
  {journal} {Phys. Lett.}\ }\textbf {\bibinfo {volume} {B666}},\ \bibinfo
  {pages} {66} (\bibinfo {year} {2008}{\natexlab{a}})},\ \Eprint
  {http://arxiv.org/abs/hep-ph/0608248} {arXiv:hep-ph/0608248 [hep-ph]}
  \BibitemShut {NoStop}%
\bibitem [{\citenamefont {Abel}\ \emph
  {et~al.}(2008{\natexlab{b}})\citenamefont {Abel}, \citenamefont {Goodsell},
  \citenamefont {Jaeckel}, \citenamefont {Khoze},\ and\ \citenamefont
  {Ringwald}}]{Abel:2008ai}%
  \BibitemOpen
  \bibfield  {author} {\bibinfo {author} {\bibfnamefont {S.~A.}\ \bibnamefont
  {Abel}}, \bibinfo {author} {\bibfnamefont {M.~D.}\ \bibnamefont {Goodsell}},
  \bibinfo {author} {\bibfnamefont {J.}~\bibnamefont {Jaeckel}}, \bibinfo
  {author} {\bibfnamefont {V.~V.}\ \bibnamefont {Khoze}}, \ and\ \bibinfo
  {author} {\bibfnamefont {A.}~\bibnamefont {Ringwald}},\ }\href {\doibase
  10.1088/1126-6708/2008/07/124} {\bibfield  {journal} {\bibinfo  {journal}
  {JHEP}\ }\textbf {\bibinfo {volume} {07}},\ \bibinfo {pages} {124} (\bibinfo
  {year} {2008}{\natexlab{b}})},\ \Eprint {http://arxiv.org/abs/0803.1449}
  {arXiv:0803.1449 [hep-ph]} \BibitemShut {NoStop}%
\bibitem [{\citenamefont {Mirizzi}\ \emph
  {et~al.}(2009{\natexlab{a}})\citenamefont {Mirizzi}, \citenamefont
  {Redondo},\ and\ \citenamefont {Sigl}}]{Mirizzi:2009iz}%
  \BibitemOpen
  \bibfield  {author} {\bibinfo {author} {\bibfnamefont {A.}~\bibnamefont
  {Mirizzi}}, \bibinfo {author} {\bibfnamefont {J.}~\bibnamefont {Redondo}}, \
  and\ \bibinfo {author} {\bibfnamefont {G.}~\bibnamefont {Sigl}},\ }\href
  {\doibase 10.1088/1475-7516/2009/03/026} {\bibfield  {journal} {\bibinfo
  {journal} {JCAP}\ }\textbf {\bibinfo {volume} {0903}},\ \bibinfo {pages}
  {026} (\bibinfo {year} {2009}{\natexlab{a}})},\ \Eprint
  {http://arxiv.org/abs/0901.0014} {arXiv:0901.0014 [hep-ph]} \BibitemShut
  {NoStop}%
\bibitem [{\citenamefont {Kunze}\ and\ \citenamefont
  {V{\'a}zquez-Mozo}(2015)}]{Kunze:2015noa}%
  \BibitemOpen
  \bibfield  {author} {\bibinfo {author} {\bibfnamefont {K.~E.}\ \bibnamefont
  {Kunze}}\ and\ \bibinfo {author} {\bibfnamefont {M.~{\'A}.}\ \bibnamefont
  {V{\'a}zquez-Mozo}},\ }\href {\doibase 10.1088/1475-7516/2015/12/028}
  {\bibfield  {journal} {\bibinfo  {journal} {JCAP}\ }\textbf {\bibinfo
  {volume} {1512}},\ \bibinfo {pages} {028} (\bibinfo {year} {2015})},\ \Eprint
  {http://arxiv.org/abs/1507.02614} {arXiv:1507.02614 [astro-ph.CO]}
  \BibitemShut {NoStop}%
\bibitem [{\citenamefont {Fixsen}\ \emph {et~al.}(1996)\citenamefont {Fixsen},
  \citenamefont {Cheng}, \citenamefont {Gales}, \citenamefont {Mather},
  \citenamefont {Shafer},\ and\ \citenamefont {Wright}}]{Fixsen:1996nj}%
  \BibitemOpen
  \bibfield  {author} {\bibinfo {author} {\bibfnamefont {D.~J.}\ \bibnamefont
  {Fixsen}}, \bibinfo {author} {\bibfnamefont {E.~S.}\ \bibnamefont {Cheng}},
  \bibinfo {author} {\bibfnamefont {J.~M.}\ \bibnamefont {Gales}}, \bibinfo
  {author} {\bibfnamefont {J.~C.}\ \bibnamefont {Mather}}, \bibinfo {author}
  {\bibfnamefont {R.~A.}\ \bibnamefont {Shafer}}, \ and\ \bibinfo {author}
  {\bibfnamefont {E.~L.}\ \bibnamefont {Wright}},\ }\href {\doibase
  10.1086/178173} {\bibfield  {journal} {\bibinfo  {journal} {Astrophys. J.}\
  }\textbf {\bibinfo {volume} {473}},\ \bibinfo {pages} {576} (\bibinfo {year}
  {1996})},\ \Eprint {http://arxiv.org/abs/astro-ph/9605054}
  {arXiv:astro-ph/9605054 [astro-ph]} \BibitemShut {NoStop}%
\bibitem [{\citenamefont {McDermott}\ and\ \citenamefont
  {Witte}(2019)}]{McDermott:2019lch}%
  \BibitemOpen
  \bibfield  {author} {\bibinfo {author} {\bibfnamefont {S.~D.}\ \bibnamefont
  {McDermott}}\ and\ \bibinfo {author} {\bibfnamefont {S.~J.}\ \bibnamefont
  {Witte}},\ }\href@noop {} {\  (\bibinfo {year} {2019})},\ \Eprint
  {http://arxiv.org/abs/1911.05086} {arXiv:1911.05086 [hep-ph]} \BibitemShut
  {NoStop}%
\bibitem [{\citenamefont {Dubovsky}\ and\ \citenamefont
  {Hern\'andez-Chifflet}(2015)}]{Dubovsky:2015cca}%
  \BibitemOpen
  \bibfield  {author} {\bibinfo {author} {\bibfnamefont {S.}~\bibnamefont
  {Dubovsky}}\ and\ \bibinfo {author} {\bibfnamefont {G.}~\bibnamefont
  {Hern\'andez-Chifflet}},\ }\href {\doibase 10.1088/1475-7516/2015/12/054}
  {\bibfield  {journal} {\bibinfo  {journal} {JCAP}\ }\textbf {\bibinfo
  {volume} {1512}},\ \bibinfo {pages} {054} (\bibinfo {year} {2015})},\ \Eprint
  {http://arxiv.org/abs/1509.00039} {arXiv:1509.00039 [hep-ph]} \BibitemShut
  {NoStop}%
\bibitem [{\citenamefont {Kovetz}\ \emph {et~al.}(2019)\citenamefont {Kovetz},
  \citenamefont {Cholis},\ and\ \citenamefont {Kaplan}}]{Kovetz:2018zes}%
  \BibitemOpen
  \bibfield  {author} {\bibinfo {author} {\bibfnamefont {E.~D.}\ \bibnamefont
  {Kovetz}}, \bibinfo {author} {\bibfnamefont {I.}~\bibnamefont {Cholis}}, \
  and\ \bibinfo {author} {\bibfnamefont {D.~E.}\ \bibnamefont {Kaplan}},\
  }\href {\doibase 10.1103/PhysRevD.99.123511} {\bibfield  {journal} {\bibinfo
  {journal} {Phys. Rev.}\ }\textbf {\bibinfo {volume} {D99}},\ \bibinfo {pages}
  {123511} (\bibinfo {year} {2019})},\ \Eprint
  {http://arxiv.org/abs/1809.01139} {arXiv:1809.01139 [astro-ph.CO]}
  \BibitemShut {NoStop}%
\bibitem [{\citenamefont {Wadekar}\ and\ \citenamefont
  {Farrar}(2019)}]{Wadekar:2019xnf}%
  \BibitemOpen
  \bibfield  {author} {\bibinfo {author} {\bibfnamefont {D.}~\bibnamefont
  {Wadekar}}\ and\ \bibinfo {author} {\bibfnamefont {G.~R.}\ \bibnamefont
  {Farrar}},\ }\href@noop {} {\  (\bibinfo {year} {2019})},\ \Eprint
  {http://arxiv.org/abs/1903.12190} {arXiv:1903.12190 [hep-ph]} \BibitemShut
  {NoStop}%
\bibitem [{\citenamefont {Chaudhuri}\ \emph {et~al.}(2015)\citenamefont
  {Chaudhuri}, \citenamefont {Graham}, \citenamefont {Irwin}, \citenamefont
  {Mardon}, \citenamefont {Rajendran},\ and\ \citenamefont
  {Zhao}}]{Chaudhuri:2014dla}%
  \BibitemOpen
  \bibfield  {author} {\bibinfo {author} {\bibfnamefont {S.}~\bibnamefont
  {Chaudhuri}}, \bibinfo {author} {\bibfnamefont {P.~W.}\ \bibnamefont
  {Graham}}, \bibinfo {author} {\bibfnamefont {K.}~\bibnamefont {Irwin}},
  \bibinfo {author} {\bibfnamefont {J.}~\bibnamefont {Mardon}}, \bibinfo
  {author} {\bibfnamefont {S.}~\bibnamefont {Rajendran}}, \ and\ \bibinfo
  {author} {\bibfnamefont {Y.}~\bibnamefont {Zhao}},\ }\href {\doibase
  10.1103/PhysRevD.92.075012} {\bibfield  {journal} {\bibinfo  {journal} {Phys.
  Rev.}\ }\textbf {\bibinfo {volume} {D92}},\ \bibinfo {pages} {075012}
  (\bibinfo {year} {2015})},\ \Eprint {http://arxiv.org/abs/1411.7382}
  {arXiv:1411.7382 [hep-ph]} \BibitemShut {NoStop}%
\bibitem [{\citenamefont {Silva-Feaver}\ \emph {et~al.}(2017)\citenamefont
  {Silva-Feaver} \emph {et~al.}}]{Silva-Feaver:2016qhh}%
  \BibitemOpen
  \bibfield  {author} {\bibinfo {author} {\bibfnamefont {M.}~\bibnamefont
  {Silva-Feaver}} \emph {et~al.},\ }\bibfield  {booktitle} {\emph {\bibinfo
  {booktitle} {{Proceedings, Applied Superconductivity Conference (ASC 2016):
  Denver, Colorado, September 4-9, 2016}}},\ }\href {\doibase
  10.1109/TASC.2016.2631425} {\bibfield  {journal} {\bibinfo  {journal} {IEEE
  Trans. Appl. Supercond.}\ }\textbf {\bibinfo {volume} {27}},\ \bibinfo
  {pages} {1400204} (\bibinfo {year} {2017})},\ \Eprint
  {http://arxiv.org/abs/1610.09344} {arXiv:1610.09344 [astro-ph.IM]}
  \BibitemShut {NoStop}%
\bibitem [{\citenamefont {Harnik}(2019)}]{HarnikSRF}%
  \BibitemOpen
  \bibfield  {author} {\bibinfo {author} {\bibfnamefont {R.}~\bibnamefont
  {Harnik}},\ }\href
  {https://indico.fnal.gov/event/19433/session/2/contribution/1/material/slides/0.pdf}
  {\enquote {\bibinfo {title} {Dark {SRF} - theory},}\ } (\bibinfo {year}
  {2019}),\ \bibinfo {note} {2019 January PAC Meeting}\BibitemShut {NoStop}%
\bibitem [{\citenamefont {Grassellino}(2019)}]{GrassellinoSRF}%
  \BibitemOpen
  \bibfield  {author} {\bibinfo {author} {\bibfnamefont {A.}~\bibnamefont
  {Grassellino}},\ }\href
  {https://indico.fnal.gov/event/19433/session/2/contribution/2/material/slides/0.pdf}
  {\enquote {\bibinfo {title} {Dark {SRF} - experiment},}\ } (\bibinfo {year}
  {2019}),\ \bibinfo {note} {2019 January PAC Meeting}\BibitemShut {NoStop}%
\bibitem [{\citenamefont {Graham}\ \emph {et~al.}(2014)\citenamefont {Graham},
  \citenamefont {Mardon}, \citenamefont {Rajendran},\ and\ \citenamefont
  {Zhao}}]{Graham:2014sha}%
  \BibitemOpen
  \bibfield  {author} {\bibinfo {author} {\bibfnamefont {P.~W.}\ \bibnamefont
  {Graham}}, \bibinfo {author} {\bibfnamefont {J.}~\bibnamefont {Mardon}},
  \bibinfo {author} {\bibfnamefont {S.}~\bibnamefont {Rajendran}}, \ and\
  \bibinfo {author} {\bibfnamefont {Y.}~\bibnamefont {Zhao}},\ }\href {\doibase
  10.1103/PhysRevD.90.075017} {\bibfield  {journal} {\bibinfo  {journal} {Phys.
  Rev.}\ }\textbf {\bibinfo {volume} {D90}},\ \bibinfo {pages} {075017}
  (\bibinfo {year} {2014})},\ \Eprint {http://arxiv.org/abs/1407.4806}
  {arXiv:1407.4806 [hep-ph]} \BibitemShut {NoStop}%
\bibitem [{\citenamefont {Caputo}\ \emph
  {et~al.}(2020{\natexlab{a}})\citenamefont {Caputo}, \citenamefont {Liu},
  \citenamefont {Mishra-Sharma},\ and\ \citenamefont
  {Ruderman}}]{OurLongPaper}%
  \BibitemOpen
  \bibfield  {author} {\bibinfo {author} {\bibfnamefont {A.}~\bibnamefont
  {Caputo}}, \bibinfo {author} {\bibfnamefont {H.}~\bibnamefont {Liu}},
  \bibinfo {author} {\bibfnamefont {S.}~\bibnamefont {Mishra-Sharma}}, \ and\
  \bibinfo {author} {\bibfnamefont {J.~T.}\ \bibnamefont {Ruderman}},\
  }\href@noop {} {\  (\bibinfo {year} {2020}{\natexlab{a}})},\ \Eprint
  {http://arxiv.org/abs/2004.06733} {arXiv:2004.06733 [astro-ph.CO]}
  \BibitemShut {NoStop}%
\bibitem [{\citenamefont {Aghanim}\ \emph {et~al.}(2019)\citenamefont {Aghanim}
  \emph {et~al.}}]{Aghanim:2019ame}%
  \BibitemOpen
  \bibfield  {author} {\bibinfo {author} {\bibfnamefont {N.}~\bibnamefont
  {Aghanim}} \emph {et~al.} (\bibinfo {collaboration} {Planck}),\ }\href@noop
  {} {\  (\bibinfo {year} {2019})},\ \Eprint {http://arxiv.org/abs/1907.12875}
  {arXiv:1907.12875 [astro-ph.CO]} \BibitemShut {NoStop}%
\bibitem [{Ghe(2019)}]{Gherghetta:2019coi}%
  \BibitemOpen
  \href {\doibase 10.1103/PhysRevD.100.095001} {\bibfield  {journal} {\bibinfo
  {journal} {Phys. Rev.}\ }\textbf {\bibinfo {volume} {D100}},\ \bibinfo
  {pages} {095001} (\bibinfo {year} {2019})},\ \Eprint
  {http://arxiv.org/abs/1909.00696} {arXiv:1909.00696 [hep-ph]} \BibitemShut
  {NoStop}%
\bibitem [{\citenamefont {Blas}\ \emph {et~al.}(2011)\citenamefont {Blas},
  \citenamefont {Lesgourgues},\ and\ \citenamefont {Tram}}]{Blas:2011rf}%
  \BibitemOpen
  \bibfield  {author} {\bibinfo {author} {\bibfnamefont {D.}~\bibnamefont
  {Blas}}, \bibinfo {author} {\bibfnamefont {J.}~\bibnamefont {Lesgourgues}}, \
  and\ \bibinfo {author} {\bibfnamefont {T.}~\bibnamefont {Tram}},\ }\href
  {\doibase 10.1088/1475-7516/2011/07/034} {\bibfield  {journal} {\bibinfo
  {journal} {JCAP}\ }\textbf {\bibinfo {volume} {1107}},\ \bibinfo {pages}
  {034} (\bibinfo {year} {2011})},\ \Eprint {http://arxiv.org/abs/1104.2933}
  {arXiv:1104.2933 [astro-ph.CO]} \BibitemShut {NoStop}%
\bibitem [{\citenamefont {Ali-Haimoud}\ and\ \citenamefont
  {Hirata}(2011)}]{AliHaimoud:2010dx}%
  \BibitemOpen
  \bibfield  {author} {\bibinfo {author} {\bibfnamefont {Y.}~\bibnamefont
  {Ali-Haimoud}}\ and\ \bibinfo {author} {\bibfnamefont {C.~M.}\ \bibnamefont
  {Hirata}},\ }\href {\doibase 10.1103/PhysRevD.83.043513} {\bibfield
  {journal} {\bibinfo  {journal} {Phys. Rev.}\ }\textbf {\bibinfo {volume}
  {D83}},\ \bibinfo {pages} {043513} (\bibinfo {year} {2011})},\ \Eprint
  {http://arxiv.org/abs/1011.3758} {arXiv:1011.3758 [astro-ph.CO]} \BibitemShut
  {NoStop}%
\bibitem [{\citenamefont {Kuo}\ and\ \citenamefont
  {Pantaleone}(1989)}]{Kuo:1989qe}%
  \BibitemOpen
  \bibfield  {author} {\bibinfo {author} {\bibfnamefont {T.-K.}\ \bibnamefont
  {Kuo}}\ and\ \bibinfo {author} {\bibfnamefont {J.~T.}\ \bibnamefont
  {Pantaleone}},\ }\href {\doibase 10.1103/RevModPhys.61.937} {\bibfield
  {journal} {\bibinfo  {journal} {Rev. Mod. Phys.}\ }\textbf {\bibinfo {volume}
  {61}},\ \bibinfo {pages} {937} (\bibinfo {year} {1989})}\BibitemShut
  {NoStop}%
\bibitem [{\citenamefont {Parke}(1986)}]{Parke:1986jy}%
  \BibitemOpen
  \bibfield  {author} {\bibinfo {author} {\bibfnamefont {S.~J.}\ \bibnamefont
  {Parke}},\ }\href {\doibase 10.1103/PhysRevLett.57.1275} {\bibfield
  {journal} {\bibinfo  {journal} {Phys. Rev. Lett.}\ }\textbf {\bibinfo
  {volume} {57}},\ \bibinfo {pages} {1275} (\bibinfo {year}
  {1986})}\BibitemShut {NoStop}%
\bibitem [{\citenamefont {Dasgupta}\ and\ \citenamefont
  {Dighe}(2007)}]{Dasgupta:2005wn}%
  \BibitemOpen
  \bibfield  {author} {\bibinfo {author} {\bibfnamefont {B.}~\bibnamefont
  {Dasgupta}}\ and\ \bibinfo {author} {\bibfnamefont {A.}~\bibnamefont
  {Dighe}},\ }\href {\doibase 10.1103/PhysRevD.75.093002} {\bibfield  {journal}
  {\bibinfo  {journal} {Phys. Rev.}\ }\textbf {\bibinfo {volume} {D75}},\
  \bibinfo {pages} {093002} (\bibinfo {year} {2007})},\ \Eprint
  {http://arxiv.org/abs/hep-ph/0510219} {arXiv:hep-ph/0510219 [hep-ph]}
  \BibitemShut {NoStop}%
\bibitem [{\citenamefont {Friedland}\ and\ \citenamefont
  {Gruzinov}(2006)}]{Friedland:2006ta}%
  \BibitemOpen
  \bibfield  {author} {\bibinfo {author} {\bibfnamefont {A.}~\bibnamefont
  {Friedland}}\ and\ \bibinfo {author} {\bibfnamefont {A.}~\bibnamefont
  {Gruzinov}},\ }\href@noop {} {\  (\bibinfo {year} {2006})},\ \Eprint
  {http://arxiv.org/abs/astro-ph/0607244} {arXiv:astro-ph/0607244 [astro-ph]}
  \BibitemShut {NoStop}%
\bibitem [{\citenamefont {Fogli}\ \emph {et~al.}(2006)\citenamefont {Fogli},
  \citenamefont {Lisi}, \citenamefont {Mirizzi},\ and\ \citenamefont
  {Montanino}}]{Fogli:2006xy}%
  \BibitemOpen
  \bibfield  {author} {\bibinfo {author} {\bibfnamefont {G.~L.}\ \bibnamefont
  {Fogli}}, \bibinfo {author} {\bibfnamefont {E.}~\bibnamefont {Lisi}},
  \bibinfo {author} {\bibfnamefont {A.}~\bibnamefont {Mirizzi}}, \ and\
  \bibinfo {author} {\bibfnamefont {D.}~\bibnamefont {Montanino}},\ }\href
  {\doibase 10.1088/1475-7516/2006/06/012} {\bibfield  {journal} {\bibinfo
  {journal} {JCAP}\ }\textbf {\bibinfo {volume} {0606}},\ \bibinfo {pages}
  {012} (\bibinfo {year} {2006})},\ \Eprint
  {http://arxiv.org/abs/hep-ph/0603033} {arXiv:hep-ph/0603033 [hep-ph]}
  \BibitemShut {NoStop}%
\bibitem [{\citenamefont {{Rice}}(1944)}]{1944BSTJ...23..282R}%
  \BibitemOpen
  \bibfield  {author} {\bibinfo {author} {\bibfnamefont {S.~O.}\ \bibnamefont
  {{Rice}}},\ }\href {\doibase 10.1002/j.1538-7305.1944.tb00874.x} {\bibfield
  {journal} {\bibinfo  {journal} {Bell System Technical Journal}\ }\textbf
  {\bibinfo {volume} {23}},\ \bibinfo {pages} {282} (\bibinfo {year}
  {1944})}\BibitemShut {NoStop}%
\bibitem [{\citenamefont {Lindgren}(2012)}]{lindgren2012stationary}%
  \BibitemOpen
  \bibfield  {author} {\bibinfo {author} {\bibfnamefont {G.}~\bibnamefont
  {Lindgren}},\ }\href {https://books.google.com/books?id=XROYk50wZCMC} {\emph
  {\bibinfo {title} {Stationary Stochastic Processes: Theory and
  Applications}}},\ Chapman \& Hall/CRC Texts in Statistical Science\ (\bibinfo
   {publisher} {Taylor \& Francis},\ \bibinfo {year} {2012})\BibitemShut
  {NoStop}%
\bibitem [{\citenamefont {Kogut}\ \emph {et~al.}(2011)\citenamefont {Kogut}
  \emph {et~al.}}]{Kogut:2011xw}%
  \BibitemOpen
  \bibfield  {author} {\bibinfo {author} {\bibfnamefont {A.}~\bibnamefont
  {Kogut}} \emph {et~al.},\ }\href {\doibase 10.1088/1475-7516/2011/07/025}
  {\bibfield  {journal} {\bibinfo  {journal} {JCAP}\ }\textbf {\bibinfo
  {volume} {1107}},\ \bibinfo {pages} {025} (\bibinfo {year} {2011})},\ \Eprint
  {http://arxiv.org/abs/1105.2044} {arXiv:1105.2044 [astro-ph.CO]} \BibitemShut
  {NoStop}%
\bibitem [{\citenamefont {Davis}\ \emph {et~al.}(1975)\citenamefont {Davis},
  \citenamefont {Goldhaber},\ and\ \citenamefont {Nieto}}]{Davis:1975mn}%
  \BibitemOpen
  \bibfield  {author} {\bibinfo {author} {\bibfnamefont {L.}~\bibnamefont
  {Davis}, \bibfnamefont {Jr.}}, \bibinfo {author} {\bibfnamefont {A.~S.}\
  \bibnamefont {Goldhaber}}, \ and\ \bibinfo {author} {\bibfnamefont {M.~M.}\
  \bibnamefont {Nieto}},\ }\href {\doibase 10.1103/PhysRevLett.35.1402}
  {\bibfield  {journal} {\bibinfo  {journal} {Phys. Rev. Lett.}\ }\textbf
  {\bibinfo {volume} {35}},\ \bibinfo {pages} {1402} (\bibinfo {year}
  {1975})}\BibitemShut {NoStop}%
\bibitem [{\citenamefont {Ahlers}\ \emph {et~al.}(2008)\citenamefont {Ahlers},
  \citenamefont {Jaeckel}, \citenamefont {Redondo},\ and\ \citenamefont
  {Ringwald}}]{Ahlers:2008qc}%
  \BibitemOpen
  \bibfield  {author} {\bibinfo {author} {\bibfnamefont {M.}~\bibnamefont
  {Ahlers}}, \bibinfo {author} {\bibfnamefont {J.}~\bibnamefont {Jaeckel}},
  \bibinfo {author} {\bibfnamefont {J.}~\bibnamefont {Redondo}}, \ and\
  \bibinfo {author} {\bibfnamefont {A.}~\bibnamefont {Ringwald}},\ }\href
  {\doibase 10.1103/PhysRevD.78.075005} {\bibfield  {journal} {\bibinfo
  {journal} {Phys. Rev.}\ }\textbf {\bibinfo {volume} {D78}},\ \bibinfo {pages}
  {075005} (\bibinfo {year} {2008})},\ \Eprint {http://arxiv.org/abs/0807.4143}
  {arXiv:0807.4143 [hep-ph]} \BibitemShut {NoStop}%
\bibitem [{\citenamefont {Battaglieri}\ \emph {et~al.}(2017)\citenamefont
  {Battaglieri} \emph {et~al.}}]{Battaglieri:2017aum}%
  \BibitemOpen
  \bibfield  {author} {\bibinfo {author} {\bibfnamefont {M.}~\bibnamefont
  {Battaglieri}} \emph {et~al.},\ }in\ \href
  {http://lss.fnal.gov/archive/2017/conf/fermilab-conf-17-282-ae-ppd-t.pdf}
  {\emph {\bibinfo {booktitle} {{U.S. Cosmic Visions: New Ideas in Dark Matter
  College Park, MD, USA, March 23-25, 2017}}}}\ (\bibinfo {year} {2017})\
  \Eprint {http://arxiv.org/abs/1707.04591} {arXiv:1707.04591 [hep-ph]}
  \BibitemShut {NoStop}%
\bibitem [{\citenamefont {{Hubble}}(1934)}]{1934ApJ....79....8H}%
  \BibitemOpen
  \bibfield  {author} {\bibinfo {author} {\bibfnamefont {E.}~\bibnamefont
  {{Hubble}}},\ }\href {\doibase 10.1086/143517} {\bibfield  {journal}
  {\bibinfo  {journal} {\apj}\ }\textbf {\bibinfo {volume} {79}},\ \bibinfo
  {pages} {8} (\bibinfo {year} {1934})}\BibitemShut {NoStop}%
\bibitem [{\citenamefont {Coles}\ and\ \citenamefont
  {Jones}(1991)}]{Coles:1991if}%
  \BibitemOpen
  \bibfield  {author} {\bibinfo {author} {\bibfnamefont {P.}~\bibnamefont
  {Coles}}\ and\ \bibinfo {author} {\bibfnamefont {B.}~\bibnamefont {Jones}},\
  }\href@noop {} {\bibfield  {journal} {\bibinfo  {journal} {Mon. Not. Roy.
  Astron. Soc.}\ }\textbf {\bibinfo {volume} {248}},\ \bibinfo {pages} {1}
  (\bibinfo {year} {1991})}\BibitemShut {NoStop}%
\bibitem [{\citenamefont {Kayo}\ \emph {et~al.}(2001)\citenamefont {Kayo},
  \citenamefont {Taruya},\ and\ \citenamefont {Suto}}]{Kayo:2001gu}%
  \BibitemOpen
  \bibfield  {author} {\bibinfo {author} {\bibfnamefont {I.}~\bibnamefont
  {Kayo}}, \bibinfo {author} {\bibfnamefont {A.}~\bibnamefont {Taruya}}, \ and\
  \bibinfo {author} {\bibfnamefont {Y.}~\bibnamefont {Suto}},\ }\href {\doibase
  10.1086/323227} {\bibfield  {journal} {\bibinfo  {journal} {Astrophys. J.}\
  }\textbf {\bibinfo {volume} {561}},\ \bibinfo {pages} {22} (\bibinfo {year}
  {2001})},\ \Eprint {http://arxiv.org/abs/astro-ph/0105218}
  {arXiv:astro-ph/0105218 [astro-ph]} \BibitemShut {NoStop}%
\bibitem [{\citenamefont {Wild}\ \emph {et~al.}(2005)\citenamefont {Wild} \emph
  {et~al.}}]{Wild:2004me}%
  \BibitemOpen
  \bibfield  {author} {\bibinfo {author} {\bibfnamefont {V.}~\bibnamefont
  {Wild}} \emph {et~al.} (\bibinfo {collaboration} {2dFGRS}),\ }\href {\doibase
  10.1111/j.1365-2966.2004.08447.x} {\bibfield  {journal} {\bibinfo  {journal}
  {Mon. Not. Roy. Astron. Soc.}\ }\textbf {\bibinfo {volume} {356}},\ \bibinfo
  {pages} {247} (\bibinfo {year} {2005})},\ \Eprint
  {http://arxiv.org/abs/astro-ph/0404275} {arXiv:astro-ph/0404275 [astro-ph]}
  \BibitemShut {NoStop}%
\bibitem [{\citenamefont {Nelson}\ \emph {et~al.}(2018)\citenamefont {Nelson}
  \emph {et~al.}}]{Nelson:2018uso}%
  \BibitemOpen
  \bibfield  {author} {\bibinfo {author} {\bibfnamefont {D.}~\bibnamefont
  {Nelson}} \emph {et~al.},\ }\href@noop {} {\  (\bibinfo {year} {2018})},\
  \Eprint {http://arxiv.org/abs/1812.05609} {arXiv:1812.05609 [astro-ph.GA]}
  \BibitemShut {NoStop}%
\bibitem [{\citenamefont {McAlpine}\ \emph {et~al.}(2016)\citenamefont
  {McAlpine} \emph {et~al.}}]{McAlpine:2015tma}%
  \BibitemOpen
  \bibfield  {author} {\bibinfo {author} {\bibfnamefont {S.}~\bibnamefont
  {McAlpine}} \emph {et~al.},\ }\href {\doibase 10.1016/j.ascom.2016.02.004}
  {\bibfield  {journal} {\bibinfo  {journal} {Astron. Comput.}\ }\textbf
  {\bibinfo {volume} {15}},\ \bibinfo {pages} {72} (\bibinfo {year} {2016})},\
  \Eprint {http://arxiv.org/abs/1510.01320} {arXiv:1510.01320 [astro-ph.GA]}
  \BibitemShut {NoStop}%
\bibitem [{\citenamefont {McCarthy}\ \emph {et~al.}(2017)\citenamefont
  {McCarthy}, \citenamefont {Schaye}, \citenamefont {Bird},\ and\ \citenamefont
  {Le~Brun}}]{McCarthy:2016mry}%
  \BibitemOpen
  \bibfield  {author} {\bibinfo {author} {\bibfnamefont {I.~G.}\ \bibnamefont
  {McCarthy}}, \bibinfo {author} {\bibfnamefont {J.}~\bibnamefont {Schaye}},
  \bibinfo {author} {\bibfnamefont {S.}~\bibnamefont {Bird}}, \ and\ \bibinfo
  {author} {\bibfnamefont {A.~M.~C.}\ \bibnamefont {Le~Brun}},\ }\href
  {\doibase 10.1093/mnras/stw2792} {\bibfield  {journal} {\bibinfo  {journal}
  {Mon. Not. Roy. Astron. Soc.}\ }\textbf {\bibinfo {volume} {465}},\ \bibinfo
  {pages} {2936} (\bibinfo {year} {2017})},\ \Eprint
  {http://arxiv.org/abs/1603.02702} {arXiv:1603.02702 [astro-ph.CO]}
  \BibitemShut {NoStop}%
\bibitem [{\citenamefont {Genel}\ \emph {et~al.}(2014)\citenamefont {Genel},
  \citenamefont {Vogelsberger}, \citenamefont {Springel}, \citenamefont
  {Sijacki}, \citenamefont {Nelson}, \citenamefont {Snyder}, \citenamefont
  {Rodriguez-Gomez}, \citenamefont {Torrey},\ and\ \citenamefont
  {Hernquist}}]{Genel:2014lma}%
  \BibitemOpen
  \bibfield  {author} {\bibinfo {author} {\bibfnamefont {S.}~\bibnamefont
  {Genel}}, \bibinfo {author} {\bibfnamefont {M.}~\bibnamefont {Vogelsberger}},
  \bibinfo {author} {\bibfnamefont {V.}~\bibnamefont {Springel}}, \bibinfo
  {author} {\bibfnamefont {D.}~\bibnamefont {Sijacki}}, \bibinfo {author}
  {\bibfnamefont {D.}~\bibnamefont {Nelson}}, \bibinfo {author} {\bibfnamefont
  {G.}~\bibnamefont {Snyder}}, \bibinfo {author} {\bibfnamefont
  {V.}~\bibnamefont {Rodriguez-Gomez}}, \bibinfo {author} {\bibfnamefont
  {P.}~\bibnamefont {Torrey}}, \ and\ \bibinfo {author} {\bibfnamefont
  {L.}~\bibnamefont {Hernquist}},\ }\href {\doibase 10.1093/mnras/stu1654}
  {\bibfield  {journal} {\bibinfo  {journal} {Mon. Not. Roy. Astron. Soc.}\
  }\textbf {\bibinfo {volume} {445}},\ \bibinfo {pages} {175} (\bibinfo {year}
  {2014})},\ \Eprint {http://arxiv.org/abs/1405.3749} {arXiv:1405.3749
  [astro-ph.CO]} \BibitemShut {NoStop}%
\bibitem [{\citenamefont {Foreman}\ \emph {et~al.}(2019)\citenamefont
  {Foreman}, \citenamefont {Coulton}, \citenamefont {Villaescusa-Navarro},\
  and\ \citenamefont {Barreira}}]{Foreman:2019ahr}%
  \BibitemOpen
  \bibfield  {author} {\bibinfo {author} {\bibfnamefont {S.}~\bibnamefont
  {Foreman}}, \bibinfo {author} {\bibfnamefont {W.}~\bibnamefont {Coulton}},
  \bibinfo {author} {\bibfnamefont {F.}~\bibnamefont {Villaescusa-Navarro}}, \
  and\ \bibinfo {author} {\bibfnamefont {A.}~\bibnamefont {Barreira}},\
  }\href@noop {} {\  (\bibinfo {year} {2019})},\ \Eprint
  {http://arxiv.org/abs/1910.03597} {arXiv:1910.03597 [astro-ph.CO]}
  \BibitemShut {NoStop}%
\bibitem [{\citenamefont {van Daalen}\ \emph {et~al.}(2020)\citenamefont {van
  Daalen}, \citenamefont {McCarthy},\ and\ \citenamefont
  {Schaye}}]{vanDaalen:2019pst}%
  \BibitemOpen
  \bibfield  {author} {\bibinfo {author} {\bibfnamefont {M.~P.}\ \bibnamefont
  {van Daalen}}, \bibinfo {author} {\bibfnamefont {I.~G.}\ \bibnamefont
  {McCarthy}}, \ and\ \bibinfo {author} {\bibfnamefont {J.}~\bibnamefont
  {Schaye}},\ }\href {\doibase 10.1093/mnras/stz3199} {\bibfield  {journal}
  {\bibinfo  {journal} {Mon. Not. Roy. Astron. Soc.}\ }\textbf {\bibinfo
  {volume} {491}},\ \bibinfo {pages} {2424} (\bibinfo {year} {2020})},\ \Eprint
  {http://arxiv.org/abs/1906.00968} {arXiv:1906.00968 [astro-ph.CO]}
  \BibitemShut {NoStop}%
\bibitem [{\citenamefont {Ivanov}\ \emph {et~al.}(2019)\citenamefont {Ivanov},
  \citenamefont {Kaurov},\ and\ \citenamefont {Sibiryakov}}]{Ivanov:2018lcg}%
  \BibitemOpen
  \bibfield  {author} {\bibinfo {author} {\bibfnamefont {M.~M.}\ \bibnamefont
  {Ivanov}}, \bibinfo {author} {\bibfnamefont {A.~A.}\ \bibnamefont {Kaurov}},
  \ and\ \bibinfo {author} {\bibfnamefont {S.}~\bibnamefont {Sibiryakov}},\
  }\href {\doibase 10.1088/1475-7516/2019/03/009} {\bibfield  {journal}
  {\bibinfo  {journal} {JCAP}\ }\textbf {\bibinfo {volume} {1903}},\ \bibinfo
  {pages} {009} (\bibinfo {year} {2019})},\ \Eprint
  {http://arxiv.org/abs/1811.07913} {arXiv:1811.07913 [astro-ph.CO]}
  \BibitemShut {NoStop}%
\bibitem [{\citenamefont {Valageas}(2002{\natexlab{a}})}]{Valageas:2001zr}%
  \BibitemOpen
  \bibfield  {author} {\bibinfo {author} {\bibfnamefont {P.}~\bibnamefont
  {Valageas}},\ }\href {\doibase 10.1051/0004-6361:20011663} {\bibfield
  {journal} {\bibinfo  {journal} {Astron. Astrophys.}\ }\textbf {\bibinfo
  {volume} {382}},\ \bibinfo {pages} {412} (\bibinfo {year}
  {2002}{\natexlab{a}})},\ \Eprint {http://arxiv.org/abs/astro-ph/0107126}
  {arXiv:astro-ph/0107126 [astro-ph]} \BibitemShut {NoStop}%
\bibitem [{\citenamefont {Valageas}(2002{\natexlab{b}})}]{Valageas:2001td}%
  \BibitemOpen
  \bibfield  {author} {\bibinfo {author} {\bibfnamefont {P.}~\bibnamefont
  {Valageas}},\ }\href {\doibase 10.1051/0004-6361:20011584} {\bibfield
  {journal} {\bibinfo  {journal} {Astron. Astrophys.}\ }\textbf {\bibinfo
  {volume} {382}},\ \bibinfo {pages} {477} (\bibinfo {year}
  {2002}{\natexlab{b}})},\ \Eprint {http://arxiv.org/abs/astro-ph/0109408}
  {arXiv:astro-ph/0109408 [astro-ph]} \BibitemShut {NoStop}%
\bibitem [{\citenamefont {Mather}\ \emph {et~al.}(1999)\citenamefont {Mather},
  \citenamefont {Fixsen}, \citenamefont {Shafer}, \citenamefont {Mosier},\ and\
  \citenamefont {Wilkinson}}]{Mather:1998gm}%
  \BibitemOpen
  \bibfield  {author} {\bibinfo {author} {\bibfnamefont {J.~C.}\ \bibnamefont
  {Mather}}, \bibinfo {author} {\bibfnamefont {D.}~\bibnamefont {Fixsen}},
  \bibinfo {author} {\bibfnamefont {R.}~\bibnamefont {Shafer}}, \bibinfo
  {author} {\bibfnamefont {C.}~\bibnamefont {Mosier}}, \ and\ \bibinfo {author}
  {\bibfnamefont {D.}~\bibnamefont {Wilkinson}},\ }\href {\doibase
  10.1086/306805} {\bibfield  {journal} {\bibinfo  {journal} {Astrophys. J.}\
  }\textbf {\bibinfo {volume} {512}},\ \bibinfo {pages} {511} (\bibinfo {year}
  {1999})},\ \Eprint {http://arxiv.org/abs/astro-ph/9810373}
  {arXiv:astro-ph/9810373} \BibitemShut {NoStop}%
\bibitem [{\citenamefont {Pani}\ \emph {et~al.}(2012)\citenamefont {Pani},
  \citenamefont {Cardoso}, \citenamefont {Gualtieri}, \citenamefont {Berti},\
  and\ \citenamefont {Ishibashi}}]{Pani:2012vp}%
  \BibitemOpen
  \bibfield  {author} {\bibinfo {author} {\bibfnamefont {P.}~\bibnamefont
  {Pani}}, \bibinfo {author} {\bibfnamefont {V.}~\bibnamefont {Cardoso}},
  \bibinfo {author} {\bibfnamefont {L.}~\bibnamefont {Gualtieri}}, \bibinfo
  {author} {\bibfnamefont {E.}~\bibnamefont {Berti}}, \ and\ \bibinfo {author}
  {\bibfnamefont {A.}~\bibnamefont {Ishibashi}},\ }\href {\doibase
  10.1103/PhysRevLett.109.131102} {\bibfield  {journal} {\bibinfo  {journal}
  {Phys. Rev. Lett.}\ }\textbf {\bibinfo {volume} {109}},\ \bibinfo {pages}
  {131102} (\bibinfo {year} {2012})},\ \Eprint {http://arxiv.org/abs/1209.0465}
  {arXiv:1209.0465 [gr-qc]} \BibitemShut {NoStop}%
\bibitem [{\citenamefont {Baryakhtar}\ \emph {et~al.}(2017)\citenamefont
  {Baryakhtar}, \citenamefont {Lasenby},\ and\ \citenamefont
  {Teo}}]{Baryakhtar:2017ngi}%
  \BibitemOpen
  \bibfield  {author} {\bibinfo {author} {\bibfnamefont {M.}~\bibnamefont
  {Baryakhtar}}, \bibinfo {author} {\bibfnamefont {R.}~\bibnamefont {Lasenby}},
  \ and\ \bibinfo {author} {\bibfnamefont {M.}~\bibnamefont {Teo}},\ }\href
  {\doibase 10.1103/PhysRevD.96.035019} {\bibfield  {journal} {\bibinfo
  {journal} {Phys. Rev.}\ }\textbf {\bibinfo {volume} {D96}},\ \bibinfo {pages}
  {035019} (\bibinfo {year} {2017})},\ \Eprint
  {http://arxiv.org/abs/1704.05081} {arXiv:1704.05081 [hep-ph]} \BibitemShut
  {NoStop}%
\bibitem [{\citenamefont {Cardoso}\ \emph {et~al.}(2018)\citenamefont
  {Cardoso}, \citenamefont {Dias}, \citenamefont {Hartnett}, \citenamefont
  {Middleton}, \citenamefont {Pani},\ and\ \citenamefont
  {Santos}}]{Cardoso:2018tly}%
  \BibitemOpen
  \bibfield  {author} {\bibinfo {author} {\bibfnamefont {V.}~\bibnamefont
  {Cardoso}}, \bibinfo {author} {\bibfnamefont {O.~J.~C.}\ \bibnamefont
  {Dias}}, \bibinfo {author} {\bibfnamefont {G.~S.}\ \bibnamefont {Hartnett}},
  \bibinfo {author} {\bibfnamefont {M.}~\bibnamefont {Middleton}}, \bibinfo
  {author} {\bibfnamefont {P.}~\bibnamefont {Pani}}, \ and\ \bibinfo {author}
  {\bibfnamefont {J.~E.}\ \bibnamefont {Santos}},\ }\href {\doibase
  10.1088/1475-7516/2018/03/043} {\bibfield  {journal} {\bibinfo  {journal}
  {JCAP}\ }\textbf {\bibinfo {volume} {1803}},\ \bibinfo {pages} {043}
  (\bibinfo {year} {2018})},\ \Eprint {http://arxiv.org/abs/1801.01420}
  {arXiv:1801.01420 [gr-qc]} \BibitemShut {NoStop}%
\bibitem [{\citenamefont {Becker}\ \emph {et~al.}(2011)\citenamefont {Becker},
  \citenamefont {Bolton}, \citenamefont {Haehnelt},\ and\ \citenamefont
  {Sargent}}]{Becker:2010cu}%
  \BibitemOpen
  \bibfield  {author} {\bibinfo {author} {\bibfnamefont {G.~D.}\ \bibnamefont
  {Becker}}, \bibinfo {author} {\bibfnamefont {J.~S.}\ \bibnamefont {Bolton}},
  \bibinfo {author} {\bibfnamefont {M.~G.}\ \bibnamefont {Haehnelt}}, \ and\
  \bibinfo {author} {\bibfnamefont {W.~L.~W.}\ \bibnamefont {Sargent}},\ }\href
  {\doibase 10.1111/j.1365-2966.2010.17507.x} {\bibfield  {journal} {\bibinfo
  {journal} {Mon. Not. Roy. Astron. Soc.}\ }\textbf {\bibinfo {volume} {410}},\
  \bibinfo {pages} {1096} (\bibinfo {year} {2011})},\ \Eprint
  {http://arxiv.org/abs/1008.2622} {arXiv:1008.2622 [astro-ph.CO]} \BibitemShut
  {NoStop}%
\bibitem [{\citenamefont {Bolton}\ \emph {et~al.}(2014)\citenamefont {Bolton},
  \citenamefont {Becker}, \citenamefont {Haehnelt},\ and\ \citenamefont
  {Viel}}]{Bolton:2013cba}%
  \BibitemOpen
  \bibfield  {author} {\bibinfo {author} {\bibfnamefont {J.~S.}\ \bibnamefont
  {Bolton}}, \bibinfo {author} {\bibfnamefont {G.~D.}\ \bibnamefont {Becker}},
  \bibinfo {author} {\bibfnamefont {M.~G.}\ \bibnamefont {Haehnelt}}, \ and\
  \bibinfo {author} {\bibfnamefont {M.}~\bibnamefont {Viel}},\ }\href {\doibase
  10.1093/mnras/stt2374} {\bibfield  {journal} {\bibinfo  {journal} {Mon. Not.
  Roy. Astron. Soc.}\ }\textbf {\bibinfo {volume} {438}},\ \bibinfo {pages}
  {2499} (\bibinfo {year} {2014})},\ \Eprint {http://arxiv.org/abs/1308.4411}
  {arXiv:1308.4411 [astro-ph.CO]} \BibitemShut {NoStop}%
\bibitem [{\citenamefont {Boera}\ \emph {et~al.}(2014)\citenamefont {Boera},
  \citenamefont {Murphy}, \citenamefont {Becker},\ and\ \citenamefont
  {Bolton}}]{Boera:2014sia}%
  \BibitemOpen
  \bibfield  {author} {\bibinfo {author} {\bibfnamefont {E.}~\bibnamefont
  {Boera}}, \bibinfo {author} {\bibfnamefont {M.~T.}\ \bibnamefont {Murphy}},
  \bibinfo {author} {\bibfnamefont {G.~D.}\ \bibnamefont {Becker}}, \ and\
  \bibinfo {author} {\bibfnamefont {J.~S.}\ \bibnamefont {Bolton}},\ }\href
  {\doibase 10.1093/mnras/stu660} {\bibfield  {journal} {\bibinfo  {journal}
  {Mon. Not. Roy. Astron. Soc.}\ }\textbf {\bibinfo {volume} {441}},\ \bibinfo
  {pages} {1916} (\bibinfo {year} {2014})},\ \Eprint
  {http://arxiv.org/abs/1404.1083} {arXiv:1404.1083 [astro-ph.CO]} \BibitemShut
  {NoStop}%
\bibitem [{\citenamefont {Rorai}\ \emph {et~al.}(2018)\citenamefont {Rorai},
  \citenamefont {Carswell}, \citenamefont {Haehnelt}, \citenamefont {Becker},
  \citenamefont {Bolton},\ and\ \citenamefont {Murphy}}]{Rorai:2017qft}%
  \BibitemOpen
  \bibfield  {author} {\bibinfo {author} {\bibfnamefont {A.}~\bibnamefont
  {Rorai}}, \bibinfo {author} {\bibfnamefont {R.~F.}\ \bibnamefont {Carswell}},
  \bibinfo {author} {\bibfnamefont {M.~G.}\ \bibnamefont {Haehnelt}}, \bibinfo
  {author} {\bibfnamefont {G.~D.}\ \bibnamefont {Becker}}, \bibinfo {author}
  {\bibfnamefont {J.~S.}\ \bibnamefont {Bolton}}, \ and\ \bibinfo {author}
  {\bibfnamefont {M.~T.}\ \bibnamefont {Murphy}},\ }\href {\doibase
  10.1093/mnras/stx2862} {\bibfield  {journal} {\bibinfo  {journal} {Mon. Not.
  Roy. Astron. Soc.}\ }\textbf {\bibinfo {volume} {474}},\ \bibinfo {pages}
  {2871} (\bibinfo {year} {2018})},\ \Eprint {http://arxiv.org/abs/1711.00930}
  {arXiv:1711.00930 [astro-ph.CO]} \BibitemShut {NoStop}%
\bibitem [{\citenamefont {Hiss}\ \emph {et~al.}(2018)\citenamefont {Hiss},
  \citenamefont {Walther}, \citenamefont {Hennawi}, \citenamefont {O\~{n}orbe},
  \citenamefont {O'Meara}, \citenamefont {Rorai},\ and\ \citenamefont
  {Luki\'{c}}}]{Hiss:2017qyw}%
  \BibitemOpen
  \bibfield  {author} {\bibinfo {author} {\bibfnamefont {H.}~\bibnamefont
  {Hiss}}, \bibinfo {author} {\bibfnamefont {M.}~\bibnamefont {Walther}},
  \bibinfo {author} {\bibfnamefont {J.~F.}\ \bibnamefont {Hennawi}}, \bibinfo
  {author} {\bibfnamefont {J.}~\bibnamefont {O\~{n}orbe}}, \bibinfo {author}
  {\bibfnamefont {J.~M.}\ \bibnamefont {O'Meara}}, \bibinfo {author}
  {\bibfnamefont {A.}~\bibnamefont {Rorai}}, \ and\ \bibinfo {author}
  {\bibfnamefont {Z.}~\bibnamefont {Luki\'{c}}},\ }\href {\doibase
  10.3847/1538-4357/aada86} {\bibfield  {journal} {\bibinfo  {journal}
  {Astrophys. J.}\ }\textbf {\bibinfo {volume} {865}},\ \bibinfo {pages} {42}
  (\bibinfo {year} {2018})},\ \Eprint {http://arxiv.org/abs/1710.00700}
  {arXiv:1710.00700 [astro-ph.CO]} \BibitemShut {NoStop}%
\bibitem [{\citenamefont {Walther}\ \emph {et~al.}(2019)\citenamefont
  {Walther}, \citenamefont {O\~{n}orbe}, \citenamefont {Hennawi},\ and\
  \citenamefont {Luki\'{c}}}]{Walther:2018pnn}%
  \BibitemOpen
  \bibfield  {author} {\bibinfo {author} {\bibfnamefont {M.}~\bibnamefont
  {Walther}}, \bibinfo {author} {\bibfnamefont {J.}~\bibnamefont {O\~{n}orbe}},
  \bibinfo {author} {\bibfnamefont {J.~F.}\ \bibnamefont {Hennawi}}, \ and\
  \bibinfo {author} {\bibfnamefont {Z.}~\bibnamefont {Luki\'{c}}},\ }\href
  {\doibase 10.3847/1538-4357/aafad1} {\bibfield  {journal} {\bibinfo
  {journal} {Astrophys. J.}\ }\textbf {\bibinfo {volume} {872}},\ \bibinfo
  {pages} {13} (\bibinfo {year} {2019})},\ \Eprint
  {http://arxiv.org/abs/1808.04367} {arXiv:1808.04367 [astro-ph.CO]}
  \BibitemShut {NoStop}%
\bibitem [{\citenamefont {Mirizzi}\ \emph
  {et~al.}(2009{\natexlab{b}})\citenamefont {Mirizzi}, \citenamefont
  {Redondo},\ and\ \citenamefont {Sigl}}]{Mirizzi:2009nq}%
  \BibitemOpen
  \bibfield  {author} {\bibinfo {author} {\bibfnamefont {A.}~\bibnamefont
  {Mirizzi}}, \bibinfo {author} {\bibfnamefont {J.}~\bibnamefont {Redondo}}, \
  and\ \bibinfo {author} {\bibfnamefont {G.}~\bibnamefont {Sigl}},\ }\href
  {\doibase 10.1088/1475-7516/2009/08/001} {\bibfield  {journal} {\bibinfo
  {journal} {JCAP}\ }\textbf {\bibinfo {volume} {0908}},\ \bibinfo {pages}
  {001} (\bibinfo {year} {2009}{\natexlab{b}})},\ \Eprint
  {http://arxiv.org/abs/0905.4865} {arXiv:0905.4865 [hep-ph]} \BibitemShut
  {NoStop}%
\bibitem [{\citenamefont {Choi}\ \emph {et~al.}(2019)\citenamefont {Choi},
  \citenamefont {Seong},\ and\ \citenamefont {Yun}}]{Choi:2019jwx}%
  \BibitemOpen
  \bibfield  {author} {\bibinfo {author} {\bibfnamefont {K.}~\bibnamefont
  {Choi}}, \bibinfo {author} {\bibfnamefont {H.}~\bibnamefont {Seong}}, \ and\
  \bibinfo {author} {\bibfnamefont {S.}~\bibnamefont {Yun}},\ }\href@noop {} {\
   (\bibinfo {year} {2019})},\ \Eprint {http://arxiv.org/abs/1911.00532}
  {arXiv:1911.00532 [hep-ph]} \BibitemShut {NoStop}%
\bibitem [{\citenamefont {Pospelov}\ \emph {et~al.}(2018)\citenamefont
  {Pospelov}, \citenamefont {Pradler}, \citenamefont {Ruderman},\ and\
  \citenamefont {Urbano}}]{Pospelov:2018kdh}%
  \BibitemOpen
  \bibfield  {author} {\bibinfo {author} {\bibfnamefont {M.}~\bibnamefont
  {Pospelov}}, \bibinfo {author} {\bibfnamefont {J.}~\bibnamefont {Pradler}},
  \bibinfo {author} {\bibfnamefont {J.~T.}\ \bibnamefont {Ruderman}}, \ and\
  \bibinfo {author} {\bibfnamefont {A.}~\bibnamefont {Urbano}},\ }\href
  {\doibase 10.1103/PhysRevLett.121.031103} {\bibfield  {journal} {\bibinfo
  {journal} {Phys. Rev. Lett.}\ }\textbf {\bibinfo {volume} {121}},\ \bibinfo
  {pages} {031103} (\bibinfo {year} {2018})},\ \Eprint
  {http://arxiv.org/abs/1803.07048} {arXiv:1803.07048 [hep-ph]} \BibitemShut
  {NoStop}%
\bibitem [{\citenamefont {Moroi}\ \emph {et~al.}(2018)\citenamefont {Moroi},
  \citenamefont {Nakayama},\ and\ \citenamefont {Tang}}]{Moroi:2018vci}%
  \BibitemOpen
  \bibfield  {author} {\bibinfo {author} {\bibfnamefont {T.}~\bibnamefont
  {Moroi}}, \bibinfo {author} {\bibfnamefont {K.}~\bibnamefont {Nakayama}}, \
  and\ \bibinfo {author} {\bibfnamefont {Y.}~\bibnamefont {Tang}},\ }\href
  {\doibase 10.1016/j.physletb.2018.07.002} {\bibfield  {journal} {\bibinfo
  {journal} {Phys. Lett.}\ }\textbf {\bibinfo {volume} {B783}},\ \bibinfo
  {pages} {301} (\bibinfo {year} {2018})},\ \Eprint
  {http://arxiv.org/abs/1804.10378} {arXiv:1804.10378 [hep-ph]} \BibitemShut
  {NoStop}%
\bibitem [{\citenamefont {Abazajian}\ \emph {et~al.}(2016)\citenamefont
  {Abazajian} \emph {et~al.}}]{Abazajian:2016yjj}%
  \BibitemOpen
  \bibfield  {author} {\bibinfo {author} {\bibfnamefont {K.~N.}\ \bibnamefont
  {Abazajian}} \emph {et~al.} (\bibinfo {collaboration} {CMB-S4}),\ }\href@noop
  {} {\  (\bibinfo {year} {2016})},\ \Eprint {http://arxiv.org/abs/1610.02743}
  {arXiv:1610.02743 [astro-ph.CO]} \BibitemShut {NoStop}%
\bibitem [{\citenamefont {Ade}\ \emph {et~al.}(2019)\citenamefont {Ade} \emph
  {et~al.}}]{Ade:2018sbj}%
  \BibitemOpen
  \bibfield  {author} {\bibinfo {author} {\bibfnamefont {P.}~\bibnamefont
  {Ade}} \emph {et~al.} (\bibinfo {collaboration} {Simons Observatory}),\
  }\href {\doibase 10.1088/1475-7516/2019/02/056} {\bibfield  {journal}
  {\bibinfo  {journal} {JCAP}\ }\textbf {\bibinfo {volume} {1902}},\ \bibinfo
  {pages} {056} (\bibinfo {year} {2019})},\ \Eprint
  {http://arxiv.org/abs/1808.07445} {arXiv:1808.07445 [astro-ph.CO]}
  \BibitemShut {NoStop}%
\bibitem [{\citenamefont {Caputo}\ \emph
  {et~al.}(2020{\natexlab{b}})\citenamefont {Caputo}, \citenamefont {Liu},
  \citenamefont {Mishra-Sharma},\ and\ \citenamefont
  {Ruderman}}]{andrea_caputo_2020_4081407}%
  \BibitemOpen
  \bibfield  {author} {\bibinfo {author} {\bibfnamefont {A.}~\bibnamefont
  {Caputo}}, \bibinfo {author} {\bibfnamefont {H.}~\bibnamefont {Liu}},
  \bibinfo {author} {\bibfnamefont {S.}~\bibnamefont {Mishra-Sharma}}, \ and\
  \bibinfo {author} {\bibfnamefont {J.~T.}\ \bibnamefont {Ruderman}},\ }\href
  {\doibase 10.5281/zenodo.4081407} {\enquote {\bibinfo {title}
  {smsharma/dark-photons-perturbations: oct-2020},}\ } (\bibinfo {year}
  {2020}{\natexlab{b}})\BibitemShut {NoStop}%
\bibitem [{\citenamefont {Price-Whelan}\ \emph {et~al.}(2018)\citenamefont
  {Price-Whelan} \emph {et~al.}}]{Price-Whelan:2018hus}%
  \BibitemOpen
  \bibfield  {author} {\bibinfo {author} {\bibfnamefont {A.~M.}\ \bibnamefont
  {Price-Whelan}} \emph {et~al.},\ }\href {\doibase 10.3847/1538-3881/aabc4f}
  {\bibfield  {journal} {\bibinfo  {journal} {Astron. J.}\ }\textbf {\bibinfo
  {volume} {156}},\ \bibinfo {pages} {123} (\bibinfo {year} {2018})},\ \Eprint
  {http://arxiv.org/abs/1801.02634} {arXiv:1801.02634} \BibitemShut {NoStop}%
\bibitem [{\citenamefont {Robitaille}\ \emph {et~al.}(2013)\citenamefont
  {Robitaille} \emph {et~al.}}]{Robitaille:2013mpa}%
  \BibitemOpen
  \bibfield  {author} {\bibinfo {author} {\bibfnamefont {T.~P.}\ \bibnamefont
  {Robitaille}} \emph {et~al.} (\bibinfo {collaboration} {Astropy}),\ }\href
  {\doibase 10.1051/0004-6361/201322068} {\bibfield  {journal} {\bibinfo
  {journal} {Astron. Astrophys.}\ }\textbf {\bibinfo {volume} {558}},\ \bibinfo
  {pages} {A33} (\bibinfo {year} {2013})},\ \Eprint
  {http://arxiv.org/abs/1307.6212} {arXiv:1307.6212 [astro-ph.IM]} \BibitemShut
  {NoStop}%
\bibitem [{\citenamefont {Lewis}\ \emph {et~al.}(2000)\citenamefont {Lewis},
  \citenamefont {Challinor},\ and\ \citenamefont {Lasenby}}]{Lewis:1999bs}%
  \BibitemOpen
  \bibfield  {author} {\bibinfo {author} {\bibfnamefont {A.}~\bibnamefont
  {Lewis}}, \bibinfo {author} {\bibfnamefont {A.}~\bibnamefont {Challinor}}, \
  and\ \bibinfo {author} {\bibfnamefont {A.}~\bibnamefont {Lasenby}},\ }\href
  {\doibase 10.1086/309179} {\bibfield  {journal} {\bibinfo  {journal}
  {Astrophys. J.}\ }\textbf {\bibinfo {volume} {538}},\ \bibinfo {pages} {473}
  (\bibinfo {year} {2000})},\ \Eprint {http://arxiv.org/abs/astro-ph/9911177}
  {arXiv:astro-ph/9911177 [astro-ph]} \BibitemShut {NoStop}%
\bibitem [{\citenamefont {Lewis}\ and\ \citenamefont
  {Bridle}(2002)}]{Lewis:2002ah}%
  \BibitemOpen
  \bibfield  {author} {\bibinfo {author} {\bibfnamefont {A.}~\bibnamefont
  {Lewis}}\ and\ \bibinfo {author} {\bibfnamefont {S.}~\bibnamefont {Bridle}},\
  }\href {\doibase 10.1103/PhysRevD.66.103511} {\bibfield  {journal} {\bibinfo
  {journal} {Phys. Rev.}\ }\textbf {\bibinfo {volume} {D66}},\ \bibinfo {pages}
  {103511} (\bibinfo {year} {2002})},\ \Eprint
  {http://arxiv.org/abs/astro-ph/0205436} {arXiv:astro-ph/0205436 [astro-ph]}
  \BibitemShut {NoStop}%
\bibitem [{\citenamefont {{Perez}}\ and\ \citenamefont
  {{Granger}}(2007)}]{PER-GRA:2007}%
  \BibitemOpen
  \bibfield  {author} {\bibinfo {author} {\bibfnamefont {F.}~\bibnamefont
  {{Perez}}}\ and\ \bibinfo {author} {\bibfnamefont {B.~E.}\ \bibnamefont
  {{Granger}}},\ }\href {\doibase 10.1109/MCSE.2007.53} {\bibfield  {journal}
  {\bibinfo  {journal} {Computing in Science and Engineering}\ }\textbf
  {\bibinfo {volume} {9}},\ \bibinfo {pages} {21} (\bibinfo {year}
  {2007})}\BibitemShut {NoStop}%
\bibitem [{\citenamefont {Kluyver}\ \emph {et~al.}(2016)\citenamefont {Kluyver}
  \emph {et~al.}}]{Kluyver2016JupyterN}%
  \BibitemOpen
  \bibfield  {author} {\bibinfo {author} {\bibfnamefont {T.}~\bibnamefont
  {Kluyver}} \emph {et~al.},\ }in\ \href@noop {} {\emph {\bibinfo {booktitle}
  {ELPUB}}}\ (\bibinfo {year} {2016})\BibitemShut {NoStop}%
\bibitem [{\citenamefont {Hunter}(2007)}]{Hunter:2007}%
  \BibitemOpen
  \bibfield  {author} {\bibinfo {author} {\bibfnamefont {J.~D.}\ \bibnamefont
  {Hunter}},\ }\href@noop {} {\bibfield  {journal} {\bibinfo  {journal}
  {Computing In Science \& Engineering}\ }\textbf {\bibinfo {volume} {9}},\
  \bibinfo {pages} {90} (\bibinfo {year} {2007})}\BibitemShut {NoStop}%
\bibitem [{\citenamefont {Hand}\ \emph {et~al.}(2018)\citenamefont {Hand},
  \citenamefont {Feng}, \citenamefont {Beutler}, \citenamefont {Li},
  \citenamefont {Modi}, \citenamefont {Seljak},\ and\ \citenamefont
  {Slepian}}]{Hand:2017pqn}%
  \BibitemOpen
  \bibfield  {author} {\bibinfo {author} {\bibfnamefont {N.}~\bibnamefont
  {Hand}}, \bibinfo {author} {\bibfnamefont {Y.}~\bibnamefont {Feng}}, \bibinfo
  {author} {\bibfnamefont {F.}~\bibnamefont {Beutler}}, \bibinfo {author}
  {\bibfnamefont {Y.}~\bibnamefont {Li}}, \bibinfo {author} {\bibfnamefont
  {C.}~\bibnamefont {Modi}}, \bibinfo {author} {\bibfnamefont {U.}~\bibnamefont
  {Seljak}}, \ and\ \bibinfo {author} {\bibfnamefont {Z.}~\bibnamefont
  {Slepian}},\ }\href {\doibase 10.3847/1538-3881/aadae0} {\bibfield  {journal}
  {\bibinfo  {journal} {Astron. J.}\ }\textbf {\bibinfo {volume} {156}},\
  \bibinfo {pages} {160} (\bibinfo {year} {2018})},\ \Eprint
  {http://arxiv.org/abs/1712.05834} {arXiv:1712.05834 [astro-ph.IM]}
  \BibitemShut {NoStop}%
\bibitem [{\citenamefont {{van der Walt}}\ \emph {et~al.}(2011)\citenamefont
  {{van der Walt}}, \citenamefont {{Colbert}},\ and\ \citenamefont
  {{Varoquaux}}}]{numpy:2011}%
  \BibitemOpen
  \bibfield  {author} {\bibinfo {author} {\bibfnamefont {S.}~\bibnamefont {{van
  der Walt}}}, \bibinfo {author} {\bibfnamefont {S.~C.}\ \bibnamefont
  {{Colbert}}}, \ and\ \bibinfo {author} {\bibfnamefont {G.}~\bibnamefont
  {{Varoquaux}}},\ }\href {\doibase 10.1109/MCSE.2011.37} {\bibfield  {journal}
  {\bibinfo  {journal} {Computing in Science and Engineering}\ }\textbf
  {\bibinfo {volume} {13}},\ \bibinfo {pages} {22} (\bibinfo {year} {2011})},\
  \Eprint {http://arxiv.org/abs/1102.1523} {arXiv:1102.1523 [cs.MS]}
  \BibitemShut {NoStop}%
\bibitem [{\citenamefont {Waskom}\ \emph {et~al.}(2017)\citenamefont {Waskom}
  \emph {et~al.}}]{seaborn}%
  \BibitemOpen
  \bibfield  {author} {\bibinfo {author} {\bibfnamefont {M.}~\bibnamefont
  {Waskom}} \emph {et~al.},\ }\href {\doibase 10.5281/zenodo.883859} {\enquote
  {\bibinfo {title} {mwaskom/seaborn: v0.8.1 (september 2017)},}\ } (\bibinfo
  {year} {2017})\BibitemShut {NoStop}%
\bibitem [{\citenamefont {McKinney}(2010)}]{pandas:2010}%
  \BibitemOpen
  \bibfield  {author} {\bibinfo {author} {\bibfnamefont {W.}~\bibnamefont
  {McKinney}},\ }in\ \href@noop {} {\emph {\bibinfo {booktitle} {Proceedings of
  the 9th Python in Science Conference}}},\ \bibinfo {editor} {edited by\
  \bibinfo {editor} {\bibfnamefont {S.}~\bibnamefont {van~der Walt}}\ and\
  \bibinfo {editor} {\bibfnamefont {J.}~\bibnamefont {Millman}}}\ (\bibinfo
  {year} {2010})\ pp.\ \bibinfo {pages} {51 -- 56}\BibitemShut {NoStop}%
\bibitem [{\citenamefont {{Virtanen}}\ \emph {et~al.}(2020)\citenamefont
  {{Virtanen}} \emph {et~al.}}]{2020SciPy-NMeth}%
  \BibitemOpen
  \bibfield  {author} {\bibinfo {author} {\bibfnamefont {P.}~\bibnamefont
  {{Virtanen}}} \emph {et~al.},\ }\href {\doibase
  https://doi.org/10.1038/s41592-019-0686-2} {\bibfield  {journal} {\bibinfo
  {journal} {Nature Methods}\ } (\bibinfo {year} {2020}),\
  https://doi.org/10.1038/s41592-019-0686-2}\BibitemShut {NoStop}%
\bibitem [{\citenamefont {da~Costa-Luis}(2019)}]{da2019tqdm}%
  \BibitemOpen
  \bibfield  {author} {\bibinfo {author} {\bibfnamefont {C.~O.}\ \bibnamefont
  {da~Costa-Luis}},\ }\href@noop {} {\bibfield  {journal} {\bibinfo  {journal}
  {JOSS}\ }\textbf {\bibinfo {volume} {4}},\ \bibinfo {pages} {1277} (\bibinfo
  {year} {2019})}\BibitemShut {NoStop}%
\bibitem [{\citenamefont {Fixsen}\ \emph {et~al.}(1998)\citenamefont {Fixsen},
  \citenamefont {Bennett},\ and\ \citenamefont {Mather}}]{Fixsen:1998js}%
  \BibitemOpen
  \bibfield  {author} {\bibinfo {author} {\bibfnamefont {D.~J.}\ \bibnamefont
  {Fixsen}}, \bibinfo {author} {\bibfnamefont {C.~L.}\ \bibnamefont {Bennett}},
  \ and\ \bibinfo {author} {\bibfnamefont {J.~C.}\ \bibnamefont {Mather}},\
  }\href@noop {} {\bibfield  {journal} {\bibinfo  {journal} {Submitted to:
  Astrophys. J.}\ } (\bibinfo {year} {1998})},\ \Eprint
  {http://arxiv.org/abs/astro-ph/9810466} {arXiv:astro-ph/9810466 [astro-ph]}
  \BibitemShut {NoStop}%
\bibitem [{\citenamefont {Bernardeau}\ and\ \citenamefont
  {Reimberg}(2016)}]{Bernardeau:2015khs}%
  \BibitemOpen
  \bibfield  {author} {\bibinfo {author} {\bibfnamefont {F.}~\bibnamefont
  {Bernardeau}}\ and\ \bibinfo {author} {\bibfnamefont {P.}~\bibnamefont
  {Reimberg}},\ }\href {\doibase 10.1103/PhysRevD.94.063520} {\bibfield
  {journal} {\bibinfo  {journal} {Phys. Rev.}\ }\textbf {\bibinfo {volume}
  {D94}},\ \bibinfo {pages} {063520} (\bibinfo {year} {2016})},\ \Eprint
  {http://arxiv.org/abs/1511.08641} {arXiv:1511.08641 [astro-ph.CO]}
  \BibitemShut {NoStop}%
\bibitem [{\citenamefont {Uhlemann}\ \emph {et~al.}(2016)\citenamefont
  {Uhlemann}, \citenamefont {Codis}, \citenamefont {Pichon}, \citenamefont
  {Bernardeau},\ and\ \citenamefont {Reimberg}}]{Uhlemann:2015npz}%
  \BibitemOpen
  \bibfield  {author} {\bibinfo {author} {\bibfnamefont {C.}~\bibnamefont
  {Uhlemann}}, \bibinfo {author} {\bibfnamefont {S.}~\bibnamefont {Codis}},
  \bibinfo {author} {\bibfnamefont {C.}~\bibnamefont {Pichon}}, \bibinfo
  {author} {\bibfnamefont {F.}~\bibnamefont {Bernardeau}}, \ and\ \bibinfo
  {author} {\bibfnamefont {P.}~\bibnamefont {Reimberg}},\ }\href {\doibase
  10.1093/mnras/stw1074} {\bibfield  {journal} {\bibinfo  {journal} {Mon. Not.
  Roy. Astron. Soc.}\ }\textbf {\bibinfo {volume} {460}},\ \bibinfo {pages}
  {1529} (\bibinfo {year} {2016})},\ \Eprint {http://arxiv.org/abs/1512.05793}
  {arXiv:1512.05793 [astro-ph.CO]} \BibitemShut {NoStop}%
\bibitem [{\citenamefont {Betancort-Rijo}\ and\ \citenamefont
  {Lopez-Corredoira}(2002)}]{BetancortRijo:2001ge}%
  \BibitemOpen
  \bibfield  {author} {\bibinfo {author} {\bibfnamefont {J.}~\bibnamefont
  {Betancort-Rijo}}\ and\ \bibinfo {author} {\bibfnamefont {M.}~\bibnamefont
  {Lopez-Corredoira}},\ }\href {\doibase 10.1086/338328} {\bibfield  {journal}
  {\bibinfo  {journal} {Astrophys. J.}\ }\textbf {\bibinfo {volume} {566}},\
  \bibinfo {pages} {623} (\bibinfo {year} {2002})},\ \Eprint
  {http://arxiv.org/abs/astro-ph/0110624} {arXiv:astro-ph/0110624 [astro-ph]}
  \BibitemShut {NoStop}%
\bibitem [{\citenamefont {Lam}\ and\ \citenamefont {Sheth}(2008)}]{Lam:2007qw}%
  \BibitemOpen
  \bibfield  {author} {\bibinfo {author} {\bibfnamefont {T.~Y.}\ \bibnamefont
  {Lam}}\ and\ \bibinfo {author} {\bibfnamefont {R.~K.}\ \bibnamefont
  {Sheth}},\ }\href {\doibase 10.1111/j.1365-2966.2008.13038.x} {\bibfield
  {journal} {\bibinfo  {journal} {Mon. Not. Roy. Astron. Soc.}\ }\textbf
  {\bibinfo {volume} {386}},\ \bibinfo {pages} {407} (\bibinfo {year}
  {2008})},\ \Eprint {http://arxiv.org/abs/0711.5029} {arXiv:0711.5029
  [astro-ph]} \BibitemShut {NoStop}%
\bibitem [{\citenamefont {Adermann}\ \emph {et~al.}(2018)\citenamefont
  {Adermann}, \citenamefont {Elahi}, \citenamefont {Lewis},\ and\ \citenamefont
  {Power}}]{Adermann:2018jba}%
  \BibitemOpen
  \bibfield  {author} {\bibinfo {author} {\bibfnamefont {E.}~\bibnamefont
  {Adermann}}, \bibinfo {author} {\bibfnamefont {P.~J.}\ \bibnamefont {Elahi}},
  \bibinfo {author} {\bibfnamefont {G.~F.}\ \bibnamefont {Lewis}}, \ and\
  \bibinfo {author} {\bibfnamefont {C.}~\bibnamefont {Power}},\ }\href
  {\doibase 10.1093/mnras/sty1824} {\bibfield  {journal} {\bibinfo  {journal}
  {Mon. Not. Roy. Astron. Soc.}\ }\textbf {\bibinfo {volume} {479}},\ \bibinfo
  {pages} {4861} (\bibinfo {year} {2018})},\ \Eprint
  {http://arxiv.org/abs/1807.02938} {arXiv:1807.02938 [astro-ph.CO]}
  \BibitemShut {NoStop}%
\bibitem [{\citenamefont {Dekel}\ and\ \citenamefont
  {Lahav}(1999)}]{Dekel:1998eq}%
  \BibitemOpen
  \bibfield  {author} {\bibinfo {author} {\bibfnamefont {A.}~\bibnamefont
  {Dekel}}\ and\ \bibinfo {author} {\bibfnamefont {O.}~\bibnamefont {Lahav}},\
  }\href {\doibase 10.1086/307428} {\bibfield  {journal} {\bibinfo  {journal}
  {Astrophys. J.}\ }\textbf {\bibinfo {volume} {520}},\ \bibinfo {pages} {24}
  (\bibinfo {year} {1999})},\ \Eprint {http://arxiv.org/abs/astro-ph/9806193}
  {arXiv:astro-ph/9806193 [astro-ph]} \BibitemShut {NoStop}%
\bibitem [{\citenamefont {Hurtado-Gil}\ \emph {et~al.}(2017)\citenamefont
  {Hurtado-Gil}, \citenamefont {Mart{\'\i}nez}, \citenamefont {Arnalte-Mur},
  \citenamefont {Pons-Border{\'\i}a}, \citenamefont {Pareja-Flores},\ and\
  \citenamefont {Paredes}}]{Hurtado-Gil:2017dbm}%
  \BibitemOpen
  \bibfield  {author} {\bibinfo {author} {\bibfnamefont {L.}~\bibnamefont
  {Hurtado-Gil}}, \bibinfo {author} {\bibfnamefont {V.~J.}\ \bibnamefont
  {Mart{\'\i}nez}}, \bibinfo {author} {\bibfnamefont {P.}~\bibnamefont
  {Arnalte-Mur}}, \bibinfo {author} {\bibfnamefont {M.~J.}\ \bibnamefont
  {Pons-Border{\'\i}a}}, \bibinfo {author} {\bibfnamefont {C.}~\bibnamefont
  {Pareja-Flores}}, \ and\ \bibinfo {author} {\bibfnamefont {S.}~\bibnamefont
  {Paredes}},\ }\href {\doibase 10.1051/0004-6361/201629097} {\bibfield
  {journal} {\bibinfo  {journal} {Astron. Astrophys.}\ }\textbf {\bibinfo
  {volume} {601}},\ \bibinfo {pages} {A40} (\bibinfo {year} {2017})},\ \Eprint
  {http://arxiv.org/abs/1703.01087} {arXiv:1703.01087 [astro-ph.CO]}
  \BibitemShut {NoStop}%
\bibitem [{\citenamefont {Mead}\ \emph {et~al.}(2016)\citenamefont {Mead},
  \citenamefont {Heymans}, \citenamefont {Lombriser}, \citenamefont {Peacock},
  \citenamefont {Steele},\ and\ \citenamefont {Winther}}]{Mead:2016zqy}%
  \BibitemOpen
  \bibfield  {author} {\bibinfo {author} {\bibfnamefont {A.}~\bibnamefont
  {Mead}}, \bibinfo {author} {\bibfnamefont {C.}~\bibnamefont {Heymans}},
  \bibinfo {author} {\bibfnamefont {L.}~\bibnamefont {Lombriser}}, \bibinfo
  {author} {\bibfnamefont {J.}~\bibnamefont {Peacock}}, \bibinfo {author}
  {\bibfnamefont {O.}~\bibnamefont {Steele}}, \ and\ \bibinfo {author}
  {\bibfnamefont {H.}~\bibnamefont {Winther}},\ }\href {\doibase
  10.1093/mnras/stw681} {\bibfield  {journal} {\bibinfo  {journal} {Mon. Not.
  Roy. Astron. Soc.}\ }\textbf {\bibinfo {volume} {459}},\ \bibinfo {pages}
  {1468} (\bibinfo {year} {2016})},\ \Eprint {http://arxiv.org/abs/1602.02154}
  {arXiv:1602.02154 [astro-ph.CO]} \BibitemShut {NoStop}%
\bibitem [{\citenamefont {Zeldovich}\ \emph {et~al.}(1982)\citenamefont
  {Zeldovich}, \citenamefont {Einasto},\ and\ \citenamefont
  {Shandarin}}]{Zeldovich:1982zz}%
  \BibitemOpen
  \bibfield  {author} {\bibinfo {author} {\bibfnamefont {{\relax Ya}.~B.}\
  \bibnamefont {Zeldovich}}, \bibinfo {author} {\bibfnamefont {J.}~\bibnamefont
  {Einasto}}, \ and\ \bibinfo {author} {\bibfnamefont {S.~F.}\ \bibnamefont
  {Shandarin}},\ }\href {\doibase 10.1038/300407a0} {\bibfield  {journal}
  {\bibinfo  {journal} {Nature}\ }\textbf {\bibinfo {volume} {300}},\ \bibinfo
  {pages} {407} (\bibinfo {year} {1982})}\BibitemShut {NoStop}%
\bibitem [{\citenamefont {Pisani}\ \emph {et~al.}(2019)\citenamefont {Pisani}
  \emph {et~al.}}]{Pisani:2019cvo}%
  \BibitemOpen
  \bibfield  {author} {\bibinfo {author} {\bibfnamefont {A.}~\bibnamefont
  {Pisani}} \emph {et~al.},\ }\href@noop {} {\  (\bibinfo {year} {2019})},\
  \Eprint {http://arxiv.org/abs/1903.05161} {arXiv:1903.05161 [astro-ph.CO]}
  \BibitemShut {NoStop}%
\bibitem [{\citenamefont {Einasto}\ \emph {et~al.}(2011)\citenamefont {Einasto}
  \emph {et~al.}}]{Einasto:2011eu}%
  \BibitemOpen
  \bibfield  {author} {\bibinfo {author} {\bibfnamefont {J.}~\bibnamefont
  {Einasto}} \emph {et~al.},\ }\href {\doibase 10.1051/0004-6361/201117248}
  {\bibfield  {journal} {\bibinfo  {journal} {Astron. Astrophys.}\ }\textbf
  {\bibinfo {volume} {534}},\ \bibinfo {pages} {A128} (\bibinfo {year}
  {2011})},\ \Eprint {http://arxiv.org/abs/1105.2464} {arXiv:1105.2464
  [astro-ph.CO]} \BibitemShut {NoStop}%
\bibitem [{\citenamefont {Jennings}\ \emph {et~al.}(2013)\citenamefont
  {Jennings}, \citenamefont {Li},\ and\ \citenamefont {Hu}}]{Jennings:2013nsa}%
  \BibitemOpen
  \bibfield  {author} {\bibinfo {author} {\bibfnamefont {E.}~\bibnamefont
  {Jennings}}, \bibinfo {author} {\bibfnamefont {Y.}~\bibnamefont {Li}}, \ and\
  \bibinfo {author} {\bibfnamefont {W.}~\bibnamefont {Hu}},\ }\href {\doibase
  10.1093/mnras/stt1169} {\  (\bibinfo {year} {2013}),\
  10.1093/mnras/stt1169},\ \bibinfo {note} {[Mon. Not. Roy. Astron.
  Soc.434,2167(2013)]},\ \Eprint {http://arxiv.org/abs/1304.6087}
  {arXiv:1304.6087 [astro-ph.CO]} \BibitemShut {NoStop}%
\bibitem [{\citenamefont {Chan}\ \emph {et~al.}(2014)\citenamefont {Chan},
  \citenamefont {Hamaus},\ and\ \citenamefont {Desjacques}}]{Chan:2014qka}%
  \BibitemOpen
  \bibfield  {author} {\bibinfo {author} {\bibfnamefont {K.~C.}\ \bibnamefont
  {Chan}}, \bibinfo {author} {\bibfnamefont {N.}~\bibnamefont {Hamaus}}, \ and\
  \bibinfo {author} {\bibfnamefont {V.}~\bibnamefont {Desjacques}},\ }\href
  {\doibase 10.1103/PhysRevD.90.103521} {\bibfield  {journal} {\bibinfo
  {journal} {Phys. Rev.}\ }\textbf {\bibinfo {volume} {D90}},\ \bibinfo {pages}
  {103521} (\bibinfo {year} {2014})},\ \Eprint {http://arxiv.org/abs/1409.3849}
  {arXiv:1409.3849 [astro-ph.CO]} \BibitemShut {NoStop}%
\bibitem [{\citenamefont {Plionis}\ and\ \citenamefont
  {Basilakos}(2002)}]{Plionis:2001gr}%
  \BibitemOpen
  \bibfield  {author} {\bibinfo {author} {\bibfnamefont {M.}~\bibnamefont
  {Plionis}}\ and\ \bibinfo {author} {\bibfnamefont {S.}~\bibnamefont
  {Basilakos}},\ }\href {\doibase 10.1046/j.1365-8711.2002.05069.x} {\bibfield
  {journal} {\bibinfo  {journal} {Mon. Not. Roy. Astron. Soc.}\ }\textbf
  {\bibinfo {volume} {330}},\ \bibinfo {pages} {399} (\bibinfo {year}
  {2002})},\ \Eprint {http://arxiv.org/abs/astro-ph/0106491}
  {arXiv:astro-ph/0106491 [astro-ph]} \BibitemShut {NoStop}%
\bibitem [{\citenamefont {Schuster}\ \emph {et~al.}(2019)\citenamefont
  {Schuster}, \citenamefont {Hamaus}, \citenamefont {Pisani}, \citenamefont
  {Carbone}, \citenamefont {Kreisch}, \citenamefont {Pollina},\ and\
  \citenamefont {Weller}}]{Schuster:2019hyl}%
  \BibitemOpen
  \bibfield  {author} {\bibinfo {author} {\bibfnamefont {N.}~\bibnamefont
  {Schuster}}, \bibinfo {author} {\bibfnamefont {N.}~\bibnamefont {Hamaus}},
  \bibinfo {author} {\bibfnamefont {A.}~\bibnamefont {Pisani}}, \bibinfo
  {author} {\bibfnamefont {C.}~\bibnamefont {Carbone}}, \bibinfo {author}
  {\bibfnamefont {C.~D.}\ \bibnamefont {Kreisch}}, \bibinfo {author}
  {\bibfnamefont {G.}~\bibnamefont {Pollina}}, \ and\ \bibinfo {author}
  {\bibfnamefont {J.}~\bibnamefont {Weller}},\ }\href {\doibase
  10.1088/1475-7516/2019/12/055} {\bibfield  {journal} {\bibinfo  {journal}
  {JCAP}\ }\textbf {\bibinfo {volume} {1912}},\ \bibinfo {pages} {055}
  (\bibinfo {year} {2019})},\ \Eprint {http://arxiv.org/abs/1905.00436}
  {arXiv:1905.00436 [astro-ph.CO]} \BibitemShut {NoStop}%
\bibitem [{\citenamefont {Aghanim}\ \emph {et~al.}(2018)\citenamefont {Aghanim}
  \emph {et~al.}}]{Aghanim:2018eyx}%
  \BibitemOpen
  \bibfield  {author} {\bibinfo {author} {\bibfnamefont {N.}~\bibnamefont
  {Aghanim}} \emph {et~al.} (\bibinfo {collaboration} {Planck}),\ }\href@noop
  {} {\  (\bibinfo {year} {2018})},\ \Eprint {http://arxiv.org/abs/1807.06209}
  {arXiv:1807.06209 [astro-ph.CO]} \BibitemShut {NoStop}%
\bibitem [{\citenamefont {Witte}\ \emph {et~al.}(2020)\citenamefont {Witte},
  \citenamefont {Rosauro-Alcaraz}, \citenamefont {McDermott},\ and\
  \citenamefont {Poulin}}]{Witte:2020rvb}%
  \BibitemOpen
  \bibfield  {author} {\bibinfo {author} {\bibfnamefont {S.~J.}\ \bibnamefont
  {Witte}}, \bibinfo {author} {\bibfnamefont {S.}~\bibnamefont
  {Rosauro-Alcaraz}}, \bibinfo {author} {\bibfnamefont {S.~D.}\ \bibnamefont
  {McDermott}}, \ and\ \bibinfo {author} {\bibfnamefont {V.}~\bibnamefont
  {Poulin}},\ }\href@noop {} {\  (\bibinfo {year} {2020})},\ \Eprint
  {http://arxiv.org/abs/2003.13698} {arXiv:2003.13698 [astro-ph.CO]}
  \BibitemShut {NoStop}%
\end{thebibliography}%

\end{document}